\documentclass[twocolumn]{aastex62}
\usepackage{amsmath}
\hypersetup{
	colorlinks	= true,
	linkcolor	= red,
	urlcolor	= cyan,
	citecolor	= blue
}
\usepackage{amssymb}
\usepackage{lipsum}
\usepackage{multirow}
\usepackage[version=4]{mhchem}
\usepackage{ragged2e}
\usepackage{gensymb}
\usepackage{tikz}
\usepackage{graphics}
\usepackage[figure,figure*]{hypcap}
\usepackage[para,online,flushleft]{threeparttable}
\usepackage{comment}
\usepackage{tablefootnote}
\usepackage{subscript}
\usepackage{graphicx}
\usepackage{subfigure}
\usepackage{booktabs}

\makeatletter
\newcounter{reaction}
\renewcommand\thereaction{R\arabic{reaction}}
\@addtoreset{reaction}{chapter}
\newcommand\reactiontag%
{\refstepcounter{reaction}\tag{\thereaction}}
\newcommand\reaction@[2][]%
{\begin{equation}\ce{#2}%
\ifx\@empty#1\@empty\else\label{#1}\fi%
\reactiontag\end{equation}}
\newcommand\reaction@nonumber[1]%
{\begin{equation*}\ce{#1}\end{equation*}}
\newcommand\reaction%
{\@ifstar{\reaction@nonumber}{\reaction@}}
\makeatother

\graphicspath{{./images/}}

\newcommand{\exotr}{\mbox{\texttt{ExoTR}}}
\newcommand{\aura}{\mbox{\texttt{AURA}}}

\turnoffediting

\shortauthors{Hu et al.}

\begin{document}

    \title{A water-rich interior in the temperate sub-Neptune K2-18~b revealed by JWST}
	
	\correspondingauthor{Renyu Hu}
	\email{renyu.hu@jpl.nasa.gov \\ @2025 California Institute of Technology. \\ Government sponsorship acknowledged.}
	
	\author[0000-0003-2215-8485]{Renyu Hu}
	\affiliation{Jet Propulsion Laboratory, California Institute of Technology, Pasadena, CA 91109, USA}
	\affiliation{Division of Geological and Planetary Sciences, California Institute of Technology, Pasadena, CA 91125, USA}

    \author[0000-0003-3355-1223]{Aaron Bello-Arufe}
    \altaffiliation{These three authors contributed equally to this work}
    \affiliation{Jet Propulsion Laboratory, California Institute of Technology, Pasadena, CA 91109, USA}

    \author[0000-0002-4675-9069]{Armen Tokadjian}
    \altaffiliation{These three authors contributed equally to this work}
    \affiliation{Jet Propulsion Laboratory, California Institute of Technology, Pasadena, CA 91109, USA}

    \author[0000-0002-1551-2610]{Jeehyun Yang}
    \altaffiliation{These three authors contributed equally to this work}
    \affiliation{Jet Propulsion Laboratory, California Institute of Technology, Pasadena, CA 91109, USA}
    \affiliation{Division of Geological and Planetary Sciences, California Institute of Technology, Pasadena, CA 91125, USA}
    
    \author[0000-0002-1830-8260]{Mario Damiano}
    \affiliation{Jet Propulsion Laboratory, California Institute of Technology, Pasadena, CA 91109, USA}

    \author[0000-0001-6809-3520]{Pierre-Alexis Roy}
    \affiliation{Department of Physics and Trottier Institute for Research on Exoplanets, Université de Montréal, Montreal, QC H3C 3J7, Canada}

    \author[0000-0002-2195-735X]{Louis-Philippe Coulombe}
    \affiliation{Department of Physics and Trottier Institute for Research on Exoplanets, Université de Montréal, Montreal, QC H3C 3J7, Canada}

    \author[0000-0002-4869-000X]{Nikku Madhusudhan}
    \affiliation{Institute of Astronomy, University of Cambridge, Madingley Road, Cambridge CB3 0HA, UK}

    \author[0000-0001-6839-4569]{Savvas Constantinou}
    \affiliation{Institute of Astronomy, University of Cambridge, Madingley Road, Cambridge CB3 0HA, UK}

    \author[0000-0001-5578-1498]{Björn Benneke}
    \affiliation{Department of Earth, Planetary, and Space Sciences, University of California, Los Angeles, Los Angeles, CA, USA}
    \affiliation{Department of Physics and Trottier Institute for Research on Exoplanets, Université de Montréal, Montreal, QC H3C 3J7, Canada}
	
	\begin{abstract}

Temperate sub-Neptunes are compelling targets for detecting liquid-water oceans beyond the Solar System. If water-rich and lacking massive hydrogen-helium envelopes, these planets could sustain liquid layers beneath their atmospheres despite sizes larger than Earth. Previous observations of the temperate sub-Neptune K2-18~b revealed an H$_2$-dominated atmosphere rich in CH$_4$, with moderate evidence for CO$_2$ and tentative signs of dimethyl sulfide (DMS). Here we present four new JWST/NIRSpec transit observations of K2-18~b. The resulting high-precision transmission spectrum robustly detects both CH$_4$ and CO$_2$, precisely measuring their abundances and firmly establishing the planet's water-rich nature -- either a thick envelope with $\geq10\%$ H$_2$O by volume or a thin atmosphere above a liquid-water ocean.
The spectrum reveals no detectable H$_2$O, NH$_3$, or CO. The absence of atmospheric water vapor suggests an efficient cold trap, while the nondetections of NH$_3$ and CO support the scenario of a small H$_2$-rich atmosphere overlying a liquid reservoir. However, alternative models that include these gases can also reproduce the spectrum within uncertainties, highlighting the need for deeper observations. The spectrum only contains marginal signals of DMS, methyl mercaptan (\ce{CH3SH}), and nitrous oxide (\ce{N2O}), with none exceeding $3\sigma$ in model preference and all falling below $\sim2\sigma$ without imposing a strong super-Rayleigh haze. Meanwhile, our self-consistent photochemical models show that DMS and \ce{CH3SH} may form abiotically in massive H$_2$-rich atmospheres of high metallicity, making it important to consider additional indicators for their potential use as biosignatures.
K2-18~b -- a cool, water-rich world -- stands out as one of the most promising temperate sub-Neptunes for exploring the emergence of liquid-water environments in non-Earth-like planets, motivating further characterization of its atmosphere and interior.

	\end{abstract}
	
	
\section{Introduction} \label{sec:intro}
	
The discovery of extrasolar planets has opened the door to the identification of planets with liquid-water oceans shrouded by thin atmospheres, extending the search for habitability beyond Earth. This pursuit of bona fide ocean worlds is a central driver of exoplanet exploration. While Earth-sized planets with Earth-like temperatures are natural targets, exoplanet surveys have revealed a diverse and ubiquitous population of planets significantly larger than Earth and richer in volatiles \citep[e.g.,][]{petigura2022california}. Among these, some sub-Neptune-sized planets are potentially water-rich \citep[e.g.,][]{venturini2020nature,luque2022density} and could host liquid-water oceans if they receive insolation comparable to or lower than Earth's \citep[e.g., the ``hycean world'' concept,][]{madhusudhan2021habitability,hu2021unveiling}. These ``temperate sub-Neptunes'' are especially compelling for study with JWST because they not only serve as laboratories for understanding atmospheric physics and chemistry in low-temperature planetary environments, but also offer a promising pathway to detect and characterize liquid-water oceans on exoplanets.

K2-18~b -- a planet with a mass of 8.6 $M_{\oplus}$ and a radius of 2.6 $R_{\oplus}$ \citep{montet2015,benneke2017,cloutier2017,benneke2019water} orbiting an M2.5V star with slightly supersolar elemental abundances \citep{hejazi2024high} -- offers a unique opportunity to conduct detailed characterization of temperate sub-Neptunes. Internal structure models indicate that the planet harbors substantial volatile-rich layers, although it remains uncertain whether these layers are predominantly composed of water or \ce{H2}/He \citep[e.g.,][]{madhusudhan2020interior,mousis2020irradiated}. Among temperate sub-Neptunes favorable for transit observations, K2-18~b stands out for its relatively low insolation -- lower, for example, than that of TOI-270~d \citep{benneke2024jwst}. According to planetary climate models, this lower insolation provides a more promising possibility for the planet to host a liquid-water ocean \citep{innes2023runaway,leconte20243d}.

Transmission spectra of K2-18~b observed by the Hubble Space Telescope revealed spectral features in the $1.1-1.7$ $\mu$m range. Atmospheric retrievals at that time consistently attributed these features to \ce{H2O} vapor absorption in an atmosphere dominated by \ce{H2} \citep{tsiaras2019water,benneke2019water,madhusudhan2020interior}. However, subsequent self-consistent atmospheric models suggested that \ce{CH4}, rather than \ce{H2O}, was the primary contributor to the spectral variations detected by Hubble \citep{Blain2020,hu2021photochemistry,hu2021unveiling,bezard2022methane}.

Initial JWST observations supported the interpretation of self-consistent atmospheric models for K2-18~b. The resulting transmission spectrum, spanning $0.9–5.2\ \mu$m, revealed the presence of \ce{CH4}, suggested the presence of \ce{CO2}, and placed upper limits on \ce{H2O} and \ce{NH3} in an \ce{H2}-rich atmosphere \citep{madhusudhan2023carbon}. Subsequent analyses of the same data showed that the significance of the \ce{CO2} detection is sensitive to specific assumptions in data reduction and spectral retrieval \citep{schmidt2025comprehensive}. In addition, \cite{madhusudhan2023carbon} proposed a tentative signal of dimethyl sulfide (DMS) -- a gas produced primarily by life on Earth. The existence of this gas appeared to be supported by follow-up observations covering the $5–12\ \mu$m range \citep{madhusudhan2025miri}, but the signal could also be attributed to a wide range of other organic molecules \citep{pica2025systematic}. However, this tentative detection has not been corroborated by independent studies, which challenged the statistical significance of any signal \citep{taylor2025there,welbanks2025challenges,luque2025}.

Before any credible interpretation of potential biogenic gases in K2-18~b's atmosphere can be made, it is essential to first assess the planet’s potential habitability -- specifically, whether it harbors a liquid-water ocean. To this end, self-consistent atmospheric models have shown that transmission spectra can provide insights into the planet’s internal composition and constrain the presence of a massive versus shallow atmosphere. In a massive \ce{H2}-dominated atmosphere, the dominant carriers of oxygen, carbon, and nitrogen should be \ce{H2O}, \ce{CH4}, and \ce{NH3}, respectively, along with \ce{CO} and \ce{CO2} at high metallicity and \ce{HCN} produced via photochemistry \citep{hu2021photochemistry,yu2021identify,tsai2021inferring}. In contrast, a small H$_2$-dominated atmosphere implies a massive, ice-rich interior based on the planet’s mass and radius. To maintain a distinct atmosphere above the ice layer, the interface must be in the liquid phase -- otherwise, the atmosphere and interior would mix into a single, undifferentiated envelope \citep{gupta2025miscibility}. This atmosphere would lack \ce{NH3}, either due to dissolution into liquid water \citep{hu2021unveiling} or destruction by photochemistry \citep{yu2021identify,tsai2021inferring}. A water-rich interior would also favor \ce{CO2} as the dominant carbon carrier \citep{hu2021unveiling}. The initial JWST observations, including the suggestion of \ce{CO2} and the nondetection of \ce{NH3}, aligned with predictions for a small atmosphere overlying a water-rich interior \citep[see, for example, the characterization roadmap in][]{hu2021unveiling}.

Several challenges to this interpretation have been raised, along with alternative scenarios. First, updated atmospheric chemistry models suggested that photolysis of \ce{CO2} in a small \ce{H2}-dominated atmosphere may not generate sufficient \ce{CH4} to match observations \citep{wogan2024jwst}. However, the predicted abundance of \ce{CH4} depends strongly on photochemical model assumptions, and the high detected levels of \ce{CH4} could alternatively be explained as either a primordial remnant \citep{yu2021identify,bergin2023exoplanet,cooke2024considerations} or the result of biogenic production \citep{wogan2024jwst}. Second, maintaining a liquid-water ocean below the runaway greenhouse threshold would likely require a high planetary albedo \citep{innes2023runaway,leconte20243d} and a finely tuned, small \ce{H2} atmosphere \citep{koll2019hot,scheucher2020consistently,hu2021unveiling}. Strong absorption of \ce{CH4} may imply that the albedo of the planet cannot be arbitrarily high \citep{jordan2025planetary}.

To explain the published observations without invoking a liquid-water ocean, detailed mechanisms involving a massive gas or steam envelope have been explored. One hypothesis suggests that the preferential partitioning of nitrogen in the mantle could explain the apparent absence of \ce{NH3} within the context of a massive atmosphere dominated by \ce{H2} \citep{shorttle2024distinguishing}. However, achieving severe nitrogen depletion via an underlying magma ocean may require fine-tuned volatile abundances and redox conditions \citep{rigby2024towards}. Regardless of the conditions at the lower boundary, a massive \ce{H2}-dominated atmosphere would exhibit low \ce{CO2}-to-CO ratios, a characteristic that could be tested through future observations \citep{hu2021unveiling,glein2024geochemical}. At the same time, any observed high \ce{CO2}-to-\ce{CH4} ratio could naturally arise in a massive envelope rich in water, without requiring extremely high metallicities for carbon or nitrogen \citep{yang2024chemical}. Such a ``mixed steam'' envelope could plausibly form through the accretion of both gas and ices during the planet’s formation \citep[e.g.,][]{burn2024}, making this scenario a potential explanation for reconciling the atmospheric constraints of K2-18~b.

It is evident that repeated observations are crucial to advancing our understanding of whether a liquid-water ocean exists on the iconic sub-Neptune K2-18~b. In this work, we present new transit observations of K2-18~b obtained with JWST, covering the wavelength between 1.7 -- 5.2 $\mu$m (Section~\ref{sec:obs}). Combined with previously published data, these new observations significantly improve constraints on the planet's atmospheric composition and provide deeper insights into its interior structure. We analyze the complete set of observations using multiple spectral retrieval frameworks and advanced atmospheric photochemical-thermochemical models (Section~\ref{sec:model}). Based on this comprehensive analysis, we report robust results on the atmospheric and internal composition of K2-18~b in Section~\ref{sec:cratio}, and discuss several key but model-dependent implications in Section~\ref{sec:discussion}. We propose an updated roadmap for characterizing similar temperate sub-Neptunes through atmospheric observations in Section~\ref{sec:roadmap}, and conclude with a forward-looking perspective on future observations in Section~\ref{sec:conclusion}.

\section{Observation and Data Reduction} \label{sec:obs}

\subsection{Observations}

\begin{table}
\caption{Observations of K2-18~b by JWST.}
\centering
\begin{tabular}{l|c|c|c}
\hline\hline
Visit & Instrument mode & Program & Date (UT) \\ \hline
 A1   &  NIRISS SOSS  &  2722  &  Jun 1, 2023 \\
 B1           &  NIRSpec G235H  &  2372 &  Jan 18, 2024  \\
 B2           &  NIRSpec G235H  & 2372   &  Jan 14, 2025 \\
 C1           &  NIRSpec G395H  &  2722  &  Jan 20, 2023 \\
 C2          &  NIRSpec G395H  &  2372  & May 28, 2024  \\
 C3           &  NIRSpec G395H  & 2372  &  Dec 12, 2024  \\
 D1           &  MIRI LRS  &  2722 & Apr 25, 2024  \\ \hline
\end{tabular}
\label{table:visitlabels}
\end{table}

We observed four transits of K2-18~b using JWST's Near InfraRed Spectrograph \citep[NIRSpec,][]{jakobsen2022near,birkmann2022near}, with two observations using the G235H grating and the other two using the G395H grating, as part of the JWST GO Program 2372 (PI: Renyu Hu). In each of these observations, photons are dispersed across two detectors, NRS1 and NRS2. Together, these observations covered wavelengths from $1.67-5.16$ $\mu$m with a native spectral resolving power ranging from $2000-3000$, except for $\sim0.1$-$\mu$m gaps between the two detectors at 2.2 and 3.7 $\mu$m. As summarized in Table~\ref{table:visitlabels}, the repeated observations were conducted over the course of approximately a year, with each visit covering the full 2.7-hour transit and $>5.5$ hours of out-of-transit baseline to precisely measure the transit depth and model instrumental systematic noise.

In addition, we analyzed the two transits of K2-18~b observed as part of the JWST GO Program 2722 (PI: Nikku Madhusudhan). One of these visits used NIRSpec/G395H and a similar setup as our G395H visit, except for a substantially shorter out-of-transit baseline. The other visit was performed using the Near Infrared Imager and Slitless Spectrograph \citep[NIRISS,][]{doyon2023near} in its Single Object Slitless Spectroscopy mode \citep[SOSS,][]{albert2023near}. The NIRISS/SOSS visit covered wavelengths from $0.85-2.85$ $\mu$m by the first spectral order. The results of these observations have been published in \cite{madhusudhan2023carbon}, and here we reanalyzed these data and combined them with our new observations to characterize the atmosphere of K2-18~b. We also evaluated the impact of the recently published spectrum obtained by JWST's Mid-Infrared Instrument Low Resolution Spectrometer \citep[MIRI/LRS,][]{madhusudhan2025miri} on our spectral retrievals.

\subsection{NIRSpec Data Analysis -- Eureka!}
\label{sec:nirspec}

We extracted the transmission spectra of K2-18~b from the five NIRSpec transits using \texttt{Eureka!} \citep[v0.10,][]{bell2022eureka}. We started the reduction from the uncalibrated raw files and applied all the default steps of stage 1, keeping the jump detection threshold at 4$\sigma$. As part of stage 1, we ran a group-level background subtraction using the average of the 16 top and 16 bottom pixels in each detector column. During this background subtraction, we masked a region with a half-width of 8 pixels centered on the trace. In stage 2, we ran all the default steps except for the flat-field correction and the photometric correction, since we are only interested in relative fluxes.

We then performed optimal extraction on the calibrated data following a setup similar to the one presented in \citet{damiano2024} and \citet{belloarufe2025l9859}. We removed columns outside the [$545,2041$] and [$6,2044$] ranges in the NRS1 and NRS2 detectors, respectively, where the signal from the target is negligible. We masked pixels with an odd Data Quality (DQ) value and aligned the trace by vertically shifting each column by an integer number of pixels. We applied another round of column-by-column background subtraction, this time at the integration level, using the mean value of pixels located farther than nine rows away from the trace. We performed optimal extraction on a window centered around the trace with a half-width of three pixels. For optimal extraction, we used
the median integration as the spatial profile image after
removing 10$\sigma$ outliers along the time axis and
applying a smoothing filter with a length of 13 pixels.

We then generated the NIRSpec light curves. Following a visual inspection of the data, we masked the detector columns with excessive noise. We generated the white light curves by binning the data in the following ranges: $1.67-2.19~\mu$m (G235H/NRS1), $2.27-3.07~\mu$m (G235H/NRS2), $2.87-3.71~\mu$m (G395H/NRS1), and $3.83-5.17~\mu$m (G395H/NRS2). We extracted spectroscopic light curves at two different resolutions: $\Delta\lambda=0.02~\mu$m and  $\Delta\lambda=0.004~\mu$m. We clipped 3$\sigma$-outliers in time from the light curves using a rolling median filter. 

\begin{figure*}[!htbp]
\centering
\includegraphics[width=\textwidth]{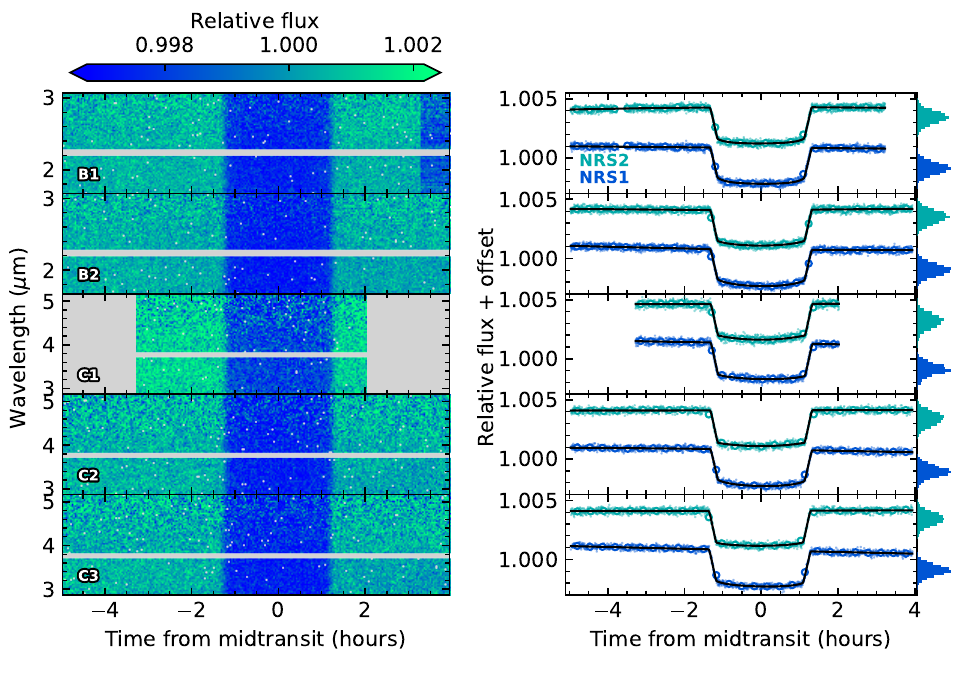}
\caption{\textbf{Left:} Spectroscopic light curves of the five NIRSpec visits of K2-18\,b as extracted with \texttt{Eureka!} at a resolution of $\Delta\lambda=0.02~\mu$m. The observations from JWST GO Program 2722 \citep{madhusudhan2023carbon} have a shorter out-of-transit baseline. The horizontal gray stripe in each transit corresponds to the gap between the NRS1 and NRS2 detectors. There is a jump in flux visible near the end of Visit B1, which we mask in our light curve fits. \textbf{Right:} White light curves (unbinned and binned to $\sim20$ minutes) from the five NIRSpec transits, including the best-fit models and histograms of the unbinned residuals. 
}
\label{fig:lcs}
\end{figure*}

Figure~\ref{fig:lcs} shows the spectroscopic and white light curves from the five transits observed with NIRSpec. The light curves of Visit B1 show a significant jump in flux. This jump is also visible in the time series of the trace location and point spectral function (PSF) width along both the spatial and spectral directions. Fortunately, this jump, which we assign to a mirror tilt event \citep{alderson2023}, happened outside the transit and close to the end of the observation. Since we had a sufficient out-of-transit baseline, we decided to discard the integrations taken after the jump. Those data also show a flare-like event in both detectors approximately 1.3 hours after the start of the observations, and so we masked integrations $498-591$ and $484-578$ in NRS1 and NRS2, respectively. We also masked the spot crossing event in Visit C1 as \cite{madhusudhan2023carbon}. 

We fit the ten NIRSpec white light curves independently using a combination of a \texttt{batman} transit model \citep{kreidberg2015} and a systematics model. In these fits, we assigned uniform priors of $\mathcal{U}$(60096.719368, 60096.739368), $\mathcal{U}(89^\circ,90^\circ)$, and $\mathcal{U}(70,90)$ to the mid-transit time ($T_{0, \rm{BJD-2,400,000.5}}$), orbital inclination ($i_p$), and the scaled semi-major axis ($a/R_s$), respectively. The prior on the mid-transit time corresponds to the epoch of the NIRISS/SOSS transit from \citet{madhusudhan2023carbon}. We assumed circular orbits, fixed the orbital period to 32.940045 days \citep{benneke2019water}, and kept the two quadratic limb-darkening parameters free \citep[$q_1$ and $q_2$,][]{kipping2013}. We included a white noise multiplier to scale the uncertainties of the data according to the scatter of the residuals. The systematics model consisted of a polynomial in time. We initially attempted to model all visits using a quadratic polynomial, but we found that the quadratic term was generally insignificant for most cases. In such instances, we opted for a linear polynomial instead. The exceptions were Visit B1 NRS1, Visit B2, and Visit C2 NRS1, which required a quadratic polynomial (Figure~\ref{fig:lcs}). The best-fit parameters from the white light curve fits are shown in Table~\ref{table:bestfitwlc}.

\begin{table*}[]
\caption{Best-fit parameters from the NIRSpec white light curve fits. In bold, we include the weighted average of the mid-transit time, inclination and scaled semi-major axis, which are kept fixed in the spectroscopic light curve fits. The weighted average of the mid-transit time is calculated transit by transit, and that of the inclination and the scaled semi-major axis are calculated from all transits.}
\centering
\resizebox{\textwidth}{!}{\begin{tabular}{lcccccccccc}
\hline\hline
\multirow{2}{*}{Parameter}             & \multicolumn{2}{c}{G395H (Visit C1)} & \multicolumn{2}{c}{G235H (Visit B1)} & \multicolumn{2}{c}{G235H (Visit B2)} & \multicolumn{2}{c}{G395H (Visit C2)} & \multicolumn{2}{c}{G395H (Visit C3)}\\
                                       & NRS1            & NRS2            & NRS1        & NRS2        & NRS1           & NRS2 & NRS1        & NRS2        & NRS1           & NRS2         \\ \hline
\multirow{2}{*}{Mid-transit time, $T_0$ (BJD$–$2,400,000.5)} &     59964.969458$^{+4.8e-5}_{-4.8e-5}$            &           59964.969552$^{+5.6e-5}_{-5.6e-5}$  & 60327.309098$^{+3.8e-5}_{-4.0e-5}$            &    60327.309265$^{+4.1e-5}_{-4.1e-5}$  & 60689.649937$^{+3.8e-5}_{-3.9e-5}$ &  60689.649942$^{+3.9e-5}_{-4.0e-5}$   &  60459.069847$^{+4.8e-5}_{-4.9e-5}$      &    60459.069800$^{+5.4e-5}_{-5.4e-5}$  & 60656.708462$^{+4.7e-5}_{-4.6e-5}$ &  60656.708431$^{+5.4e-5}_{-5.3e-5}$   \\
                                       & \multicolumn{2}{c}{$\mathbf{59964.969498\pm3.6e-05}$}              & \multicolumn{2}{c}{$\mathbf{60327.309177\pm2.8e-05}$}    & \multicolumn{2}{c}{$\mathbf{60689.64994\pm2.7e-05}$} & \multicolumn{2}{c}{$\mathbf{60459.069826\pm3.6e-05}$}       & \multicolumn{2}{c}{$\mathbf{60656.708449\pm3.5e-05}$}    \\
\multirow{2}{*}{Inclination, $i_p$ ($^\circ$)}               &     89.546$^{+0.016}_{-0.015}$         &       89.591$^{+0.020}_{-0.020}$    &      89.565$^{+0.015}_{-0.014}$      &   89.535$^{+0.015}_{-0.014}$ & 89.567$^{+0.014}_{-0.014}$ & 89.599$^{+0.016}_{-0.015}$        &   89.585$^{+0.021}_{-0.018}$          &      89.577$^{+0.019}_{-0.018}$  & 89.556$^{+0.018}_{-0.017}$ & 89.561$^{+0.019}_{-0.017}$\\
                                       & \multicolumn{10}{c}{$\mathbf{89.5661\pm0.0052}$}                                                                           \\
\multirow{2}{*}{Scaled semi-major axis, $a/R_s$}                  &      79.5$^{+1.1}_{-1.0}$      &    82.5$^{+1.4}_{-1.4}$         &       80.86$^{+1.03}_{-0.96}$     &   78.82$^{+1.00}_{-0.96}$  & 80.98$^{+0.94}_{-0.92}$ & 83.1$^{+1.1}_{-1.0}$     &    82.0$^{+1.4}_{-1.2}$          &        81.5$^{+1.3}_{-1.2}$   & 80.3$^{+1.2}_{-1.2}$ &  80.6$^{+1.3}_{-1.2}$  \\
                                       & \multicolumn{10}{c}{$\mathbf{80.87\pm0.35}$}                                                                           \\
Planet-to-star radius ratio, $R_p/R_s$     &  0.05451$^{+1.8e-4}_{-1.8e-4}$        &       0.05414$^{+1.8e-4}_{-1.8e-4}$       & 0.05410$^{+1.0e-4}_{-1.0e-4}$      &     0.05464$^{+1.9e-4}_{-1.6e-4}$   & 0.05459$^{+1.8e-4}_{-1.8e-4}$ & 0.05397$^{+1.7e-4}_{-1.6e-4}$   &     0.05422$^{+2.1e-4}_{-2.0e-4}$        &      0.05427$^{+1.7e-4}_{-1.7e-4}$  & 0.05462$^{+2.0e-4}_{-2.0e-4}$ &  0.05399$^{+1.3e-4}_{-1.6e-4}$ \\
First limb-darkening coefficient, $q_1$                          &      0.0407$^{+0.0116}_{-0.0089}$      &         0.0497$^{+0.0106}_{-0.0075}$      &       0.112$^{+0.013}_{-0.012}$    &    0.062$^{+0.013}_{-0.012}$       & 0.099$^{+0.016}_{-0.014}$ & 0.072$^{+0.016}_{-0.014}$  &       0.121$^{+0.024}_{-0.022}$       &      0.0415$^{+0.0136}_{-0.0098}$  & 0.516$^{+0.015}_{-0.011}$ &  0.0249$^{+0.0080}_{-0.0056}$  \\
Second limb-darkening coefficient, $q_2$                         &      0.46$^{+0.29}_{-0.25}$        &      0.81$^{+0.14}_{-0.21}$     &    0.025$^{+0.042}_{-0.019}$         &      0.25$^{+0.23}_{-0.16}$   & 0.47$^{+0.16}_{-0.14}$ &   0.37$^{+0.19}_{-0.16}$  &   0.19$^{+0.15}_{-0.12}$     &       0.51$^{+0.29}_{-0.26}$   & 0.47$^{+0.28}_{-0.24}$ &  0.69$^{+0.22}_{-0.30}$\\ \hline
\end{tabular}}\label{table:bestfitwlc}
\end{table*}

We fit the spectroscopic light curves in a similar fashion to the white light curves. However, here we fixed $i_p$ and $a/R_s$ to the weighted average of the ten values derived from the NIRSpec white light curves, shown in Table~\ref{table:bestfitwlc}. Similarly, we also fixed the mid-transit time, but in this case on a transit-by-transit basis. For those light curves where we used a quadratic polynomial in time, we fixed the quadratic term to the best-fit value derived from the corresponding white light curve \citep[see e.g.,][]{moran2023,madhusudhan2023carbon,belloarufe2025l9859}. All other parameters, including the limb darkening coefficients, were kept free, with the same priors used in the white light curve fits.

\begin{figure*}[!htbp]
\centering
\includegraphics[trim=5 11 5 5, clip,width=\textwidth]{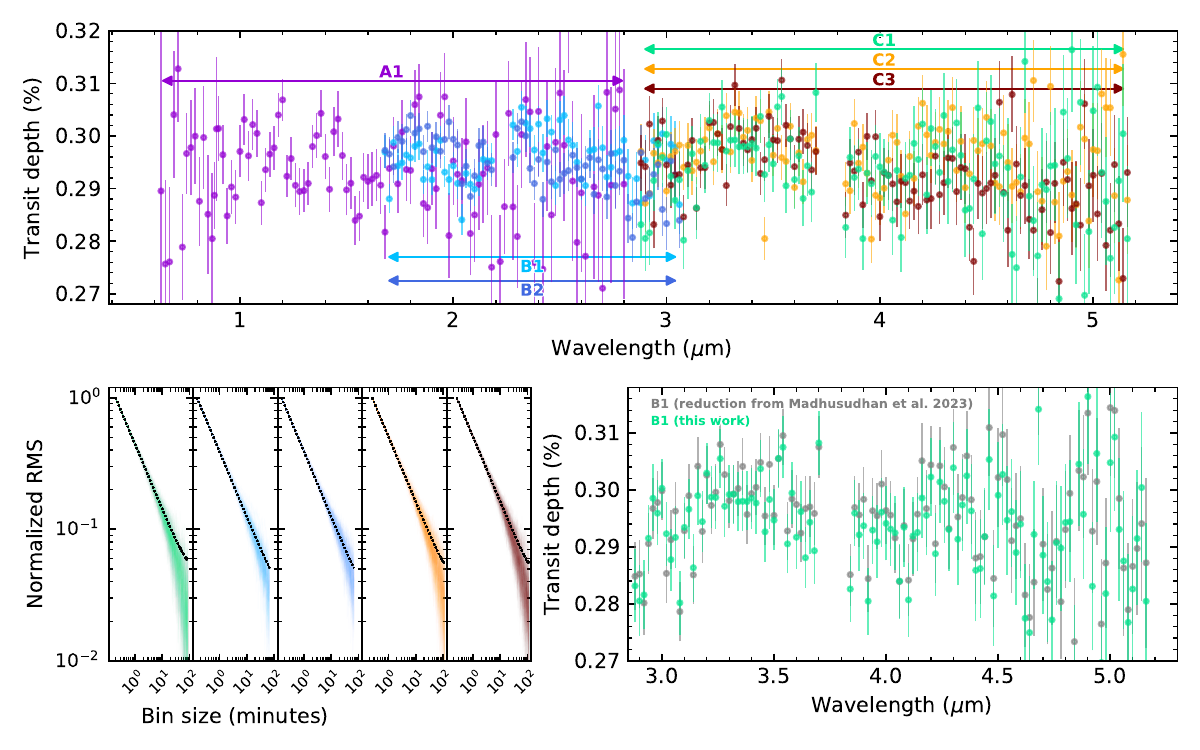}
\caption{Top: JWST transmission spectra of K2-18\,b. We show our reduction of the currently available NIRSpec and NIRISS data of K2-18\,b. In this plot, the NIRISS data was binned down to the same resolution as the NIRSpec data ($\Delta\lambda=0.02~\mu$m). Bottom left: Normalized RMS versus bin size for all NIRSpec spectroscopic light curves (the color scheme follows that of the top plot). The black lines indicate the expected RMS for white noise. Bottom right: Comparison of our reduction of the G395H data from Program 2722 against the reduction by \citet{madhusudhan2023carbon} of the same data, binned to the same resolution. No offsets were applied to any spectra in this figure.
}
\label{fig:spectra_comparison}
\end{figure*}

We compare all JWST near-infrared transmission spectra of K2-18~b in Figure~\ref{fig:spectra_comparison}, including our reduction of the NIRISS/SOSS data collected in Program 2722 (see Section~\ref{sec:niriss}). The NIRISS/SOSS and NIRSpec/G235H data overlap in the $1.67-2.81$ $\mu$m region. They both show similar features at these wavelengths, but the uncertainties of the NIRSpec/G235H data are $\sim 1.4-4\times$ smaller when binned to the same spectral resolution. The three G395H transmission spectra are apparently consistent as well, although the uncertainties from Program 2372 are $\sim20\%$ smaller than those from Program 2722 due to the longer out-of-transit baseline. Plotting the root mean square (RMS) of the spectroscopic light curve residuals as a function of bin size (shown in Figure~\ref{fig:spectra_comparison}) shows no evidence for red noise. Finally, in the same figure, we compare our reduction of the G395H data from Program 2722 against the reduction of the same data presented in \citet{madhusudhan2023carbon}. Both reductions appear consistent at all wavelengths.

\subsection{NIRSpec Data Analysis -- ExoTEDRF}
\label{sec:nirspec_ind}

We performed an independent data reduction of the five NIRSpec visits of K2-18\,b using the ExoTEDRF framework \citep{Feinstein_2023, radica_2023, radica2024} and closely following the method presented in \citet{ahrer2025escaping}. For all visits, we performed the standard Stage 1 reduction steps, and added a group-level 1/$f$ background subtraction. In that step, we subtracted the median of each detector column for each group of each integration. When evaluating the median for each pixel column, bad pixels were masked, and only pixels that are 8 pixels away from the center of the trace were considered. At the end of Stage 1, we further performed a cosmic ray detection step using a time-domain outlier flagging routine. We then performed the standard ExoTEDRF Stage 2 calibration steps. This includes the \texttt{Extract2D} step of the STScI pipeline, which updates the wavelength solution as a function of the source location in the NIRSpec slit. We further performed an additional 1/$f$ background subtraction step at the integration level using the same settings as in Stage 1. Once the integration-level frames are calibrated, we performed a detector-level PCA analysis to extract the time series of the movement of the trace on the detector. Finally, we extracted the time series of stellar spectra using a standard box extraction with an extraction aperture that is 8-pixels wide and centered on the trace. 

We performed the light-curve fitting of the ExoTEDRF-reduced observations using the ExoTEP framework \citep{benneke2017, benneke2019water}. We used all the integrations from each visit, except for the flaring event and the tilt event in Visit B1. We started by binning the light curves into the same 0.004\,$\mu$m spectroscopic bins as for the Eureka! analysis described above. To form the spectroscopic light curves, the flux from each pixel column in a given bin were added together. The white light curve was obtained independently for each detector by adding the flux from all the spectroscopic bins together. 

For each detector of each visit, the white light curve was fitted using a \texttt{batman} transit model \citep{kreidberg2015}, along with a linear trend in time for the systematics model. The parameters we fit for were the transit depth, the mid-transit time, the impact parameter, the scaled semi-major axis, the two quadratic limb darkening coefficients u$_1$ and u$_2$, a photometric scatter parameter, and finally the slope and normalization factor of the linear systematics model. All fits were performed using a Markov Chain Monte Carlo analysis via the emcee package \citep{Foreman_Mackey_2013}, and we ran the white-light-curve fits for 3500 steps, using four times as many walkers as the number of parameters. 

We performed the spectroscopic light-curve fitting independently for each detector and each visit. The orbital parameters (mid-transit time, impact parameter, scaled semi-major axis) were fixed to the best-fitting values from the corresponding white light-curve fits, leaving the transit depth, the two quadratic limb darkening coefficients u$_1$ and u$_2$, the photometric scatter parameter, and the two parameters of the linear systematics model as the fitted parameters in the spectroscopic fits. The fits were performed using the same MCMC method as that for the white light curves, and we ran the fits for 1500 steps. 

The NIRSpec transmission spectra of K2-18\,b obtained from the Eureka! and ExoTEDRF pipelines are in strong agreement (Figure~\ref{fig:compare_reduction} in Appendix). We inspected the residuals between the transmission spectra obtained for each visit, and produced a binned version (0.04\,$\mu$m) of the residuals for clarity (Figure~\ref{fig:compare_reduction}). We found that the transit depths are consistent in virtually all spectral channels between the reductions, and there are no discrepant spectral features between the spectra obtained from the two reductions. We found some $\mathcal{O}(10\,\mathrm{ppm})$ constant offsets between the reductions, which are probably due to different orbital parameters used for the spectroscopic light-curve fitting. For some detectors of some visits (e.g., Visit B2 NRS2 and Visit C1), we also found subtle $<1\sigma$ slopes in the residuals, which can be attributed to the different orbital parameters and to the different limb darkening parameterizations. We thus concluded that, despite the many differences between the reduction and light-curve-fitting techniques of the Eureka! and ExoTEDRF pipelines, we derived consistent transmission spectra for K2-18\,b, making the results presented in this work robust to the specifics of data analysis methods.

\subsection{NIRISS Data Analysis -- NAMELESS}
\label{sec:niriss}

We performed a reanalysis of the NIRISS/SOSS observations of K2-18\,b using the \texttt{NAMELESS} data reduction pipeline \citep{Feinstein_2023,Coulombe2023,Coulombe2025highlyreflectivewhiteclouds}, closely following the methods outlined in \citet{benneke2024jwst}, \citet{Piaulet2024}, and \citet{Coulombe2025highlyreflectivewhiteclouds}. We performed all Stage 1 steps of the \texttt{jwst} pipeline \citep{Bushouse2023jwst}, except for the dark subtraction step whose reference file shows signs of residual 1/$f$ noise. Once counts had been converted to rates via the ramp-fitting step, we performed a set of four custom routines that account for common noise sources.

First, bad/hot pixels were flagged systematically using the spatial second derivative of the detector images, and corrected through bicubic interpolation. Second, the non-uniform zodiacal light background was subtracted by scaling independently to the median frame the two regions of the background model provided by STScI\footnote{Available at \url{https://jwst-docs.stsci.edu/}.} that are separated by the sudden jump in flux around column $x\sim700$. Third, remaining cosmic rays were corrected by taking the individual median in time of all detector pixels and bringing any $>$4$\sigma$ outlier to the median's value. Fourth, 1/$f$ noise was computed and subtracted from each individual integration by independently scaling all individual columns of the first and second spectral orders, and fitting for a constant additive value such that the chi-square is minimized. Finally, spectral extraction was performed using a box aperture of width 36 pixels, and we use the wavelength solution provided by the \texttt{PASTASOSS} Python package \citep{Baynes2023} considering the pupil wheel position angle for this visit.

We produced a white-light curve using the first spectral order data ($\lambda$ = 0.85--2.85\,$\mu$m), along with spectroscopic light curves at fixed resolving powers of 300 and 100 for the first and second spectral orders, respectively. Light-curve fitting was performed using the ExoTEP framework \citep{Benneke_2019_methanedepleted,benneke2019water}, which couples the \texttt{batman} transit modeling tool \citep{kreidberg2015} to a Markov chain Monte Carlo sampler \citep{Foreman_Mackey_2013}. First, we fit the white-light curve, considering the planet-to-star radius ratio ($R_\mathrm{p}/R_\mathrm{s}$), scaled semi-major axis ($a/R_\mathrm{s}$), impact parameter ($b$), and mid-transit time ($T_0$) as free parameters with wide uniform priors. For limb darkening, we considered the quadratic coefficients [$u_1$,$u_2$] with wide uniform priors to avoid biasing the measured transit depths at the reddest wavelengths of NIRISS/SOSS \citep{Coulombe2024}. A linear slope was used for the systematics model. As was observed in \citet{madhusudhan2023carbon}, a spot-crossing event occurs near the end of the transit's ingress. To account for this, we also included a Gaussian with free width, amplitude, and center position in our systematics model. Second, we fit for the spectroscopic light curves, where we fixed the semi-major axis, impact parameter, mid-transit time, and Gaussian width and center position to their best-fit values from the white-light curve, and kept free the remainder of the astrophysical and systematics model parameters. The resulting transmission spectrum is shown in Figure~\ref{fig:spectra_comparison}. Finally, we also performed a test where we masked the integrations (integrations between \# [231, 356]) affected by the spot-crossing event and found both spectra to be virtually identical. The spectrum obtained using the \texttt{NAMELESS} reduction show similar structures as the one presented in \citet{madhusudhan2023carbon} (Figure~\ref{fig:compare_reduction_NIRISS} in Appendix), with slightly larger final uncertainties when binned to the same spectral resolution. This most likely arises from fitting the limb-darkening coefficients for spectroscopic light curves, as opposed to fixing the coefficients to the values from an initial fit performed at a lower resolving power of $R=20$ in \citet{madhusudhan2023carbon}.

We adopted the \texttt{Eureka!} reduction of the NIRSpec datasets (at $\Delta\lambda=0.004\ \mu$m) and the \texttt{NAMELESS} reduction of the NIRISS dataset for spectral retrievals and subsequent analyses. This choice was motivated in part by the inclusion of the Order 2 spectrum in the NIRISS dataset -- an element not incorporated in the previous analysis by \citet{madhusudhan2023carbon} -- as well as by the use of a more conservative treatment of uncertainties. Substituting alternative reductions for either the NIRSpec or NIRISS datasets yields no significant changes in the retrieval results or the overall scientific conclusions.
	
\section{Atmospheric Models} \label{sec:model}

\begin{figure*}[!htbp]
\centering
\includegraphics[width=1.0\textwidth]{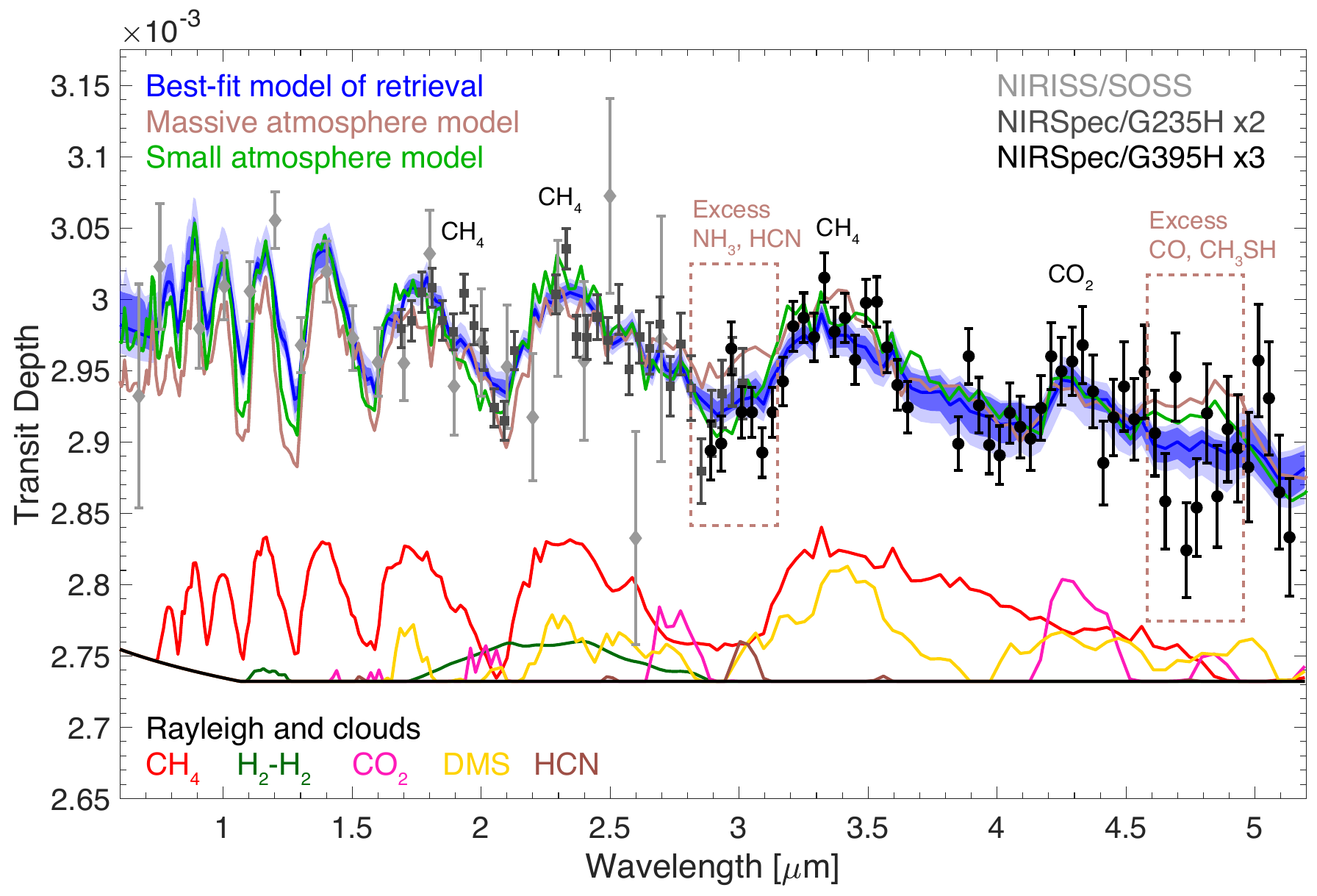}
\caption{Transmission spectrum of K2-18~b compared with models. Five transit measurements using NIRSpec are averaged with offsets taken out between visits (see Appendix~\ref{sec:shiftedaverage}), and the data are binned to $\Delta\lambda=0.1\ \mu$m and $0.04\ \mu$m for NIRISS and NIRSpec measurements for clarity. The fit shown in blue is the model that maximizes the posterior probability (MAP) in the baseline+DMS retrieval using \exotr\ and its $1\sigma$ and $2\sigma$ boundaries (Section~\ref{sec:exotr}).
Spectral contributions from individual gases that increase the transit depth by at least $0.02 \times 10^{-3}$ relative to the Rayleigh scattering and cloud baseline are shown below. The best-fit model is dominated by CH$_4$ and CO$_2$ absorption. Replacing DMS with \ce{CH3SH} or \ce{N2O}, or omitting DMS entirely, does not significantly alter the best-fit model.
The self-consistent model shown in brown assumes a massive envelope with an H$_2$-to-H$_2$O ratio of 75:25 and $100\times$ solar metallicity for carbon, nitrogen, and sulfur (Section~\ref{sec:epacris_massive}). The model shown in green assumes a 1-bar atmosphere with 10\% \ce{CH4}, $5\times10^{-4}$ \ce{CO2}, a temperature-dependent \ce{H2O} abundance (assuming a planetary Bond albedo of 0.5), and a hypothetical DMS surface flux $20\times$ that of modern Earth (Section~\ref{sec:epacris_small}). The self-consistent massive-atmosphere model provides a reasonable fit to the data within uncertainties but overproduces \ce{NH3}, HCN, CO, and \ce{CH3SH}, leading to excess absorption near 3.0 and $4.6-4.8\ \mu$m. The small-atmosphere model also shows enhanced absorption near $4.6–4.8\ \mu$m due to CO and \ce{CH3SH}.
}
\label{fig:model_fit}
\end{figure*}

We investigated the atmospheric composition and physical state of K2-18~b using the observed transmission spectra. To interpret the data, we employed three independent Bayesian spectral retrieval frameworks to explore a broad range of possible atmospheric gases, quantify detection significances, and constrain molecular abundances. In parallel, we developed physically self-consistent atmospheric models that incorporate planetary physics and chemistry, enabling us to connect the retrieved gas abundances to the planet's internal structure. The key results are summarized in Figures~\ref{fig:model_fit}--\ref{fig:hycean}, with detailed descriptions of our retrieval and modeling efforts provided in Sections~\ref{sec:retrievals} and~\ref{sec:epacris}.

\subsection{Spectral Retrievals}
\label{sec:retrievals}

\subsubsection{ExoTR retrievals}
\label{sec:exotr}

We used \exotr\ (Exoplanetary Transmission Retrieval) to statistically interpret the observed spectra of K2-18b. \exotr\ computes transmission spectra and employs a nested sampling algorithm via \texttt{MultiNest} \citep{Feroz2009} and \texttt{PyMultiNest} \citep{Buchner2014} to explore atmospheric scenarios. Key features of \exotr\ include the joint retrieval of stellar heterogeneity and planetary atmosphere parameters, as well as the ability to model physically motivated clouds and hazes \citep[][Tokadjian et al. 2025, in prep]{damiano2024}. The framework has been successfully applied to interpret the transmission spectra and derive atmospheric conditions for a wide range of exoplanets, including the cold water-rich world LHS-1140~b, the potentially volcanic rocky planet L98-59~b, and the hazy super-puff Kepler-51~d \citep{damiano2024, belloarufe2025l9859, libbyroberts2025}.

Using \exotr, we explored a broad range of atmospheric abundance scenarios for K2-18~b. The model parameters and their priors are summarized in Table~\ref{table:exotrprior}. We assumed a background composition of 80\% \ce{H2} and 20\% He, fixed the planet's mass to $M_p = 8.63$~M${\oplus}$ \citep{cloutier2019confirmation}, and adopted an isothermal pressure–temperature profile. 
Initial retrievals included all molecules listed in Table~\ref{table:exotrprior} as well as H$_2$S, SO$_2$, OCS, CH$_3$Cl, CS$_2$, C$_2$H$_2$, C$_2$H$_4$, C$_2$H$_6$, and C$_4$H$_2$. However, none of these additional species were found to be constrained by the data and were therefore excluded from subsequent retrievals.
Molecular opacities were computed line-by-line using the most up-to-date databases: HITEMP for CH$_4$ \citep{Hargreaves2020}, ExoMol for SO$_2$ and SO$_3$ \citep{Tennyson2016}, and HITRAN for all remaining molecules \citep{Gordon2022}, at a resolving power of 200,000. Opacities for DMS and CH$_3$SH were taken from the HITRAN database at 1 bar and 298 K. 
We also incorporated H$_2$–H$_2$ collision-induced absorption, Rayleigh scattering, and a gray cloud deck in all models. The opacities used in this effort were compared with those used in other retrieval codes, including Planet Spectrum Generator (PSG, \citealt{villanueva2018psg}), rfast \citep{robinson2023rfast}, and ARtful modelling Code for exoplanet Science (ARCiS, \citealt{min2020arcis}), with excellent agreement between them.

\setlength{\tabcolsep}{3pt}

       \begin{deluxetable}{cclc}
		\tablecaption{Parameters and priors in \exotr\ retrievals.}
		\tablehead{
			\colhead{Parameter} & \colhead{Symbol} & \colhead{Range} & \colhead{Type}}
		\startdata
         Offsets& offset$_n$ & [-200, 200] ppm & Linear-uniform\\
         Planet Radius & $R_p$ & [1.305, 5.22] $R_{\oplus}$ & Linear-uniform\\
         Planet Temperature & $T_p$ & [100, 500] K & Linear-uniform\\
         Cloud Top Pressure& log($P_{top}$) & [-1, 6] Pa & Log-uniform\\
         \underline{Molecules} & & & \\
         H$_2$O VMR & log(H$_2$O) & [-12, -0.3] & Log-uniform\\
         CH$_4$ VMR & log(CH$_4$) & [-12, -0.3] & Log-uniform\\
         NH$_3$ VMR & log(NH$_3$) & [-12, -0.3] & Log-uniform\\
         HCN VMR & log(HCN) & [-12, -0.3] & Log-uniform\\
         CO VMR & log(CO) & [-12, -0.3] & Log-uniform\\
         CO$_2$ VMR & log(CO$_2$) & [-12, -0.3] & Log-uniform\\
         DMS VMR & log(DMS) & [-12, -0.3] & Log-uniform\\
         CH$_3$SH VMR & log(CH$_3$SH) & [-12, -0.3] & Log-uniform\\
         N$_2$O VMR & log(N$_2$O) & [-12, -0.3] & Log-uniform\\
         \underline{Stellar Parameters} & & & \\
         Het. Fraction & $\delta_{het}$ & [0, 0.5] & Linear-uniform\\
          Het. Temperature& $T_{het}$ & [1729, 4148] K & Linear-uniform\\
          Phot. Temperature & $T_{phot}$ & [2957,3957] K & Linear-uniform\\
		\enddata
       \label{table:exotrprior}
       \tablecomments{We considered either offsets between instruments or offsets between detectors. The offsets between instruments are defined relative to the NIRSpec/G235H dataset and the offsets between detectors are defined relative to the NIRSpec/G235H NRS1 data. The prior range of the planetary radius spans from $0.5\times$ to $2\times$ the measured radius. VMR refers to the volume mixing ratio.
       }
      \end{deluxetable}
      
\begin{figure*}[!htbp]
\centering
\includegraphics[trim=5 5 5 5, clip,width=\textwidth]{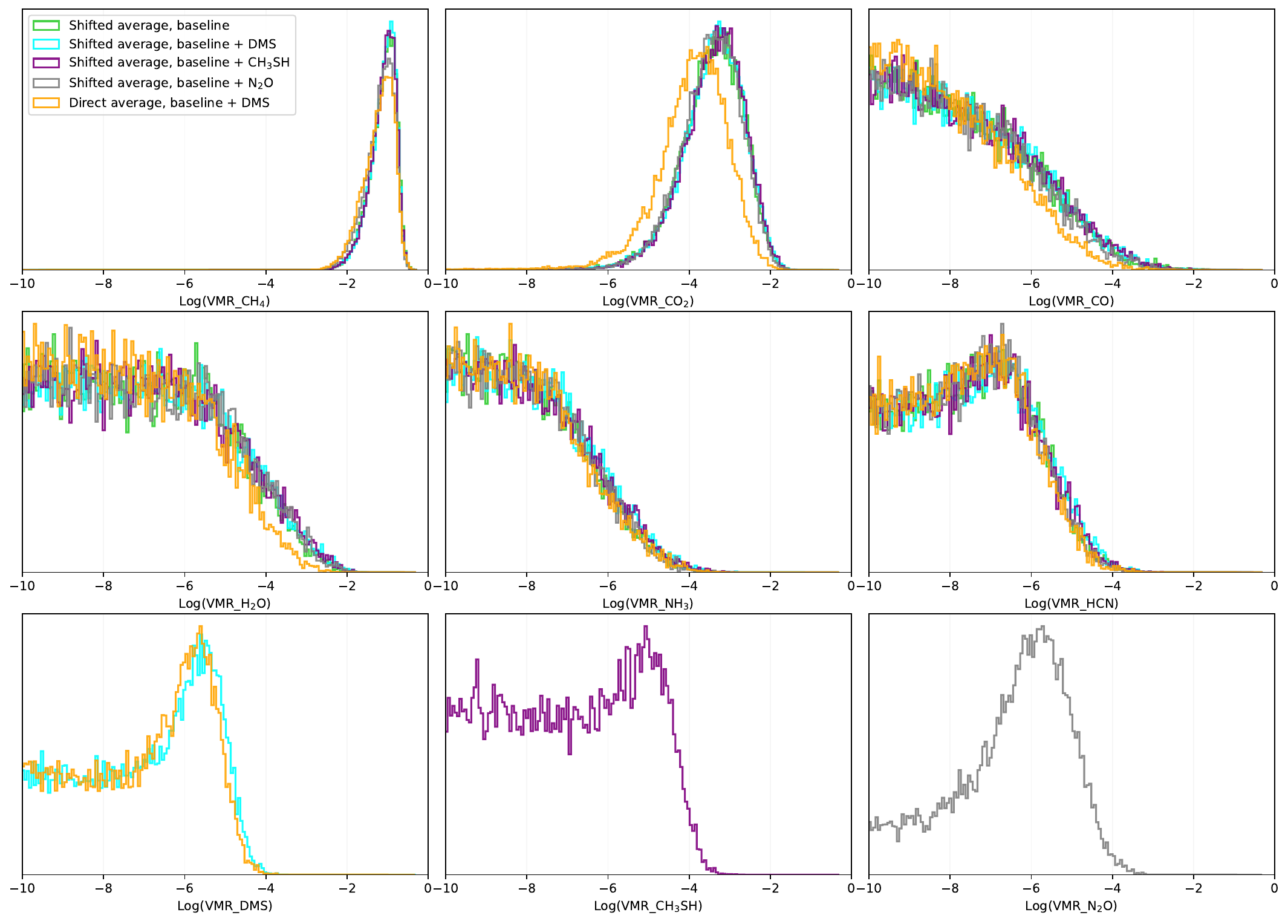}
\caption{
Posterior distributions from \exotr\ atmospheric retrievals of K2-18~b's transmission spectra using six visits: two with NIRSpec/G235H, three with NIRSpec/G395H, and one with NIRISS/SOSS. The baseline cases refer to the retrieval setup with the molecules listed in Table~\ref{table:exotrprior} except for DMS, CH$_3$SH, and N$_2$O. We added these gases one at a time to see if including them would be preferred over the baseline setup. All retrievals yield tight constraints on \ce{CH4} and \ce{CO2}, and upper limits for CO, H$_2$O, NH$_3$, and HCN. The inclusion of DMS, CH$_3$SH, or N$_2$O results in weak indications of their presence, though all show broad tails toward negligible abundances.
}
\label{fig:exotr_posterior_by_method}
\end{figure*}

As summarized in Table~\ref{table:cases_by_visits}, our standard dataset includes six visits in total: two with NIRSpec/G235H, three with NIRSpec/G395H, and one with NIRISS/SOSS. We began by directly combining the spectra by instrument and performed atmospheric retrievals using the prior ranges listed in Table~\ref{table:exotrprior} (excluding CH$_3$SH and N$_2$O). This approach included two instrument offsets—Offset 1 between G235H and G395H, and Offset 2 between G235H and SOSS. The resulting posterior distribution is shown in orange in Figure~\ref{fig:exotr_posterior_by_method}, and key atmospheric constraints are summarized in the ``direct average'' column of Table~\ref{table:keyresults_exotr}. We detected \ce{CH4} and \ce{CO2} with high significance, with $1\sigma$ uncertainties of less than 0.5 dex and 1 dex, respectively. The retrieval placed a cloud deck at pressures $>\sim0.03$ bar ($2\sigma$), i.e., a deep cloud that does not strongly interfere with the transmission spectrum, if any. We found only a hint of DMS, with a mixing ratio peaking near $10^{-5.5}$ and a broad tail toward lower abundances. In addition, the retrieval placed strong upper limits on \ce{CO}, \ce{H2O}, \ce{NH3}, and \ce{HCN}.

\begin{deluxetable}{lcc}
\tablecaption{Visit combinations adopted in spectral retrievals.}
\tablehead{
  \colhead{Case} & \colhead{Visits} & \colhead{Code}
}
\startdata
4 Visits   & A1 + B1 + C1 + C2       & \texttt{ExoTR} \\
5 Visits   & A1 + B1 + C1 + C2 + C3  & \texttt{ExoTR} \\
6 Visits   & A1 + B1 + B2 + C1 + C2 + C3 & \texttt{ExoTR} \\
6 Visits   & A1 + B1 + B2 + C1 + C2 + C3 & \texttt{AURA} \\
6 Visits   & A1 + B1 + B2 + C1 + C2 + C3 & \texttt{SCARLET} \\
7 Visits   & A1 + B1 + B2 + C1 + C2 + C3 + D1 & \texttt{ExoTR} \\
\enddata
\label{table:cases_by_visits}
\tablecomments{
  The lettered visit labels (e.g., A1, B1) correspond to individual observations as defined in Table~\ref{table:visitlabels}. 
  The ``Code'' column indicates the atmospheric retrieval frameworks used.}
\end{deluxetable}

In addition to directly combining repeated visits, we noticed visit-to-visit discrepancies in the spectra. The origin of these discrepancies remains uncertain, potentially arising from astrophysical variability, instrumental systematics, or a combination of both. The two remaining observations from our JWST GO program, scheduled for later this year, are expected to provide further insight into this issue. To mitigate the discrepancies in the repeated observations, we constructed a ``shifted average'' dataset by aligning NRS1–NRS2 offsets across visits prior to averaging (see Appendix~\ref{sec:shiftedaverage}). This method significantly improved consistency between visits before co-adding. We then performed atmospheric retrievals on the shifted average data, allowing additional offsets between NRS1 and NRS2 detectors to be treated as free parameters. The resulting posterior solutions are shown as cyan curves in Figure~\ref{fig:exotr_posterior_by_method}, with key atmospheric constraints listed in Table~\ref{table:keyresults_exotr}. Relative to the direct average, the shifted average approach yielded tighter constraints on \ce{CH4} and \ce{CO2}, a slightly higher retrieved abundance of \ce{CO2}, and also deep clouds. The shifted average approach may have better preserved the dominant spectral features. Posteriors for DMS remained unchanged. Upper limits for CO, \ce{H2O}, \ce{NH3}, and HCN were slightly weaker, likely due to the increased number of free offsets. Overall, retrieval results using the shifted average and direct average datasets are very similar. Given the observed variability between visits, we proceed with the shifted average dataset in our analysis, although our main conclusions are robust to this choice.

Also shown in Figure~\ref{fig:exotr_posterior_by_method} are retrieval results using the 6-visit shifted average dataset, but excluding DMS or replacing it by CH$_3$SH or N$_2$O. The posteriors for all other gases remain nearly identical to those from the DMS case, indicating that the atmospheric constraints are robust to the inclusion or exclusion of these additional molecules. The posterior for CH$_3$SH peaks near the same mixing ratio as DMS (around 10$^{-5}$) but exhibits a broader tail toward lower abundances. The posterior for N$_2$O peaks at 10$^{-6}$, also with a tail that extends toward lower abundances. 

Given the lack of a lower bound for either DMS, CH$_3$SH, or N$_2$O, we defined the retrieval without these molecules as our baseline case and used it to assess detection significance. Comparing the Bayesian evidence between the baseline and the cases including DMS, CH$_3$SH, or N$_2$O (Table~\ref{table:exotrposterior}), we found only a marginal increase in evidence, corresponding to a Bayes factor $<\sim2$. In contrast, removing CO$_2$ from the retrieval led to a substantial drop in evidence, indicating a strong detection with a Bayes factor $>200$ ($3.7\sigma$).

We repeated these retrievals with the addition of the possibility to include a haze layer, where the haze opacity is described by the optical properties of soots \citep{OpticalPropertiesofAerosolsandCloudsTheSoftwarePackageOPAC} and parameterized by the particle size and number density in each layer \citep[as applied in][]{libbyroberts2025}. For the retrieved molecular abundances, we found nearly identical results to the retrievals without haze. The Bayesian evidence is slightly lower for the hazy case but the difference is insignificant ($\Delta \mathrm{ln}(Z)=0.75)$.

\begin{figure*}[!htbp]
\centering
\includegraphics[trim=5 5 5 5, clip,width=\textwidth]{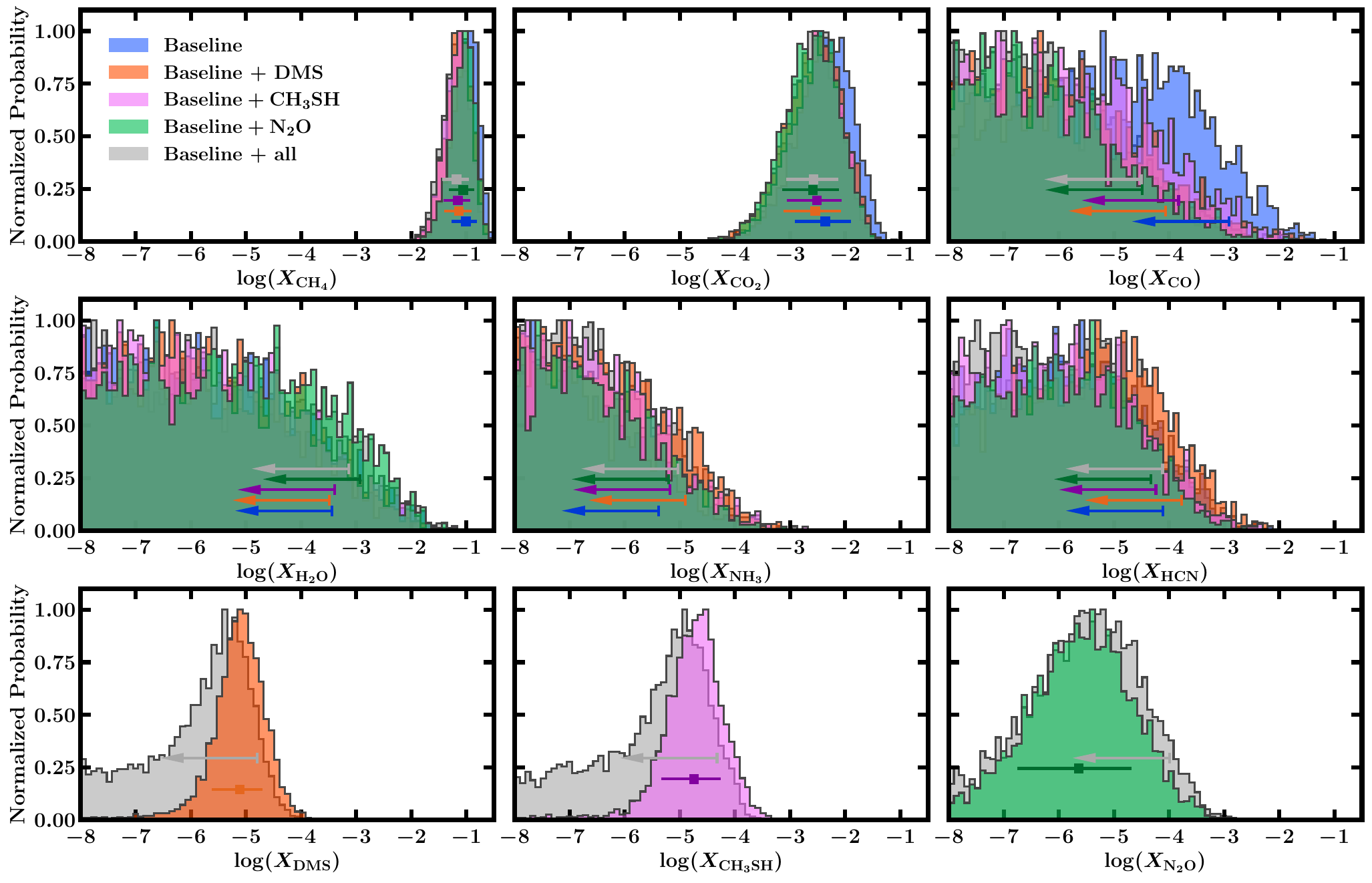}
\caption{Posterior distributions from atmospheric retrievals of K2-18~b's transmission spectra with the \texttt{AURA} retrieval framework. The baseline case refers to the retrieval setup with the six primary molecules listed in Table~\ref{table:exotrprior} and the presence of inhomogeneous clouds/hazes as discussed in Section~\ref{sec:aura}. For \ce{DMS}, \ce{CH3SH}, and \ce{N2O}, we added these gases one at a time to see if including them would be preferred over the baseline setup, followed by a retrieval including all three molecules. All the retrievals use the shifted average data with four offsets as free parameters. All retrievals yield tight constraints on \ce{CH4} and \ce{CO2}, and upper limits for CO, H$_2$O, NH$_3$, and HCN. The inclusion of DMS, CH$_3$SH, or N$_2$O results in weak-to-moderate indications of their presence, though all show tails toward negligible abundances. The points with error bars show the median and 1-$\sigma$ uncertainties, and the arrows denote 2-$\sigma$ upper limits.}
\label{fig:posterior_by_aura}

\end{figure*}

      \begin{deluxetable}{lcc}
		\tablecaption{Key atmospheric constraints for K2-18~b from \exotr\ retrievals.}
		\tablehead{
        \colhead{Parameter} & \colhead{Shifted Average} & \colhead{Direct Average}}
		\startdata
        $\log$(\ce{CH4}) & $-1.06^{+0.24}_{-0.37}$ & $-1.15^{+0.31}_{-0.46}$ \\
       $\log$(\ce{CO2}) & $-3.35^{+0.66}_{-0.80}$ & $-3.87^{+0.78}_{-1.04}$ \\
        $\log$(\ce{CO}) & $<-6.37(1\sigma)$ $-5.01(2\sigma)$ & $<-8.07(1\sigma)$ $-5.47(2\sigma)$ \\
        C/H (solar) & $122^{+93}_{-71}$ & $99^{+103}_{-65}$ \\
        \hline
        $\log$(\ce{H2O}) & $<-5.25(1\sigma)$ $-3.93(2\sigma)$ & $<-6.89(1\sigma)$ $-4.34(2\sigma)$ \\
        \hline
        $\log$(\ce{NH3}) & $<-6.87(1\sigma)$ $-5.59(2\sigma)$ & $<-7.97(1\sigma)$ $-5.62(2\sigma)$ \\
        $\log$\ce({HCN}) & $<-6.20(1\sigma)$ $-5.20(2\sigma)$ & $<-7.27(1\sigma)$ $-5.37(2\sigma)$ \\
		\enddata
\label{table:keyresults_exotr}
       \tablecomments{Median values and 1$\sigma$ uncertainties are shown for the retrieval including DMS. Retrievals excluding DMS or including \ce{CH3SH} or N$_2$O instead yield nearly identical results. Reported upper limits correspond to 68\% (1$\sigma$) and 95\% (2$\sigma$) cumulative probability levels.}
      \end{deluxetable}

\begin{deluxetable}{lcc}
\tablecaption{Bayesian evidence and model preference for atmospheric retrieval scenarios using \exotr.}
\tablehead{
    \colhead{Scenario} & \colhead{$\ln(Z)$} & \colhead{Bayes Factor}}
\startdata
\underline{Shifted average} & & \\
Baseline & 10051.54 & --- \\
Baseline without CO$_2$ & 10046.07 & 237 ($3.8\sigma$) \\
Baseline + DMS & 10051.84 & 1.3 ($<2\sigma$) \\
Baseline + CH$_3$SH & 10051.70 & 1.2 ($<2\sigma$) \\
Baseline + N$_2$O & 10052.32 & 2.2 ($<2\sigma$) \\
\underline{Shifted average with MIRI} & & \\
Baseline & 10259.94 & --- \\
Baseline + DMS & 10260.71 & 2.2 ($<2\sigma$) \\
\underline{Direct average} & & \\
Baseline & 10054.33 & --- \\
Baseline without CO$_2$ & 10050.27 & 58 ($3.3\sigma$) \\
Baseline + DMS & 10054.53 & 1.2 ($<2\sigma$) \\
\enddata
\label{table:exotrposterior}
\tablecomments{Significance is reported relative to the baseline \exotr\ retrieval, which considers a flat cloud deck as discussed in Section~\ref{sec:exotr}. The model selection preference is estimated from the Bayes factor based on \cite{Sellke2001,benneke2013how}. A preference $<2\sigma$ indicates no strong detection of DMS, CH$_3$SH, or N$_2$O. The significance does not change substantially between direct-average and shifted-average datasets. }
\end{deluxetable}

\begin{deluxetable}{lcc}
\tablecaption{Key atmospheric constraints for K2-18~b from \aura\ retrievals.}
\tablehead{
\colhead{Parameter} & \colhead{Hazes} & \colhead{SH}}
\startdata
$\log$(\ce{CH4}) & $-1.09^{+0.21}_{-0.26}$ & $-1.00^{+0.17}_{-0.19}$ \\
$\log$(\ce{CO2}) & $-2.52^{+0.48}_{-0.57}$ & $-2.51^{+0.45}_{-0.50}$ \\
$\log$(\ce{DMS}) & $-5.12^{+0.41}_{-0.49}$ & $-5.36^{+0.44}_{-1.86}$ \\
$\log$(\ce{CO})  & $-4.07~(2\sigma)$ & $-4.33~(2\sigma)$ \\
$\log$(\ce{H2O}) & $-3.52~(2\sigma)$ & $-3.57~(2\sigma)$ \\
$\log$(\ce{NH3}) & $-4.93~(2\sigma)$ & $-5.17~(2\sigma)$ \\
$\log$\ce({HCN}) & $-3.81~(2\sigma)$ & $-4.21~(2\sigma)$ \\
\enddata
\label{table:keyresultsaura}
\tablecomments{These retrievals correspond to the \aura\ retrievals including DMS, following Table~\ref{table:keyresults_exotr} for \exotr\ retrievals. The column marked ``Hazes'' corresponds to the standard \aura\ retrieval discussed in Section~\ref{sec:aura}, which includes inhomogeneous clouds/hazes. The column marked `SH' corresponds to the \aura\ retrievals with no inhomogeneous clouds/hazes but considering stellar heterogeneity and an opaque cloud deck, similar to the standard \exotr\ retrievals. The resulting constraints are comparable. Median values and 1$\sigma$ uncertainties are shown for \ce{CH4}, \ce{CO2}, and \ce{DMS}, and 2$\sigma$ upper limits are shown for the remaining species.}
\end{deluxetable}

\begin{deluxetable}{lcc}
\tablecaption{Bayesian evidence and model preference for atmospheric retrievals using the \aura\ retrieval framework.}
\tablehead{
    \colhead{Scenario} & \colhead{$\ln(Z)$} & \colhead{Bayes Factor}}
\startdata
\underline{Shifted average} & & \\
Baseline & 10056.25 & --- \\
    Baseline without CH$_4$ & 10000.24 & 2.1E+24 (10.8$\sigma$) \\
Baseline without CO$_2$ & 10047.07 & 9.8E+3 (4.7$\sigma$) \\
\hline
Baseline + DMS & 10058.59 & 10.3 (2.7$\sigma$) \\
Baseline + CH$_3$SH & 10058.60 & 10.5 (2.7$\sigma$) \\
Baseline + N$_2$O & 10058.53 & 9.8 (2.7$\sigma$) \\
Baseline + all & 10059.78 & 33.8 (3.1$\sigma$) \\
\enddata
\label{table:exotrposterior_aura}
\tablecomments{Significance is reported relative to the baseline \aura\ retrieval, which considers inhomogeneous clouds/hazes as discussed in Section~\ref{sec:aura}. The model preference is obtained similarly to that in Table~\ref{table:exotrposterior}. In the upper part of the table, the model preference for a molecule is determined by the Bayes factor of the baseline model with higher $\ln(Z)$ relative to the model without the molecule concerned. In the lower part of the table, the preference is determined by Bayes factor of a model with the molecule added relative to the baseline model. Baseline + all denotes a model that simultaneously includes DMS, CH$_3$SH, and N$_2$O in addition to the baseline model.}
\end{deluxetable}

With the recent publication of K2-18~b’s transmission spectrum from MIRI/LRS observations \citep{madhusudhan2025miri}, we performed additional retrievals incorporating this dataset alongside the six visits from our standard dataset. Including the MIRI data required fitting one additional offset (between G235H NRS1 and LRS), while the rest of the retrieval setup remained unchanged from the baseline scenarios. We found that the posterior distributions of the molecular abundances remain largely consistent with the baseline cases (Figure~\ref{fig:exotr_posterior_miri}). With the MIRI/LRS dataset, the DMS posterior shows a peak near a mixing ratio of $10^{-5.5}$ with a long tail toward zero abundance, though the peak becomes more pronounced when the MIRI/LRS dataset is included. However, the Bayes factor between the two MIRI-inclusive cases (with and without DMS) is only 2.2, corresponding to a significance of less than $2\sigma$ for DMS (Table~\ref{table:exotrposterior}).

Finally, we also performed retrievals on the K2-18~b dataset in the chronological order of observation to evaluate how the posterior distributions evolve with each additional visit (see the retrieval configurations in Table~\ref{table:cases_by_visits}). With the first 4 visits, we found a prominent posterior peak for the CO$_2$ mixing ratio near 10$^{-2.5}$, though with a long tail extending toward lower abundances. The DMS posterior also showed a more pronounced peak at this stage (Figure~\ref{fig:exotr_posterior_by_visit}). Adding a fifth visit -- an additional NIRSpec/G395H observation -- tightened the constraint on CO$_2$ and introduced a tentative signal for HCN. However, the DMS posterior weakened, exhibiting a broader tail toward negligible abundances. Adding the final visit with NIRSpec/G235H caused the HCN signal to disappear and further weakened the hint for DMS. Overall, we found that each successive visit consistently improved the upper limits on undetected species (CO, H$_2$O, NH$_3$, and potentially HCN), while reducing the apparent significance of DMS. The disappearance of tentative signals for gases by repeated visits serves as a cautionary tale for future interpretation of exoplanet spectra.

\subsubsection{AURA retrievals}
\label{sec:aura}
We also performed independent spectral retrievals using the \texttt{AURA} framework \citep{pinhas2019h2o, madhusudhan2023carbon, constantinou2024vira}. The model computes transmission spectra for a plane parallel atmosphere in hydrostatic equilibrium and LTE, with parametric chemical abundances and temperature structure. The model includes prescriptions for considering inhomogeneous clouds/hazes, non-isothermal temperature structure, and stellar heterogeneities. The \texttt{AURA} framework has been used extensively for atmospheric retrievals of exoplanets using JWST data, including previous near-infrared observations of K2-18~b, similar to those in the present work \citep{madhusudhan2023carbon}. In order to obtain a robust comparative analysis, we perform independent retrievals to accompany the \exotr\ retrievals discussed above using the same data and chemical species explored. For the chemical species, we consider the same six molecules as \exotr\ as our primary species: \ce{H2O}, \ce{CH4}, \ce{CO2}, \ce{NH3}, \ce{CO2}, \ce{HCN}. For an independent assessment, we use the cross sections for the primary molecules from a separate opacity database, publicly available with the latest version of the POSEIDON retrieval code \citep{macdonald2024}. The corresponding line lists are from the following sources: H$_2$O \citep{2018MNRAS.480.2597P}, CH$_4$ \citep{2024MNRAS.528.3719Y}, NH$_3$ \citep{2019MNRAS.490.4638C}, HCN \citep{2014MNRAS.437.1828B}, CO \citep{2015ApJS..216...15L} and CO$_2$ \citep{2020MNRAS.496.5282Y}. For the three other molecules explored, namely DMS, CH$_3$SH, and N$_2$O, we use the cross sections from the HITRAN database \citep{Gordon2022} similar to the \exotr\ retrievals discussed above. 

\begin{figure}[!htbp]
\centering
\includegraphics[width=0.45\textwidth]{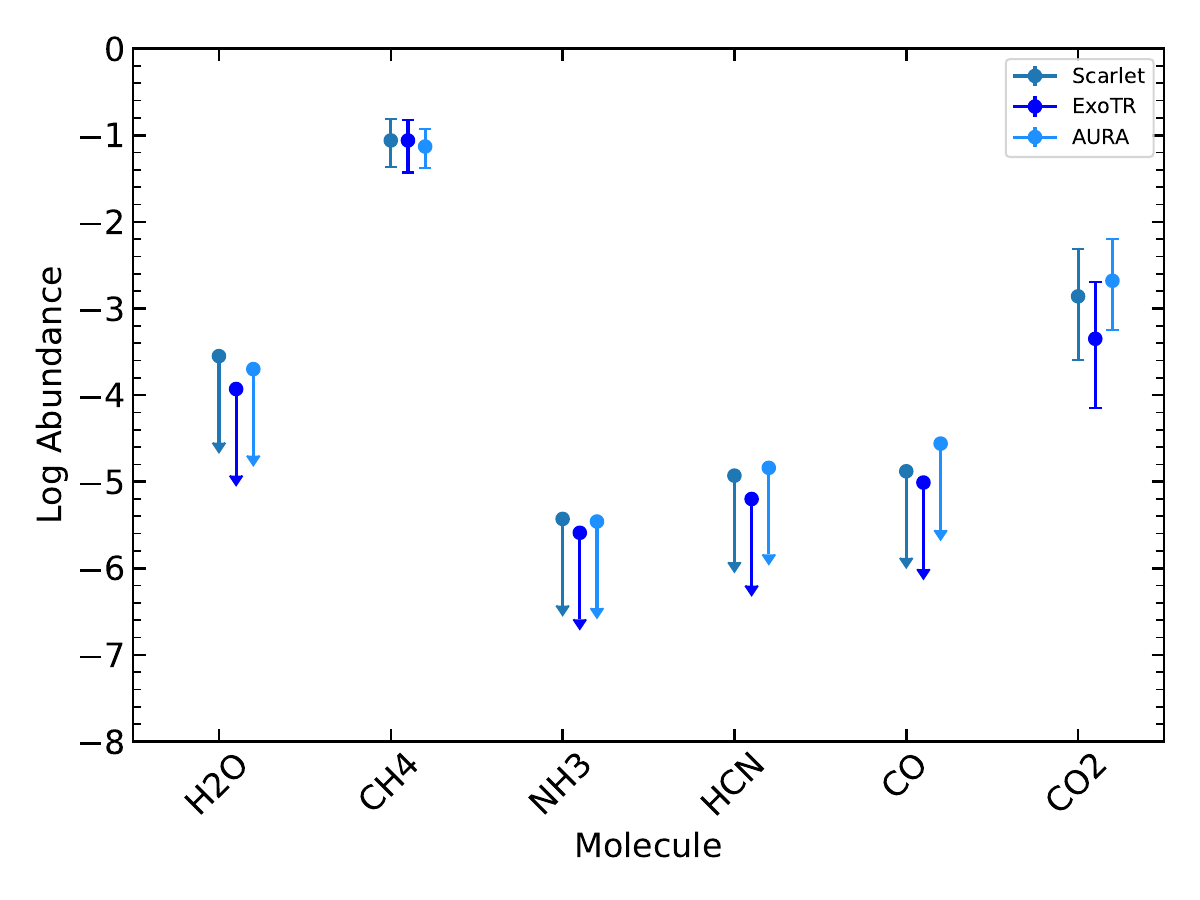}
\caption{Molecular abundance constraints obtained from the baseline retrieval analyses using \exotr, \texttt{AURA}, and \texttt{SCARLET} frameworks. The two-sided error bars show the median and $1\sigma$ uncertainties on the CH$_4$ and CO$_2$ abundances, while the arrows show the 2$\sigma$ (95\% probability) upper limits for the other molecules. Results are shown for the retrievals on the shifted average spectrum from 6 visits. We found strong agreement between \texttt{SCARLET}, \exotr, and \texttt{AURA}, with the retrieved abundances of CH$_4$ and CO$_2$ agreeing well within $1\sigma$ and the upper limits consistent within 0.5 dex for all other species.}
\label{fig:compare_constraints}
\end{figure}

We first conducted the retrievals using the same setup as the \exotr\ retrievals to verify the consistency between the retrieval frameworks. We started with a baseline model containing the six primary molecules, an isothermal temperature structure, and an opaque cloud deck with the cloud-top pressure as a free parameter. We performed the retrievals on the same NIRISS and NIRSpec data as used in the \exotr\ retrievals, considering the shifted average NIRSpec data with four free parameters for offsets between the detectors. We obtained nearly identical constraints on the six molecular abundances for this baseline retrieval. The constraints also remained comparable to the \exotr\ values when considering stellar heterogeneities and additional opacity contribution from DMS, as shown hello thank youFigures~\ref{fig:compare_constraints} and \ref{fig:baseline_posteriors_all}. 

We then proceeded to conduct atmospheric retrievals with the standard \aura\ framework described above, with considerations of inhomogeneous clouds/hazes, non-isothermal pressure-temperature profile, and stellar heterogeneities, while keeping the chemical species same as the \exotr\ retrievals. The inhomogeneous clouds/hazes are modeled as a combination of a gray cloud deck and a generic power law scattering allowing for a wide range of scattering slopes and amplitudes possible due to Mie scattering \citep{ pinhas2017,ohno2020super}.
Based on Bayesian model comparisons, we found that a model including the inhomogeneous clouds/hazes is preferred over one including no clouds/hazes at a Bayes factor of 4.6 (2.3$\sigma$). 
We also found that the effects of star spots and hazes are somewhat degenerate, as either of these can provide a blue-ward slope in the spectrum that is required to explain the NIRISS/SOSS data, with a slightly higher preference for inhomogeneous clouds/hazes. When both hazes and stellar heterogeneity were included in the model, we found no significant preference for stellar heterogeneity. This is consistent with the finding in \citet{madhusudhan2023carbon} who also found a model preferences for the inhomogeneous clouds/hazes. We found no significant preference for a non-isothermal $P$-$T$ profile. We therefore chose to proceed with a model including inhomogeneous clouds and hazes, no stellar heterogeneities, and an isothermal $P$-$T$ profile as the baseline model for \texttt{AURA} retrievals, while still verifying that the key results are consistent when including stellar heterogeneities.

We used this baseline model to determine the atmospheric constraints as well as detection significances for key molecules. The posterior distributions for various retrievals are shown in Figure~\ref{fig:posterior_by_aura}. Our abundance constraints for the six primary molecules are consistent across all the cases shown in Figure~\ref{fig:posterior_by_aura}, and with those of \exotr\ for similar data and model considerations, as shown in Table~\ref{table:keyresultsaura}. Considering the \aura\ retrieval with DMS included, we retrieved a CH$_4$ abundance of $\log$(\ce{CH4}) = -1.09$^{+0.21}_{-0.26}$ and a CO$_2$ abundance of $\log$(\ce{CO2}) = -2.52$^{+0.48}_{-0.57}$, which are consistent with previous estimates within the 1$\sigma$ uncertainties \citep{madhusudhan2023carbon}. We obtained 2$\sigma$ upper-limits on the log-mixing-ratios of the remaining molecules to be -3.52 for H$_2$O, -4.93 for NH$_3$, -4.07 for CO and -3.81 for HCN. These constraints are consistent with the \exotr\ constraints summarized in Table~\ref{table:keyresults_exotr} within $\sim0.5$ dex. The most notable difference is the abundance of CO$_2$, which is higher than the \exotr\ results by $\sim$1$\sigma$. We investigated the origin of this difference and found that it came from subtle differences in the calculations of the CO$_2$ cross sections, such as the assumptions on the far wing cutoff and the ratio of isotopologues. The Bayesian evidence for key molecules is shown in Table~\ref{table:exotrposterior_aura}. Among the primary molecules, we robustly detect \ce{CH4} with a Bayes factor of 2.1$\times$10$^{24}$ (10.8 $\sigma$) and \ce{CO2} at 9.8$\times$10$^{3}$ (4.7 $\sigma$).

As with \exotr\, we also investigated possible contributions from three other molecules: DMS, CH$_3$SH and N$_2$O (Figure~\ref{fig:posterior_by_aura}). All the three molecules show peaks in the posterior distributions with low-abundance tails and only weak-to-moderate preference for adding these molecules to the baseline model. When adding each of the molecules individually, we found a Bayes factor for including DMS, CH$_3$SH and N$_2$O of 10.3 (2.7 $\sigma$), 10.5 (2.7 $\sigma$) and 9.8 (2.7 $\sigma$), respectively. When all three are included in the model, DMS and CH$_3$SH are strongly degenerate and not individually favored at any modest significance. However, the preference for including all three molecules together is 33.8 (3.1 $\sigma$) relative to the baseline model. The corresponding log-mixing-ratios retrieved when each of them is added individually are:  $\log$(\ce{DMS}) = -5.12$^{+0.41}_{-0.49}$, $\log$(\ce{CH3SH}) = -4.75$^{+0.48}_{-0.59}$, and $\log$(\ce{N2O}) = -5.64$^{+0.95}_{-1.11}$. 

For robustness, we additionally assessed the effect of model assumptions on the significance of DMS. We considered the effect of stellar heterogeneity (SH) and found no significant difference. Furthermore, as noted above, we found minor differences in the CO$_2$ abundance, by $\sim$0.5 dex, between the \texttt{AURA} and \exotr\ retrievals due to differences in the CO$_2$ opacities. We therefore conducted retrievals using the same CO$_2$ opacities as used in \exotr\ and found slightly lower model preference for DMS at 2.5$\sigma$ considering the baseline+DMS model. Finally, when considering no inhomogeneous hazes in the model, we found that the Bayes factor is reduced to only 3.2 (2.1$\sigma$) with no SH and 2.6 (2$\sigma$) with SH. Considering a typical error in evidence estimates of $\sim$0.1-0.2 $\sigma$, the lower significance in the absence of inhomogeneous clouds/hazes is consistent with the $<2\sigma$ evidence inferred by the \exotr\ and \texttt{SCARLET} retrievals using similar setups. Overall, our results are consistent across the three retrieval codes for similar model assumptions, while the inclusion of inhomogeneous clouds/hazes increases the statistical preference for models including DMS and other minor species.

Overall, considering the model with the highest Bayesian evidence across all our cases, i.e., one with inhomogeneous clouds/hazes, we found weak to moderate evidence for excess absorption beyond the six primary molecules at 2.5-3.1$\sigma$. This excess absorption may be contributed by DMS or CH$_3$SH or N$_2$O or a combination thereof, or yet another unexplored molecule. The present signal, while still tentative, is stronger than that obtained previously in the near-infrared under a similar retrieval setup. For example, no significant evidence was found for DMS previously when considering detector-level offsets in NIRSpec \citep[i.e., the two-offset case in][]{madhusudhan2023carbon}, as the offsets introduce uncertainty to the continuum. That the present data may provide $\sim2-3\sigma$ level model preference with four detector-level offsets is notable, but still insufficient to claim a detection, similar to previous studies. Future studies with more extensive retrievals, including native resolution spectra and a more extensive set of molecules, could further constrain the evidence for one or more of these species in the atmosphere of K2-18~b. 

\subsubsection{SCARLET retrievals}

We also performed retrievals on the measured transmission spectra of K2-18~b using the \texttt{SCARLET} framework \citep{benneke2012atmospheric, benneke2013how,benneke2015strict, benneke2019water, Benneke_2019_methanedepleted,benneke2024jwst}. \texttt{SCARLET} parametrizes the molecular abundances, the cloud opacities, the temperature in the photosphere, and the stellar contamination from the transit light source effect to fit atmospheric parameters to the observed transmission spectra. The retrievals use nested sampling \citep{Skilling2004} to obtain the evidence of the models and the posterior probability distribution of the parameters that describe the atmosphere. During the exploration, for each sample of parameters, \texttt{SCARLET} models a 1D atmosphere of 60 pressure layers in hydrostatic equilibrium, computes the molecular and cloud opacities, and then calculates a simulated transit spectrum for these parameters, which is compared to the data in the likelihood evaluation. For each iteration, the planetary radius at 1 mbar is minimized by chi-square to fit the observed data. The models are created at a resolving power of 20000. 

To test robustness, we performed atmosphere retrievals on both the direct and shifted average transit spectra using the same parameterization as described with the \exotr\ framework in Section \ref{sec:exotr}; specifically, we fit for an isothermal photosphere temperature and modeled the clouds as a gray cloud deck. We used the same molecules and adopted the same priors as those described in Table~\ref{table:exotrprior}. However, the \texttt{SCARLET} implementation of TLS differs from that of \exotr\ in that both spots (cooler activity regions) and faculae (hotter activity regions) are modeled at the same time \citep{Piaulet2024}. Hence, \texttt{SCARLET} uses five parameters describing the TLS effect: the spots covering fraction, spots temperature contrast, faculae covering fraction, faculae temperature contrast, and the photosphere temperature. For the covering fraction, we used a uniform prior from 0 to 0.5 of the stellar surface for both the spots and faculae. The prior on the spots temperature contrast extends from -1000 to -50 K, and that on the faculae temperature contrast from 50 to 1000 K.

We found strong agreement between the molecular constraints derived from the \texttt{SCARLET}, \exotr, and \texttt{AURA} retrievals (Figure~\ref{fig:compare_constraints}). In retrievals both on the direct average spectrum and on the shifted average spectrum, we found that the CH$_4$ and CO$_2$ abundances are consistent well within $1\sigma$ uncertainties. The $2\sigma$ upper limits on the other molecules agree within 0.5 dex. We performed further tests using the \texttt{SCARLET} framework in which we tested retrievals without the TLS effects, with the addition of super-Rayleigh hazes, and with the use of the temperature model from \citet{Madhusudhan2009} instead of an isotherm, similar to the \texttt{AURA} retrievals. We confirmed that the results are robust to these changes, i.e., the CH$_4$ and CO$_2$ abundances agree within uncertainties in all cases. Finally, we also assessed the detection strength of DMS and CH$_3$SH by comparing the Bayesian evidence of models with and without these molecules, and found no robust evidence for their existence, with a Bayes factor of 1.97 and 1.05 for DMS and \ce{CH3SH}, respectively. Our tests with the \texttt{SCARLET} framework showed that the atmospheric inference made here is insensitive to the framework or parameterization used for the spectral retrieval.

\subsection{Self-Consistent Models}
\label{sec:epacris}

To gain deeper insights into the atmospheric physics and chemistry of K2-18~b, it is crucial to compare observations and atmospheric retrievals with predictions from self-consistent models. To this end, we performed one-dimensional photochemical and kinetic-transport simulations using the \texttt{EPACRIS} framework, exploring key parameters such as the \ce{H2O}-to-\ce{H2} ratio and the eddy diffusion coefficient ($K_{\rm zz}$). \texttt{EPACRIS} is a state-of-the-art modeling tool that simultaneously solves for the atmospheric pressure–temperature profile and chemical composition. Its climate module computes non-grey radiative-convective equilibrium temperature profiles and has been applied to model a variety of exoplanet atmospheres, including those of K2-18~b and TOI-270~d \citep{yang2024chemical}, the secondary atmosphere of the hot rocky planet 55~Cancri~e \citep{hu2024secondary}, the potential atmosphere on the cold water world LHS~1140~b \citep{damiano2024}, and a potential volcanic atmosphere on the sub-Earth L~98-59~b \citep{belloarufe2025l9859}. The chemistry module of \texttt{EPACRIS} uniquely generates chemical reaction networks automatically based on planetary conditions, enabling efficient and accurate modeling of disequilibrium chemistry in planetary atmospheres. \texttt{EPACRIS} has been applied to explain JWST observations of hot Jupiters \citep{yang2024epacris} and used to map chemical signatures in water-rich atmospheres on temperate sub-Neptunes such as K2-18~b \citep{yang2024chemical}.

\subsubsection{Massive atmosphere scenarios}
\label{sec:epacris_massive}

\begin{figure*}[!htbp]
\centering
\includegraphics[width=0.99\textwidth]{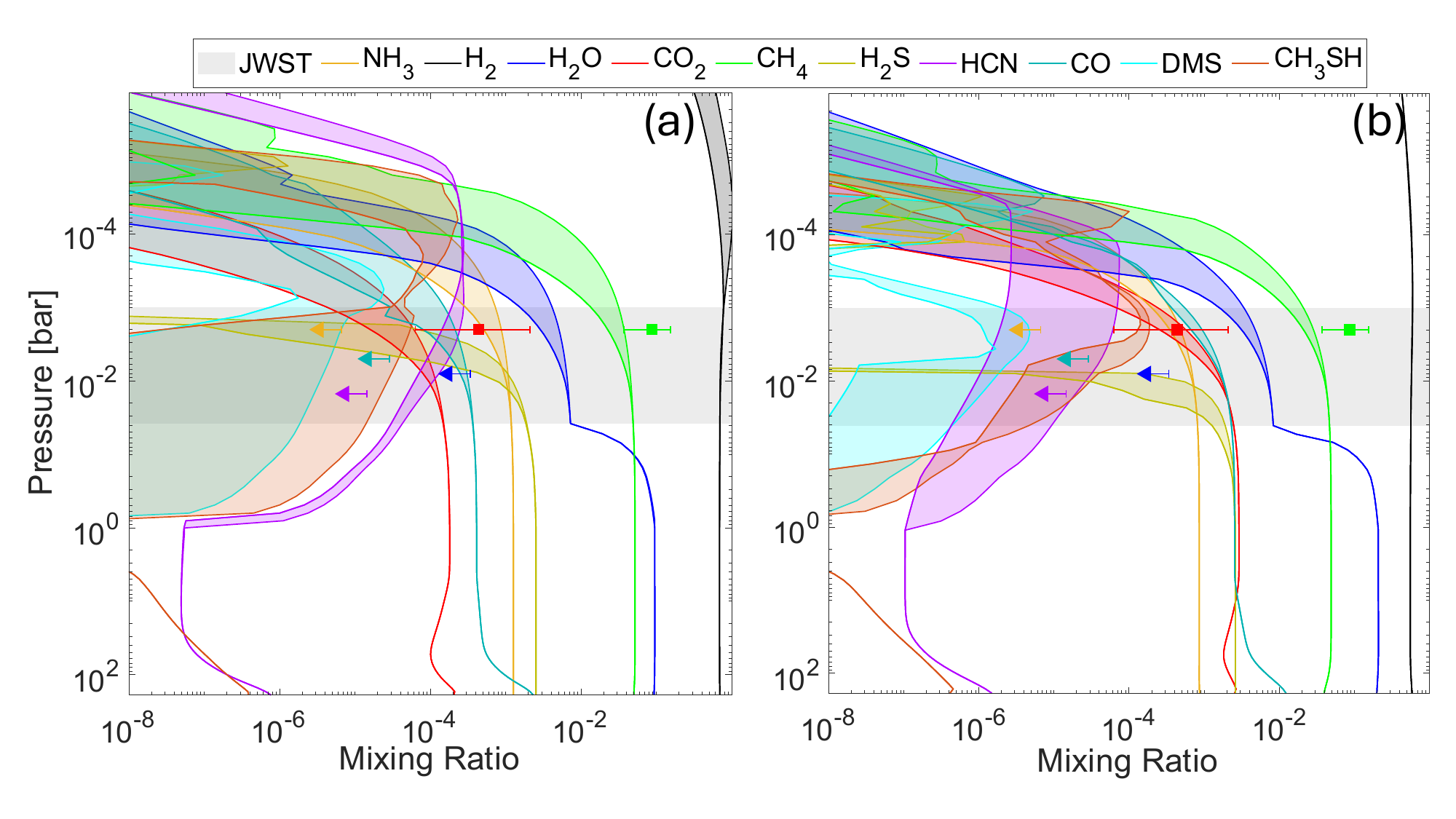}
\caption{
Self-consistent atmospheric chemistry models for K2-18~b assuming a massive gas envelope. Panel (a) corresponds to a standard $100\times$ solar metallicity envelope, while panel (b) adopts an elevated \ce{H2O}-to-\ce{H2} ratio of 25:75. The gray-shaded region marks the pressure range probed by the transmission spectra. Colored shaded areas represent the model sensitivity to variations in the eddy diffusion coefficient ($K_{\rm zz}$), as described in Figure~\ref{fig:model_inputs}. Colored square symbols with error bars indicate the retrieved $1\sigma$ abundance constraints for key species from Table~\ref{table:keyresults_exotr}, and left-pointing arrows denote $2\sigma$ upper limits. The models predict substantial photochemical production of HCN, DMS, and \ce{CH3SH} within the observable atmospheric layers.}
\label{fig:vmr}
\end{figure*}

For the massive atmosphere scenarios, we adopted the pressure–temperature profiles and chemical networks developed in \citet{yang2024chemical}. Guided by the predictions of \citet{yang2024chemical} and assuming the Bond albedo of 0.3, we considered scenarios ranging from a standard $100\times$ solar metallicity envelope to one with an elevated \ce{H2O}-to-\ce{H2} ratio of 25:75. We chose to retain the nitrogen and sulfur abundances at the $100\times$ solar baseline to evaluate whether photochemical processes alone can account for the observed depletion of atmospheric \ce{NH3} and the production of organosulfur species at detectable levels. Since our focus in this analysis is on photochemical production and loss in the upper atmosphere, we explored a range of eddy diffusion coefficients at high altitudes (Figure~\ref{fig:model_inputs}). Molecular diffusion in the upper atmosphere was modeled following \cite{chapman1990mathematical, banks2013aeronomy}, with advection discretized using a first-order upwind scheme and diffusion handled via a centered-difference scheme for stability. In addition to the chemical networks in \citet{yang2024chemical}, we incorporated reactions relevant to the formation and destruction of dimethyl sulfide (DMS) and methyl mercaptan (\ce{CH3SH}), using both our automatic reaction network generation for \ce{CH3SH}-related chemistry (see Section~\ref{sec:dms}) and recent updates with regard to DMS chemistry from \citet{tsai2024biogenic}. To assess the influence of stellar spectral variations, we substituted the host star spectrum with those of similar M dwarfs -- GJ~176, GJ~436, and GJ~876 -- using data from the MUSCLES Treasury Survey \citep{france2016muscles}, and found minimal sensitivity in the resulting atmospheric compositions. The outcomes of the massive-atmosphere simulations are shown in Figure~\ref{fig:vmr}.

As shown in Figure~\ref{fig:vmr}, the two model scenarios -- the standard $100\times$ solar metallicity envelope and the case with an elevated \ce{H2O}-to-\ce{H2} ratio -- successfully bracket the observed \ce{CO2} abundance, reproducing its lower and upper bounds, respectively. The retrieved \ce{CH4} abundance is also consistent with both models. Adjusting the \ce{H2O}-to-\ce{H2} ratio beyond this range leads to overproduction or underproduction of \ce{CO2}, suggesting that K2-18~b's envelope composition lies between these two endmembers. Assuming a Bond albedo of 0.3 when calculating the pressure-temperature profile \citep{yang2024chemical}, these models overpredict the atmospheric \ce{H2O} abundance relative to spectral retrieval constraints. A higher Bond albedo would cool the atmosphere and enhance water condensation, which can lower the atmospheric \ce{H2O} to levels consistent with the retrievals (see Section~\ref{sec:coldtrap} for further discussion).

Both model scenarios predict significantly more \ce{NH3} than the retrieved upper limit, suggesting that photochemical processes alone cannot effectively deplete \ce{NH3} in the atmosphere of K2-18~b. Then, the lack of \ce{NH3} must arise from processes operating deeper in the envelope or from an intrinsically low nitrogen abundance. We explore these possibilities further in Section~\ref{sec:nh3prob}. Our simulations also show that photochemical reactions produce substantial HCN within the pressure levels probed by transmission spectroscopy. In the $100\times$ solar envelope, the HCN mixing ratio reaches $\sim10^{-4}$, consistent with previous results \citep{hu2021photochemistry}. In the more water-rich scenario, the HCN abundance is sensitive to the choice of $K_{\rm zz}$ (Figure~\ref{fig:vmr}). The overabundance of \ce{NH3} and \ce{HCN} causes the model to overpredict the transit depth near $\sim3\ \mu$m compared to the data (Figure~\ref{fig:model_fit}). As a variant, we incorporated the HCN formation pathways involving electronically excited nitrogen atoms from \ce{N2} photolysis \citep[e.g.,][]{vuitton2019simulating}, but found minimal impact on the resulting HCN abundance. This confirms that HCN formation is primarily driven by \ce{NH3} photolysis, as described in \citet{hu2021photochemistry}. Consequently, any depletion of atmospheric \ce{NH3} would lead to a corresponding reduction in HCN production.

Interestingly, we found substantial photochemical formation of \ce{CH3SH} and DMS in these models. Using sulfur ultimately sourced from deep-atmospheric \ce{H2S}, photochemically produced \ce{CH3SH} builds up to mixing ratios of $10^{-5}$–$10^{-4}$ at the pressure levels probed by transmission spectra, and DMS reaches $10^{-6}$–$10^{-5}$, with peak mixing ratios near $\sim1$ mbar (Figure~\ref{fig:vmr}). As detailed in Section~\ref{sec:dms}, the formation of these organosulfur compounds involves methyl radicals from \ce{CH4} photolysis and CO as a catalyst. Notably, the natural formation of \ce{CH3SH} already exceeds what is allowed by the data and would produce a potentially discernible discrepancy in the transmission spectrum near $\sim4.8\ \mu$m (Figure~\ref{fig:model_fit}). The natural production of DMS is consistent with the tentative signal suggested by the data. Although \ce{SO2} and OCS are omitted from Figure~\ref{fig:vmr}, our model predicts \ce{SO2} mixing ratios of $\sim4\times10^{-7}$–$4\times10^{-5}$ near 1 mbar in both scenarios. OCS remains below $\sim10^{-6}$ throughout the $100\times$ solar-metallicity atmosphere, but can increase to $2\times10^{-5}$ in the more water-rich case. We searched for \ce{SO2} and \ce{OCS} in the transmission spectra but found no evidence of their presence; the $2\sigma$ upper limits of $\sim5\times10^{-5}$ remain consistent with our model predictions.

Finally, both model scenarios substantially overpredict the atmospheric CO mixing ratio compared to the spectral retrievals (Figure~\ref{fig:vmr}). In these models, the \ce{CO2}-to-CO ratio within the pressure levels probed by the transmission spectra ranges from 0.1 to 1. As shown in \citet{yang2024chemical}, increasing the bulk water content of the envelope above $25\%$ could make this ratio exceed unity, bringing it closer to the indication from the spectral retrievals. However, such an increase would also increase the absolute \ce{CO2} abundance and the \ce{CO2}-to-\ce{CH4} ratio, leading to inconsistencies with the observed data. We discuss the implications of the CO overprediction in Section~\ref{sec:co2co}.

\subsubsection{Small atmosphere scenarios}
\label{sec:epacris_small}

We also studied thin-atmosphere scenarios for K2-18~b in light of the observed transmission spectra. Using the climate module of EPACRIS, we calculated the pressure–temperature profile of a small atmosphere assuming a \ce{CH4} mixing ratio of $10^{-1}$ and a \ce{CO2} mixing ratio of $5 \times 10^{-4}$ at the lower boundary, as informed by the spectral retrievals. The \ce{H2O} mixing ratio was computed self-consistently based on the vapor pressure as a function of temperature. We explored surface pressures of 1, 3, and 10 bars, and Bond albedos of 0, 0.3, 0.5, and 0.7. We found that the pressure–temperature profile above 1 bar is largely insensitive to the assumed surface pressure, while the extent of water depletion due to condensation is controlled by the Bond albedo (see Section~\ref{sec:coldtrap} for details). A Bond albedo of 0.5 yields a \ce{H2O} mixing ratio consistent with the observations, and we focused on this case for our subsequent atmospheric chemistry modeling.

We then ran the chemical module of \texttt{EPACRIS} to evaluate the effects of photochemistry in this atmosphere. CO is produced via photolysis of \ce{CO2} and can build up to a mixing ratio that is lower than that of \ce{CO2} by a factor of a few, consistent with \citet{hu2021unveiling}. As such, \ce{CO} remains overpredicted relative to the spectral retrievals, but the discrepancy is less severe than in the massive-atmosphere scenarios and remains within observational uncertainties (Figure~\ref{fig:model_fit}).

\begin{figure}
    \centering    
    \includegraphics[width=0.49\textwidth]{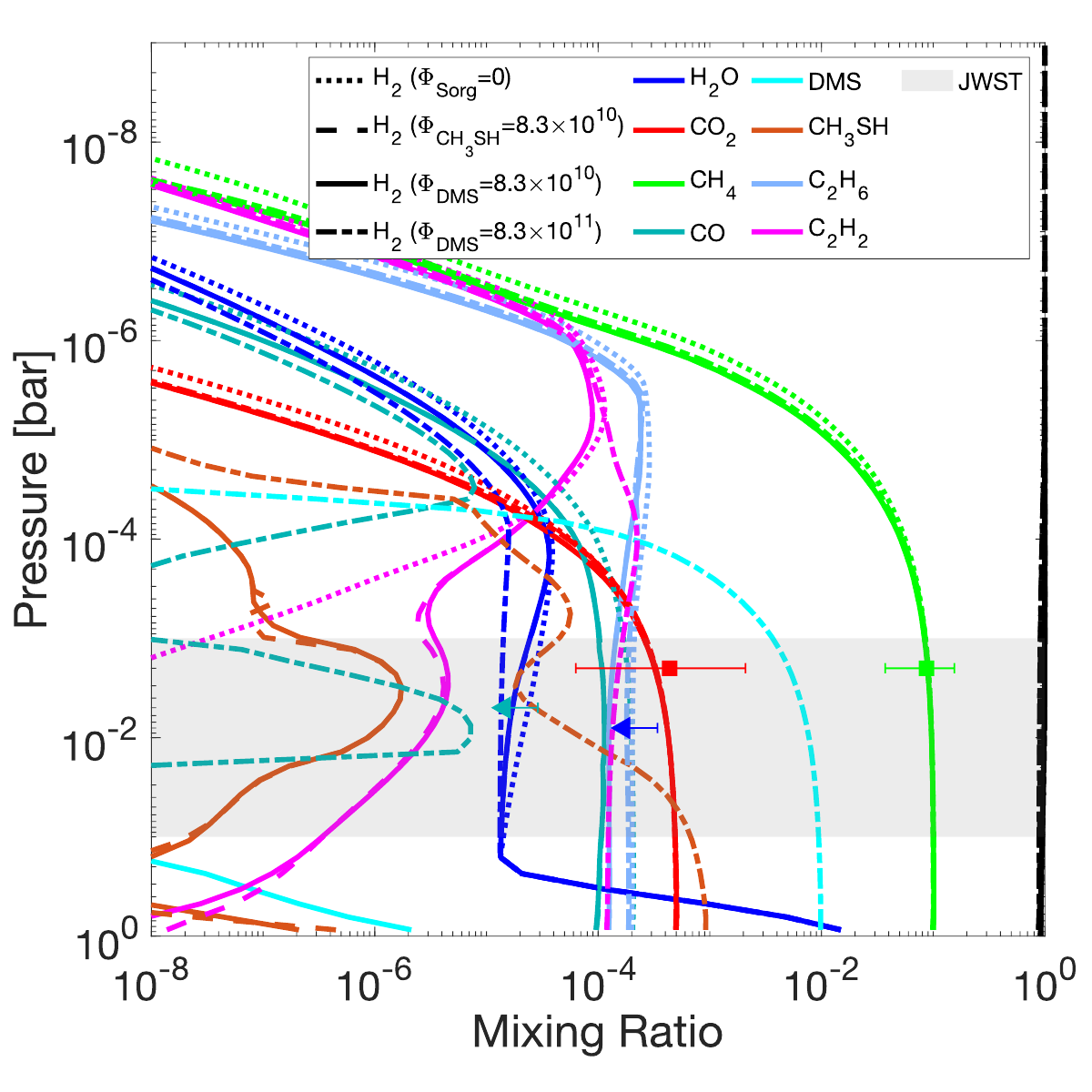}
     \caption{Self-consistent atmospheric chemistry models for K2-18~b assuming a thin atmosphere atop a liquid-water ocean. Solid and dash-dotted lines represent the model with a DMS flux that is $20\times$ and $200\times$ the modern Earth flux ($8.3\times10^{10}$ and $8.3\times10^{11}$ molecules cm$^{-2}$ s$^{-1}$, respectively). Dashed lines correspond to the model with a \ce{CH3SH} flux that is $100\times$ the modern Earth flux (also $8.3\times10^{10}$ molecules cm$^{-2}$ s$^{-1}$), and dotted lines show the model without any sulfur flux. The gray-shaded area marks the pressure range probed by the transmission spectra. Colored square symbols with error bars indicate the retrieved $1\sigma$ abundance constraints for key species from Table~\ref{table:keyresults_exotr}, while left-pointing arrows denote $2\sigma$ upper limits.}  
    \label{fig:hycean}
\end{figure}

We also explored a range of organosulfur compound fluxes at the lower boundary to represent potential biogenic emissions from the ocean \citep{domagal2011using,tsai2024biogenic}. Specifically, we examined surface emissions of DMS from $20\times$ to $200\times$ the modern Earth flux. We found that for fluxes of $20\times$ or $125\times$, the DMS mixing ratio remains very low, but \ce{CH3SH} (produced in this case by photochemistry that starts with DMS) can build up to $>10^{-6}$ in the pressure levels probed by the transmission spectra (Figure~\ref{fig:hycean}). However, when the flux is increased to $200\times$, the mixing ratio of DMS abruptly jumps to $\sim10^{-3}$ (Figure~\ref{fig:hycean}). Correspondingly, the mixing ratio of \ce{CH3SH} increases to $\sim10^{-4}$, and that of CO drops to $\sim10^{-6}$. A similar threshold behavior was reported by \citet{tsai2024biogenic}, who found the transition occurred at a somewhat lower flux, between $20\times$ and $50\times$. This difference may be due to the fact that \citet{tsai2024biogenic} included multiple organic sulfur fluxes, whereas our models include only a single sulfur source. When assuming a surface flux of \ce{CH3SH}, even a $100\times$ modern-Earth \ce{CH3SH} flux yields only a mixing ratio of $\sim10^{-6}$.

In all cases, the photolysis of DMS or \ce{CH3SH} produces \ce{CH3} radicals, which subsequently lead to the formation of \ce{C2H6} and \ce{C2H2}. Assuming a deposition velocity of $10^{-5}$ m s$^{-1}$ for \ce{C2H6} to represent carbon loss via organic haze formation and surface deposition \citep{hu2021unveiling}, our models predict a \ce{C2H6} mixing ratio of approximately $2\times10^{-4}$. When DMS accumulates in the atmosphere, \ce{C2H2} builds up to $\sim10^{-4}$. Although these hydrocarbons are not detected in the current spectra (but with upper limits permitting $\sim10^{-2}$), they contribute to a photochemical shielding effect that facilitates further accumulation of DMS and \ce{CH3SH}. Similar hydrocarbon buildup in response to elevated organosulfur emissions has been noted in previous studies \citep{domagal2011using,tsai2024biogenic}. Overall, our small-atmosphere models show that a shallow, ocean-bearing atmosphere can naturally explain the key spectral features observed on K2-18~b, and it requires high surface biogenic fluxes to build detectable levels of organosulfur compounds (most likely in the form of \ce{CH3SH} instead of DMS).

\section{Result: A Water-Rich Planet}
\label{sec:cratio}

For a temperate exoplanet like K2-18~b, the atmospheric water abundance inferred from transmission spectroscopy does not directly reveal whether the planet possesses a water-rich envelope. This is because water can condense out of the observable atmosphere to form clouds, leaving behind a relatively dry upper atmosphere. Consequently, we must rely on alternative chemical tracers -- such as \ce{CO2} -- to assess the bulk composition of the envelope. A water-rich envelope should not be assumed by default; for example, a solar-metallicity atmosphere would contain less than 0.1\% water by volume, and the proposed ``soot world'' scenario predicts an even drier atmosphere \citep{bergin2023exoplanet}.

The transmission spectrum presented in this work reveals robust detections of both \ce{CH4} and \ce{CO2}, resolving the debate between \citet{madhusudhan2023carbon} and \citet{schmidt2025comprehensive}. As summarized in Table~\ref{table:keyresults_exotr}, the retrieved \ce{CH4} volume mixing ratio lies between $10^{-1.4}$ and $10^{-0.8}$ ($1\sigma$), while \ce{CO2} ranges from $10^{-4.2}$ to $10^{-2.7}$ (based on \exotr\ retrievals).
\aura\ retrievals obtained slightly higher constraints for \ce{CO2} between $10^{-3.1}$ and $10^{-2.0}$. These constraints are remarkably consistent across the spectral retrieval frameworks used (Figure~\ref{fig:compare_constraints}), as well as different retrieval settings on the stellar heterogeneity, photochemical haze, and atmospheric pressure-temperature profiles (Section~\ref{sec:retrievals}).
With repeated JWST observations, we have significantly tightened the constraints on \ce{CH4} and confirmed the presence of \ce{CO2} in K2-18~b’s atmosphere. Relative to the findings of \citet{madhusudhan2023carbon}, the inferred \ce{CH4} abundance is higher by $0.6-0.7$ dex and the \ce{CO2} abundance is lower by $0.4-1.2$ dex, but all are consistent within $1-2\sigma$ uncertainties.

We can also derive the atmospheric metallicity in terms of the carbon-to-hydrogen ratio (C/H) by summing the retrieved abundances of \ce{CH4}, \ce{CO2}, and \ce{CO}. 
We found that the C/H ratio is approximately 100 times solar, with an uncertainty of about a factor of two (Table~\ref{table:keyresults_exotr}).

If K2-18~b has a massive hydrogen-rich envelope, the atmospheric \ce{CO2}-to-\ce{CH4} ratio serves as a valuable proxy for the bulk \ce{H2O}-to-\ce{H2} ratio \citep{yang2024chemical}. This is because, in such envelopes, a higher water content tends to drive up the atmospheric \ce{CO2}-to-\ce{CH4} ratio. This approach is especially important because, as shown in Section~\ref{sec:coldtrap}, water vapor condenses in the observable atmosphere of K2-18~b, making it an unreliable tracer of the envelope's water abundance. Comparing the $1\sigma$ bounds of the observed \ce{CO2}-to-\ce{CH4} ratio to the predictions from \citet{yang2024chemical} for a $100\times$ solar C/H atmosphere, we inferred that the bulk \ce{H2O} mixing ratio lies between 10\% and 25\% by volume (or 50\% -- 75\% by mass). This inference is also confirmed by self-consistent atmospheric models presented in Section~\ref{sec:epacris_massive}. For comparison, a standard $100\times$ solar metallicity envelope contains roughly 10\% \ce{H2O}. Thus, our results suggest that K2-18~b’s envelope likely contains more water than expected from the standard $100\times$ solar metallicity scenario.

Alternatively, if K2-18~b hosts a liquid water ocean, the measured \ce{CO2} abundance falls within the reasonable range allowed by cosmochemical and geochemical estimates \citep{hu2021unveiling}. However, the high abundance of \ce{CH4} would likely need to be primordial \citep{yu2021identify,cooke2024considerations} or sustained by biological activity \citep{wogan2024jwst}. Because a liquid water ocean necessarily implies a bulk water-dominated envelope, our observations have conclusively demonstrated that K2-18~b has at least a water-rich envelope, with water comprising more than $\sim10\%$ of its volume or $\sim50\%$ of its mass.

\section{Discussions}
\label{sec:discussion}

\subsection{A Potential Water Cold Trap}
\label{sec:coldtrap}

Retrievals of the spectra presented in this paper place stringent upper limits on the atmospheric \ce{H2O} mixing ratio: less than $10^{-5.3}$ at $1\sigma$ and $10^{-3.9}$ at $2\sigma$. Using directly combined spectra rather than pre-shifted combinations yields even tighter constraints (Table~\ref{table:keyresults_exotr}). This inference implies that atmospheric \ce{H2O} should be depleted by at least two orders of magnitude relative to the bulk envelope. Such a depletion is most naturally explained by the operation of a water cold trap, where water vapor condenses out before reaching altitudes probed by transmission spectroscopy. The observations and analysis presented here may thus provide a rare window into the water cold trap in an exoplanetary atmosphere.

The upper limits on atmospheric water vapor allow us to evaluate the efficiency of the cold trap, because it is the temperature at the cold trap that governs the mixing ratio of water vapor in the atmosphere above the cold trap. Typically,
\begin{equation}
    f_{\ce{H2O}|{\rm observed}} \ge f_{\ce{H2O}|{\rm coldtrap}} ,
\end{equation}
where $f$ denotes the mixing ratio. This inequality accounts for possible photochemical production of water in the upper atmosphere. The cold-trap mixing ratio is related to the saturation vapor pressure by
\begin{equation}
    f_{\ce{H2O}|{\rm coldtrap}} P_{\rm coldtrap} \sim p_{\rm sat}(T_{\rm coldtrap}) ,
    \label{eq:coldtrap}
\end{equation}
where $P$ and $T$ represent the atmospheric pressure and temperature, respectively, and $p_{\rm sat}$ is the saturation vapor pressure. Equation (\ref{eq:coldtrap}) is approximate, as the cold-trap efficiency can also be affected by atmospheric dynamics and transport. For K2-18~b, the cold-trap pressure is estimated to lie between 0.04 and 0.08 bar \citep{yang2024chemical} and this pressure, if approximated by the radiative-convective boundary, is largely insensitive to the planet's insolation or Bond albedo \citep{robinson2012analytic}. Using the observed upper limits on water vapor and Equation (\ref{eq:coldtrap}), we derived an upper limit on the cold-trap temperature: $T_{\rm coldtrap}<175$ K ($1\sigma$) and 215 K ($2\sigma$). This is also notionally consistent with the temperature we obtained from spectral retrievals, which is $<\sim210$ K. Furthermore, if we approximate the cold-trap temperature as the ``skin temperature'' under the assumption of a nearly isothermal stratosphere, following \citet{kasting1991co2}, we have:
\begin{equation}
    T_{\rm coldtrap} \sim  T_{\rm skin} = \frac{1}{2^{1/4}}\bigg[\frac{S}{4\sigma}(1-A_B)\bigg]^{1/4},
\end{equation}
where $S$ is the insolation, $\sigma$ is the Stefan-Boltzmann constant, and $A_B$ is the Bond albedo. Translating the upper limits on $T_{\rm coldtrap}$ into constraints on albedo, we found that K2-18~b should have a minimum Bond albedo of 0.7 ($1\sigma$) or 0.3 ($2\sigma$).

\begin{figure}[!htbp]
\centering
\includegraphics[width=0.4\textwidth]{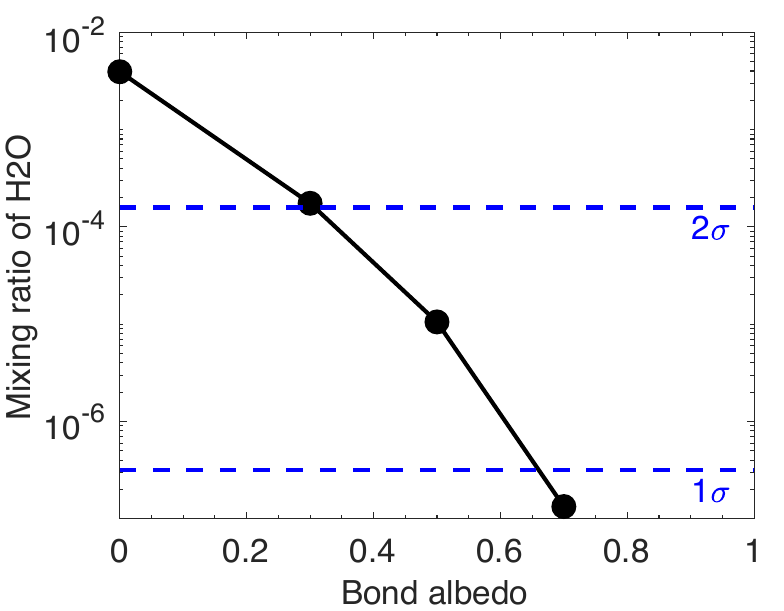}
\caption{Mixing ratios of \ce{H2O} above the cold trap, compared with the upper limits on atmospheric water vapor derived from transmission spectra. The \ce{H2O} mixing ratios are calculated from non-gray, radiative-convective simulations of K2-18~b’s atmosphere, assuming the most probable \ce{CH4} and \ce{CO2} abundances inferred from the spectra. A notional surface pressure of 3 bars was used in the simulations; varying it between 1 and 10 bars produces negligible changes in the results. The lack of detected \ce{H2O} indicates that K2-18~b should have an effective cold trap and a substantial Bond albedo.
}
\label{fig:coldtrap}
\end{figure}

We validated the analytical estimates using radiative-convective calculations of the planet's atmospheric pressure–temperature profile. Using the \texttt{EPACRIS} model (see Section~\ref{sec:epacris}), we simulated an atmosphere that has 10\% \ce{CH4}, $5 \times 10^{-4}$ \ce{CO2}, and \ce{H2O}, with the \ce{H2O} abundance set self-consistently by the vapor pressure at each temperature. We explored a range of Bond albedos from 0 to 0.7, and the resulting \ce{H2O} mixing ratio above the cold trap is shown in Figure~\ref{fig:coldtrap}. These simulations confirm that the upper-atmosphere \ce{H2O} abundance probed by transmission spectra serves as a diagnostic of the Bond albedo. The derived minimum albedo values are fully consistent with the analytical results presented above.

Because the inference of the cold trap relies on the nondetection of H$_2$O in spectral retrievals, we tested the robustness of this inference by performing an additional retrieval using \exotr, in which we constrained the atmosphere to contain more H$_2$O than CH$_4$. The best-fit model from this constrained retrieval differs markedly from the best-fit model in the standard, free retrieval: it introduces a cloud deck at $\sim0.02$ bar to ``fill in'' the lower part of the CH$_4$ absorption feature, allowing additional H$_2$O absorption to be accommodated (Figure~\ref{fig:h2oggch4}). This solution also features slightly less CH$_4$ and slightly more CO$_2$, and requires to offset the NIRSpec/G395H portion of the spectrum upwards by approximately $\sim30$ ppm compared to the free retrieval. The Bayes factor between the free and constrained retrievals is $\sim10^6$, corresponding to a model selection preference well above $5\sigma$, consistent with the strong statistical preference for the non-H$_2$O solution in the free retrieval. Nonetheless, the best-fit model from the constrained retrieval provides a visually acceptable fit to the data (Figure~\ref{fig:h2oggch4}), with a reduced $\chi^2$ of 1503/1264, compared to 1479/1264 in the free retrieval. From a frequentist perspective, the corresponding likelihood ratio translates to odds of approximately 7:1 in favor of the free retrieval -- only a weak preference. The larger difference in the Bayes factor may stem from the fact that the constrained retrieval solution requires specific cloud-top pressures, whereas the free retrieval solution remains valid across a broader range of cloud scenarios, including deep or no clouds. Additionally, the parameter space in the constrained retrieval is inherently narrower due to the imposed abundance constraints. These factors lead to a much smaller prior volume for the constrained model, making the Bayes factor potentially unreliable as a sole metric for model comparison in this context. Moreover, a cloud deck at $\sim$20 mbar is physically plausible based on self-consistent atmospheric models \citep[e.g.,][]{yang2024chemical}, but such clouds may require some degree of water depletion to form. This additional test demonstrates that alternative solutions may indeed exist, preventing us from robustly concluding that K2-18~b must have a strongly water-depleted atmosphere and an effective cold trap.

Taking the non-detection of \ce{H2O} at face value, however, one would find the implied lower bounds on the Bond albedo to be broadly consistent with those required to sustain a liquid-water ocean in climate models \citep[e.g., $A_B > 0.5$,][]{leconte20243d}. Recent work has argued that the planetary albedo cannot be arbitrarily high due to molecular absorption above cloud layers \citep{jordan2025planetary}. By applying our retrieval-derived constraints on \ce{CH4} abundance and cloud-top pressure to the models of \citet{jordan2025planetary}, we found that the maximum achievable albedo is approximately 0.5.
As a high albedo would lend crucial support to the possibility that the planet has remained in a pre-runaway greenhouse state and potentially hosts a liquid-water ocean \citep[e.g.,][]{innes2023runaway,leconte20243d}, future refinement of water vapor constraints through additional observations -- ideally constraining the offsets between instruments -- will be key to tighten the albedo estimate and enhance our ability to assess the planet's potential habitability.

\subsection{Does K2-18~b have a liquid water ocean?}
\label{sec:liquidwater}

Now that we have established that K2-18~b hosts a water-rich interior and probably water clouds in the atmosphere, a central question emerges: does it also harbor a liquid-water ocean? For K2-18~b, the presence of a small atmosphere effectively implies a liquid-water layer, since otherwise we would expect a massive atmosphere shrouding a well-mixed, supercritical \ce{H2}-\ce{H2O} interior \citep[e.g.,][]{gupta2025miscibility}. The broad wavelength coverage and high signal-to-noise ratio of our repeated transmission spectra observations enable us to probe this question through a suite of atmospheric chemistry diagnostics.

\subsubsection{The Apparent Lack of \ce{NH3}}
\label{sec:nh3prob}

The absence of \ce{NH3} has been proposed as a potential indicator of the small-atmosphere scenario for K2-18~b, due to its short chemical lifetime in thin atmospheres and high solubility in liquid water \citep{yu2021identify,hu2021unveiling}. Spectral retrievals of the transmission spectra presented in this work place a stringent upper limit on the atmospheric mixing ratio of \ce{NH3}, constraining it to below $\sim10^{-5.6}$ at the $2\sigma$ level. This non-detection is consistent with the results of \citet{madhusudhan2023carbon}, but our upper limit is tighter by approximately an order of magnitude. The apparent lack of \ce{NH3} is most naturally explained by the existence of a liquid-water ocean on K2-18~b.

Could there be scenarios and mechanisms within the massive-envelope framework that explain the observed depletion of \ce{NH3}? If we adopt a nitrogen-to-hydrogen ratio (N/H) equal to the data-constrained carbon-to-hydrogen ratio (C/H) of $100\times$ solar, and apply the observed \ce{H2O}-to-\ce{H2} ratio in the envelope, self-consistent atmospheric models predict an \ce{NH3} mixing ratio between $3\times10^{-4}$ and $2\times10^{-3}$. The estimates have already accounted for the thermochemical speciation between \ce{NH3} and \ce{N2}, and the range is mostly driven by deep atmospheric temperatures and eddy diffusivities \citep{yang2024chemical}. These predicted values far exceed the upper limit from spectral retrievals, implying that additional mechanisms must operate to further deplete \ce{NH3} by at least two orders of magnitude.

Before discussing potential physical mechanisms for the apparent NH$_3$ depletion, we first assess the robustness of this observational inference. Following the approach used for H$_2$O, we performed an additional spectral retrieval in which the NH$_3$ mixing ratio was constrained to be above $3\times10^{-4}$. As shown in Figure~\ref{fig:nh3co}, a visually acceptable fit can be achieved under this constraint. The best-fit model places NH$_3$ near the imposed lower limit and additionally incurs a cloud deck, yielding a reduced $\chi^2$ of 1491/1264, compared to 1479/1264 in the free retrieval. From a frequentist standpoint, this corresponds to a likelihood ratio with odds of only 3:1 in favor of the free retrieval. Compared to the free retrieval, the constrained retrieval offsets the NIRSpec/G235H/NRS2 and NIRSpec/G395H/NRS1 spectra upward by approximately 5–7 ppm, thereby allowing room to accommodate the additional NH$_3$ absorption near 3.0~$\mu$m (Figure~\ref{fig:nh3co}). This exercise highlights the caution required when drawing theoretical conclusions solely from free-retrieval posteriors, especially when detector offsets are allowed to vary freely. While further observations will be necessary to resolve this ambiguity, we proceed to explore several physical mechanisms that could plausibly lead to NH$_3$ depletion.

The first mechanism we considered is the photochemical conversion of \ce{NH3} into \ce{N2} and \ce{HCN}, as described by \citet{hu2021photochemistry}. Using the \texttt{EPACRIS} atmospheric chemistry model with comprehensive forward and reverse reactions, we found that \ce{NH3} remains largely unaltered down to pressures of $\sim10^{-3}$ bar, even when assuming a small eddy diffusion coefficient of $10^3$ cm$^2$ s$^{-1}$ (Figure~\ref{fig:vmr}). This result is consistent with previous studies \citep{hu2021photochemistry, cooke2024considerations}, suggesting that photochemical processes alone have a minimal effect on the observable \ce{NH3} abundance.

The second mechanism we considered is the dissolution of \ce{NH3} into liquid-water droplet clouds deep in the atmosphere \citep{hu2019information}. We evaluated this scenario for a bulk envelope containing $10–25\%$ water, following the methodology of \citet{hu2019information}, and extended it to include the co-dissolution of \ce{CO2}, which lowers droplet pH and enhances \ce{NH3} solubility. Under favorable conditions -- such as high cloud densities and low temperatures -- \ce{NH3} mixing ratios can be reduced by up to a factor of two. This mechanism alone, or combined with atmospheric photochemistry, cannot produce the inferred extent of \ce{NH3} depletion.

The third mechanism is the dissolution of \ce{NH3} into an underlying magma ocean, as proposed by \citet{shorttle2024distinguishing}. Self-consistent models of a coupled envelope and magma ocean suggest that a depletion of \ce{NH3} by approximately two orders of magnitude is feasible in the presence of a highly reduced magma ocean \citep{rigby2024towards}. However, the low oxygen fugacity required for such strong \ce{NH3} depletion would buffer the envelope's \ce{H2O}-to-\ce{H2} ratio to below $\sim1\%$ \citep[e.g.,][]{gaillard2022redox,tian2024atmospheric}. Even with non-ideal gas corrections \citep{glein2024geochemical}, this low water abundance would imply a \ce{CO2} mixing ratio below $10^{-5}$ -- a level that is difficult to reconcile with the robust \ce{CO2} detection reported in this work. An alternative possibility is that nitrogen has been depleted from the entire envelope–magma ocean–mantle system and preferentially sequestered into the planet's core \citep[e.g.,][]{roskosz2013nitrogen}. Earth's bulk nitrogen abundance is depleted by roughly an order of magnitude relative to chondritic values, likely due to core partitioning during differentiation \citep{marty2012origins}. The extent of nitrogen sequestration could reach up to two orders of magnitude under graphite-undersaturated conditions and when the core and mantle are comparable in mass \citep{grewal2021rates}. Preferential partitioning into the core thus presents a potential mechanism to explain the observed nitrogen depletion in K2-18~b's atmosphere.

Lastly, K2-18~b's envelope may be intrinsically nitrogen-poor. In protoplanetary disks, nitrogen-bearing volatiles such as \ce{N2}, \ce{NH3}, and HCN are significantly more volatile than water and can be substantially depleted at the planet's formation location \citep[e.g.,][]{schwarz2014effects, pontoppidan2019nitrogen, oberg2021astrochemistry, bergner2022hcn}. If K2-18~b accreted relatively little nitrogen-bearing ice compared to water, its primordial nitrogen inventory could have been low. For instance, if the planet's carbon and nitrogen were primarily sourced from refractory carbonaceous materials similar to primitive meteorites in the solar system, the resulting C/N ratio would be $\sim$25 -- much higher than the solar value of $\sim$3 \citep{alexander2017nature}. This initial nitrogen deficiency, compounded by further depletion via magma-ocean dissolution and core partitioning, could reduce the atmospheric \ce{NH3} mixing ratio to below $\sim10^{-6}$ and explain the non-detection in the JWST transmission spectra.

\subsubsection{CO$_2$-to-CO ratio}
\label{sec:co2co}

Complementing \ce{NH3}, the CO$_2$-to-CO ratio serves as an additional diagnostic for distinguishing between small-atmosphere and massive-envelope scenarios on planets like K2-18~b. In a small atmosphere overlying a liquid-water ocean, photochemical processes convert some \ce{CO2} to CO, typically yielding a CO$_2$-to-CO ratio greater than 3 \citep{hu2021unveiling}. In contrast, a massive \ce{H2}-dominated envelope typically maintains the CO$_2$-to-CO ratio less than $\sim0.1$, largely independent of metallicity \citep{hu2021photochemistry,wogan2024jwst}. For water-rich envelopes containing $10-25\%$ water by volume, the CO$_2$-to-CO ratio is predicted to lie between 0.1 and 1 \citep{yang2024chemical}. An even more water-rich envelope would have the CO$_2$-to-CO ratio greater than 1, overlapping with the small-atmosphere regime.

Although CO is not detected in the current data, we derived the posterior distribution of the CO$_2$-to-CO ratio from the retrievals. This posterior distribution predominantly falls within the range $\ce{CO2}/\ce{CO} > 3$. From the posterior distribution of free retrievals, $\ce{CO2}/\ce{CO} < 1$ is only allowable with a probability of $3\%$ (i.e., $>2\sigma$ excluded) and $\ce{CO2}/\ce{CO} < 0.1$ has a very low probability below $1\%$. Unlike the case of \ce{NH3}, the non-detection of CO cannot be attributed to an overall carbon deficiency, since \ce{CH4} and \ce{CO2} are both robustly detected. 

To test the robustness of the inferred high CO$_2$-to-CO ratio, we also performed an additional retrieval in which the ratio was constrained to be less than 1. As shown in Figure~\ref{fig:nh3co}, the best-fit model from this constrained retrieval also provides a visually acceptable fit to the data, with both CO and CO$_2$ present in the atmosphere. The model closely resembles the best-fit model from the free retrieval, including identical detector offsets. Based on $\chi^2$ values, the free retrieval is favored only modestly, with odds of approximately 2:1. This result indicates that the current dataset does not rule out scenarios with $\ce{CO2}/\ce{CO} < 1$, even though such compositions are not preferred in the unconstrained retrievals.

If a high CO$_2$-to-CO ratio is confirmed by future observations (e.g., more repeated transit observations in NIRSpec/G395, or ground-based high-resolution spectroscopy), it would support the presence of a relatively small atmosphere potentially overlying a liquid-water ocean. Such a result would disfavor typical massive-envelope scenarios and leave only limited room for certain water-rich massive-envelope scenarios (see Section~\ref{sec:roadmap}).

\subsection{Atmospheric Chemistry of Organosulfur Compounds}
\label{sec:dms}

\begin{figure}
    \centering    
    \includegraphics[width=0.4\textwidth]{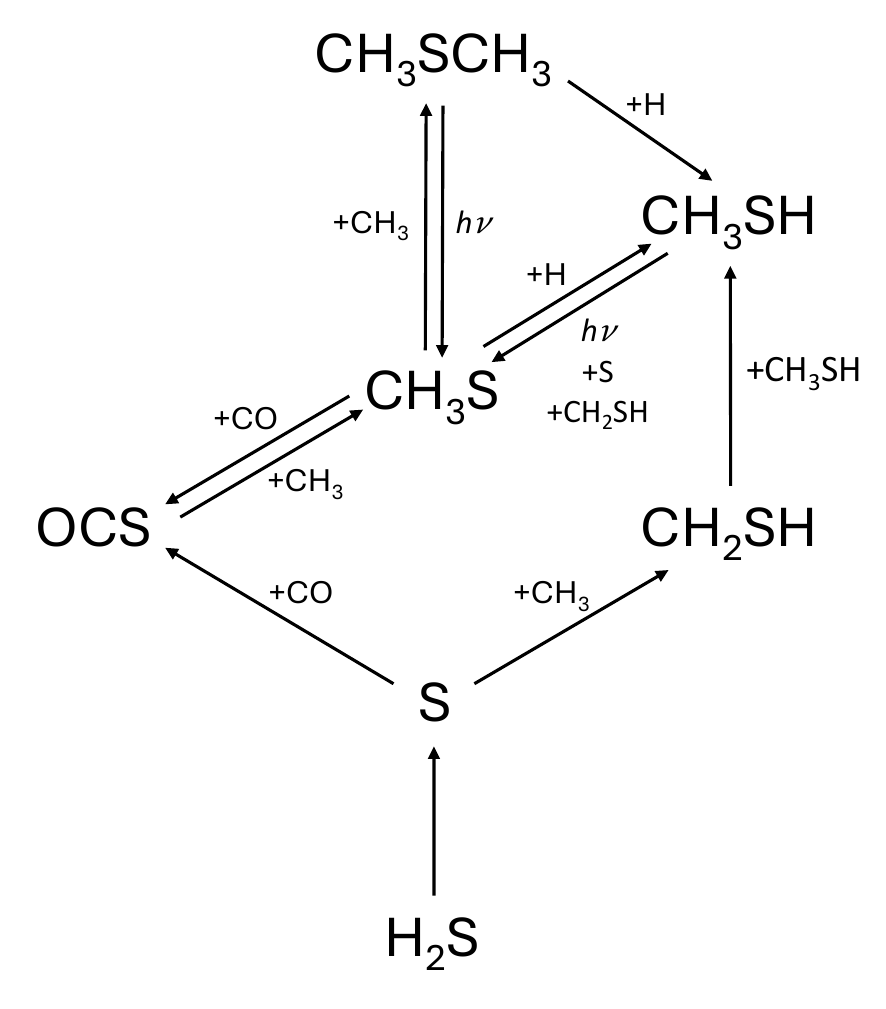}
     \caption{Major formation pathways of organosulfur species in the massive atmospheres of temperate sub-Neptunes like K2-18~b.}  
    \label{fig:sulfur_chem}
\end{figure}

A novel result from our updated atmospheric chemistry models for the massive-atmosphere scenarios of K2-18~b is the natural formation of organosulfur compounds, including DMS and \ce{CH3SH}. These gases, which are the dominant sulfur species emitted from Earth's oceans, were proposed as potential biosignatures in diverse environments \citep{domagal2011using, seager2013biosignature}, including the atmosphere of K2-18~b \citep{madhusudhan2023carbon,madhusudhan2025miri}. However, our self-consistent models demonstrate that the production of these compounds may occur readily in the atmosphere of a massive, high-metallicity envelope on a temperate sub-Neptune like K2-18~b (Figure~\ref{fig:vmr}). The abundance of \ce{CH3SH}, in particular, can reach levels that produce spectral features stronger than the weak signals suggested by current observations (Figure~\ref{fig:model_fit}) without any fine tuning. DMS and \ce{CH3SH} have overlapping molecular absorption features in the $2–5\ \mu$m range, and the current observations lack the sensitivity to distinguish between them.

The primary formation pathways (see~Figure~\ref{fig:sulfur_chem}) for \ce{CH3SH} and DMS (\ce{CH3SCH3}) originate from \ce{H2S} photo- and thermo-dissociation and proceed as follows:
\begin{equation}
    \begin{split}
        \ce{H2S}&\rightarrow\ce{H2}+\ce{S},\\
\ce{S}+\ce{CH4}&\rightarrow\ce{CH3}+\ce{SH},\\
        \ce{S}+\ce{CO}&\rightarrow\ce{OCS},\\
\ce{CH3}+\ce{OCS}&\rightarrow\ce{CH3S}+\ce{CO},\\
        \ce{CH3}+\ce{S}&\rightarrow\ce{CH2SH},\\
        \ce{CH3S}+\ce{H}&\rightarrow\ce{CH3SH},\\
\ce{CH2SH}+\ce{CH3SH}&\rightarrow\ce{CH3SH}+\ce{CH3S},\\        \ce{CH3S}+\ce{CH3}&\rightarrow\ce{CH3SCH3}.\\
        \end{split}
    \label{eqn:S_org_formation}
\end{equation}
For these reactions, we adopted the rate coefficient for \ce{CH3S} + CO $\rightarrow$ \ce{CH3} + OCS from \citet{tang2008ab}, who used the G3MP2//B3LYP/6-311++G(d,p) level of theory to compute the potential energy surfaces and applied both transition state theory (TST) and Rice–Ramsperger–Kassel–Marcus (RRKM) theory to derive the rate coefficients. The reactions \ce{CH3S} + \ce{H} $\rightarrow$ \ce{CH3SH} and \ce{CH3} + S $\rightarrow$ \ce{CH2SH} were computed by \texttt{RMG} (Reaction Mechanism Generator), using its \texttt{PDepNetwork} (pressure-dependent network) algorithm \citep{Gao_2016, liu2021reaction} based on its sulfur-specific reaction library \citep{RMG-database}. The H abstraction reaction \ce{CH2SH} + \ce{CH3SH} $\rightarrow$ \ce{CH3SH} + \ce{CH3S} was estimated by \texttt{RMG} using the \texttt{H\_Abstraction} kinetics family algorithm \citep{Gao_2016, liu2021reaction, RMG-database}. The rate coefficients for these four reactions are summarized in Table~\ref{tbl:rate_coefficients}.

\begin{deluxetable}{cccc}
\caption{Rate coefficients involving \ce{CH3SH} and DMS formation chemistry mentioned in Section~\ref{sec:dms}.}
\label{tbl:rate_coefficients}
\tablehead{
\colhead{$P$ [atm]} & 
\colhead{$A$ [cm$^3$/mol/s]}  & 
\colhead{$n$} & 
\colhead{$E_a$ [kcal/mol]}
}
\startdata
\multicolumn{4}{c}{\textbf{\ce{CH3S} + CO $\rightarrow$ \ce{CH3S} + OCS}} \\[4pt]
--         & 6.622$\times10^{7}$   & 1.57   & 6.675  \\
\midrule
\multicolumn{4}{c}{\textbf{\ce{CH3S} + H $\rightarrow$ \ce{CH3SH}}} \\[4pt]
0.000888   & 1.971$\times10^{23}$  & -3.787  & 0.767  \\
0.009041   & 4.859$\times10^{24}$  & -3.902  & 0.956  \\
0.092029   & 5.598$\times10^{27}$  & -4.508  & 2.072  \\
0.936746   & 1.274$\times10^{29}$  & -4.548  & 3.446  \\
9.534991   & 8.083$\times10^{22}$  & -2.333  & 2.419  \\
97.05521   & 6.981$\times10^{16}$  & -0.314  & 0.935  \\
\midrule
\multicolumn{4}{c}{\textbf{\ce{CH3} + S $\rightarrow$ \ce{CH2SH}}} \\[4pt]
0.000888   & 4.886$\times10^{26}$  & -4.616  & 1.818  \\
0.009041   & 2.506$\times10^{26}$  & -4.218  & 2.159  \\
0.092029   & 4.093$\times10^{24}$  & -3.386  & 2.219  \\
0.936746   & 4.489$\times10^{20}$  & -1.951  & 1.573  \\
9.534991   & 3.617$\times10^{16}$  & -0.571  & 0.626  \\
97.05521   & 4.084$\times10^{14}$  &  0.071  & 0.129  \\
\midrule
\multicolumn{4}{c}{\textbf{\ce{CH2SH} + \ce{CH3SH} $\rightarrow$ \ce{CH3SH} + \ce{CH3S}}} \\[4pt]
--         & 4.290$\times10^{1}$   & 3.060   & 0.600  \\
\midrule
\multicolumn{4}{c}{\textbf{\ce{CH3S} + \ce{SH} $\rightarrow$ \ce{CH3SH} + \ce{S}}} \\[4pt]
--         & 3.458$\times10^{-2}$   & 4.450   & 6.916  \\
\enddata
\tablecomments{The rate coefficient is expressed in the Arrhenius form: $k(T)=AT^n\exp{(-E_a/RT})$, where $R$ is the ideal gas constant. Two recombination reactions are described using a \texttt{Pdep-Arrhenius-type} expression, with formula given in Section~2.3.2 of \citet{yang2024epacris}.}
\end{deluxetable}

The primary removal pathways for these two species (i.e., \ce{CH3SH} and DMS) are
\begin{equation}
    \begin{split}
\ce{CH3SH}&\xrightarrow{\text{$h\nu$}}\ce{CH3S}+\ce{H},\\       \ce{CH3SH}+\ce{S}&\rightarrow\ce{CH3S}+\ce{SH},\\
\ce{CH3SCH3}&\xrightarrow{\text{$h\nu$}}\ce{CH3S}+\ce{CH3},\\
\ce{CH3SCH3}+\ce{H}&\rightarrow\ce{CH3SH}+\ce{CH3},\\        \ce{CH3S}+\ce{CO}&\rightarrow\ce{OCS}+\ce{CH3}.\\
        \end{split}
    \label{eqn:S_org_deformation}
\end{equation}
The H abstraction reaction \ce{CH3SH} + \ce{S} $\rightarrow$ \ce{CH3S} + \ce{SH} was estimated by \texttt{RMG} using the \texttt{H\_Abstraction} kinetics family algorithm \citep{Gao_2016, liu2021reaction, RMG-database}. The rate coefficient for this reaction is also provided in Table~\ref{tbl:rate_coefficients}.

In our mechanism, once \ce{CH3S} is produced in the reaction between \ce{CH3} and OCS, it also acts as a catalyst to facilitate the formation of \ce{CH3SH} from the reaction between \ce{CH3} and S:
\begin{equation}
    \begin{split}
       \ce{CH3} + \ce{S} &\rightarrow\ce{CH2SH},\\
        \ce{CH3S}+\ce{H}&\rightarrow\ce{CH3SH},\\
\ce{CH2SH}+\ce{CH3SH}&\rightarrow\ce{CH3SH}+\ce{CH3S}.\\
\hline
{\rm Net:}  \ce{CH3}+\ce{S}+\ce{H} &\rightarrow \ce{CH3SH}\\
        \end{split}
    \label{eqn:S_org_formation}
\end{equation}

It is also worth noting that \cite{reed2024abiotic} demonstrated the abiotic formation of organosulfur compounds through UV photochemistry in gas mixtures of \ce{H2S}, \ce{CH4}, and \ce{CO2}, producing primarily OCS and \ce{CH3SH}, with DMS forming only in much smaller amounts. Our massive-envelope scenarios, along with the photochemical pathways illustrated in Figure~\ref{fig:sulfur_chem}, are broadly consistent with these experimental findings. More generally, the interplay between sulfur and carbon chemistry in H$_2$-dominated atmospheres remains poorly understood, with many reaction rates still unconstrained. Additionally, the steady-state abundances of organosulfur species are highly sensitive to the assumed eddy diffusivity (Figure~\ref{fig:vmr}). Despite these uncertainties, the fact that \ce{CH3SH} and DMS may accumulate to potentially detectable levels -- and the general agreement between our photochemical models and laboratory data -- suggest that the presence of organosulfur species in the atmosphere of K2-18~b or other temperate sub-Neptunes does not necessarily imply biological activity. Instead, such detections could plausibly result from abiotic photochemical processes.

Interestingly, the photochemical production of DMS and \ce{CH3SH} in the massive-atmosphere scenario does not lead to a high abundance of hydrocarbons. For instance, the mixing ratios of \ce{C2H2} and \ce{C2H6} -- the most abundant hydrocarbons in our models -- remain below $10^{-4}$ in the observable part of the atmosphere. This is significantly lower than the hydrocarbon abundances expected when similar levels of DMS and \ce{CH3SH} are produced via biogenic fluxes from an ocean surface (Figure~\ref{fig:hycean}). Because hydrocarbons have distinct spectral features, their detection -- or the establishment of tighter upper limits in future observations -- could help distinguish between abiotic and biogenic origins of organosulfur compounds in the atmosphere of K2-18~b and other temperate sub-Neptunes.

\section{A Refined Roadmap for Characterizing Temperate Sub-Neptunes}
\label{sec:roadmap}

\begin{figure*}[!htbp]
\centering
\includegraphics[width=1.0\textwidth]{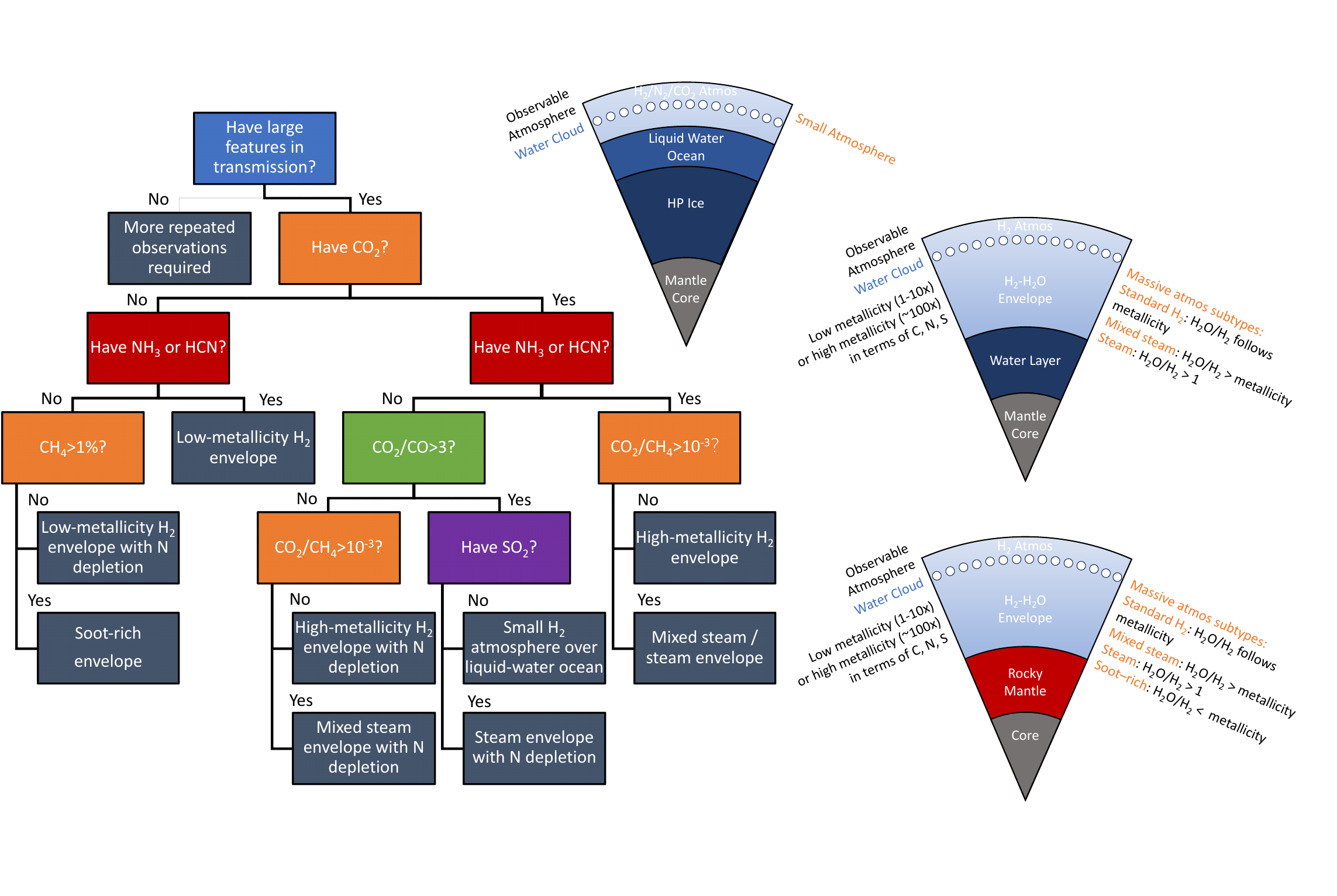}
\caption{
Illustration of the range of possible internal compositions for temperate sub-Neptunes like K2-18~b, and a roadmap to characterize them through atmospheric observations. A liquid-water ocean separating the atmosphere from the interior necessarily implies a small atmosphere, which may be composed of either \ce{H2} or heavier molecules. In contrast, a massive atmosphere is expected to be well-mixed with the volatile envelope underneath. We propose characterizing such massive envelopes along two principal axes: the \ce{H2}-to-\ce{H2O} ratio and the metallicities defined by carbon, nitrogen, and sulfur. In a standard \ce{H2}-dominated envelope, the oxygen abundance tracks that of C, N, and S with the solar abundance ratio. In a mixed or steam-rich envelope, oxygen (and thus \ce{H2O}) is more abundant; in a soot-rich envelope, it is less so. Nitrogen and sulfur may be depleted relative to carbon due to preferential partitioning into the mantle or core. The existence of a water-dominated layer between the envelope and the rocky interior remains uncertain, but is retained here as a possibility. Regardless of envelope composition, water may condense in the atmospheres of K2-18~b and cooler sub-Neptunes, rendering direct H$_2$O measurements from transmission spectra unreliable as diagnostics of the bulk composition. Instead, detecting and quantifying a suite of gases, such as \ce{CH4}, \ce{CO2}, \ce{CO}, \ce{NH3}, and \ce{SO2}, and their relative abundances offers a more robust pathway to distinguish among envelope compositions, physical states, and the potential presence of a liquid-water ocean.}
\label{fig:schematic}
\end{figure*}

Temperate sub-Neptunes -- planets with radii approximately twice that of Earth and receiving Earth-like stellar flux -- present an immediate opportunity to search for liquid-water oceans beyond the solar system. Several these planets are accessible to transmission spectroscopy with current facilities, and their atmospheric compositions offer critical clues about their internal structures. In \citet{hu2021unveiling}, we presented an initial roadmap for using atmospheric observations to distinguish among key interior and atmospheric scenarios, such as massive \ce{H2}-rich envelopes with varying metallicities versus thin atmospheres atop liquid-water oceans. That framework has since guided multiple observational campaigns and their interpretations \citep[e.g.,][]{madhusudhan2023carbon,benneke2024jwst}, including the present study.

Meanwhile, our understanding of the chemistry and physics governing sub-Neptune-sized exoplanets has advanced substantially since 2021. With JWST in operation, we can now probe exoplanetary atmospheres with unprecedented precision, including for temperate sub-Neptunes such as K2-18~b. These high-quality datasets, coupled with new insights into planet formation, volatile partitioning, and atmospheric photochemistry, motivate a refinement of the original roadmap. In what follows, we outline updated chemical diagnostics, examine key degeneracies, and propose a revised observational sequence to guide the characterization of temperate sub-Neptunes (Figure~\ref{fig:schematic}).

The foundational step remains the detection of key trace gases -- \ce{CH4}, \ce{CO2}, and \ce{NH3} -- in transmission spectra. If we adopt a practical detection threshold of $\sim$10 ppm for \ce{CO2}, its presence or absence serves as a powerful indicator of envelope metallicity: low-metallicity ($\lesssim10\times$ solar) \ce{H2}-rich atmospheres typically lack detectable \ce{CO2}, while high-metallicity envelopes, or small atmospheres on top of water-rich layers, generally produce observable levels \citep{hu2021photochemistry,hu2021unveiling,yang2024chemical,cooke2024considerations}.

Similarly, the presence or absence of \ce{NH3} at a threshold of $\sim100$ ppm provides a useful first-order discriminator between thin atmospheres and more massive envelopes. In smaller atmospheres, \ce{NH3} is naturally depleted by photolysis or dissolved into liquid-water oceans, whereas in deeper \ce{H2}-rich envelopes, \ce{NH3} is generally expected to persist due to thermochemical equilibrium. However, this diagnostic is complicated by interior processes: nitrogen may be sequestered in a planet’s molten mantle or core \citep{grewal2021rates,shorttle2024distinguishing,rigby2024towards}, or the planet may have formed with an intrinsically nitrogen-poor composition (see Section~\ref{sec:nh3prob}). Thus, while a nondetection of \ce{NH3} is consistent with the presence of a liquid-water ocean, it is not definitive.

To resolve such ambiguities, we turn to more chemically specific indicators. A robust distinction emerges from the \ce{CO2}-to-CO ratio. In small atmospheres overlaying liquid-water oceans, photochemical models consistently predict \ce{CO2}/CO $\gtrsim3$ \citep{hu2021unveiling,wogan2024jwst,cooke2024considerations}. This result is remarkably insensitive to variations in stellar UV radiation, planetary albedo, and photochemical model assumptions. In contrast, massive \ce{H2}-rich envelopes produce \ce{CO2}/CO $\lesssim1$, a result similarly robust across a wide parameter space \citep{yu2021identify,tsai2021inferring,wogan2024jwst,yang2024chemical,cooke2024considerations}. This ratio thus offers a powerful chemical signature for distinguishing small atmospheres from massive envelopes, complementary to \ce{NH3}.

However, a special class of massive envelopes with high \ce{H2O}-to-\ce{H2} ratios introduces a more ambiguous regime. When \ce{H2O} and \ce{H2} are present in comparable amounts, the \ce{CO2}-to-CO ratio can approach 3; as \ce{H2O}/\ce{H2} increases further, this ratio may rise to $10–100$ \citep{yang2024chemical}. These water-dominated envelopes also produce low \ce{NH3} mixing ratios ($<10^{-4}$). The atmospheres on such envelopes may therefore be difficult to distinguish from small atmospheres based on the \ce{NH3} and \ce{CO2}/CO diagnostics.

In this observationally degenerate regime, sulfur chemistry offers an additional diagnostic tool. Highly \ce{H2O}-rich envelopes promote the formation of detectable \ce{SO2}, particularly when the conversion to OCS is kinetically inhibited \citep{yang2024chemical}. In contrast, atmospheres over liquid-water oceans are expected to lack significant \ce{SO2} due to efficient sulfate dissolution into the ocean \citep{loftus2019sulfate}. The detection of \ce{SO2} may thus indicate a highly oxidizing, water-rich envelope, helping to break degeneracies unresolved by carbon and nitrogen species alone.

An important caveat when using the \ce{CO2}-to-CO ratio is that CO may be partially converted to OCS through sulfur chemistry. Given current uncertainties in OCS production kinetics \citep{yang2024chemical}, it is advisable to consider the combined abundance of CO and OCS when evaluating this ratio. Future laboratory measurements or quantum chemical calculations will be critical for reducing this source of uncertainty.

As discussed in Sections~\ref{sec:coldtrap} and \ref{sec:liquidwater}, caution is warranted when interpreting nondetections in the context of the characterization roadmap. As illustrated by the cases of \ce{NH3}, CO, and \ce{H2O} in this study, spectral retrievals employing nominally ``flat'' priors on gas abundances may suggest nondetections, while alternative scenarios involving those gases can still yield acceptable fits with minimal degradation in goodness-of-fit metrics. In the case of K2-18~b, our analyses suggest that higher-precision data, particularly in the NIRSpec/G395H wavelength range, are needed to confirm the elevated \ce{CO2}-to-CO ratio; two future visits from Program 2372 are expected to contribute to this. Furthermore, direct measurement of the detector-to-detector offsets with a precision better than $\sim$10 ppm could help distinguish between competing interpretations, such as scenarios involving elevated \ce{H2O} or \ce{NH3} abundances.

We also recognize that CO, OCS, and \ce{SO2} are significantly more difficult to detect in transmission spectra than \ce{CH4}, \ce{CO2}, and \ce{NH3}. The refined roadmap presented in Figure~\ref{fig:schematic} reflects this practical constraint. For any temperate sub-Neptune, the initial step is to assess the presence of \ce{CH4}, \ce{CO2}, \ce{NH3}, and \ce{H2O}, which together constrain the basic interior structure. If these detections suggest a potentially habitable or compositionally interesting planet -- such as one with a liquid-water ocean -- then a more detailed follow-up campaign can be initiated. This may include additional JWST visits or high-resolution ground-based spectroscopy to search for more subtle diagnostics like the \ce{CO2}-to-CO ratio and \ce{SO2}, which can further refine our understanding of the planet's interior, atmosphere, and surface conditions.

\section{Conclusion} \label{sec:conclusion}

In this study, we presented a comprehensive characterization of the temperate sub-Neptune K2-18~b through repeated transit observations with JWST, including two new transits with NIRSpec/G235H and two with NIRSpec/G395H. By combining these with previous data from NIRISS/SOSS and G395H, and optionally including MIRI/LRS, we constructed the most precise near- to mid-infrared transmission spectrum of K2-18~b to date.

We analyzed the dataset using multiple independent data reduction pipelines, spectral retrieval frameworks, and self-consistent atmospheric models. Our observations and analyses robustly establish that the atmosphere of K2-18~b contains abundant \ce{CH4} and \ce{CO2}, with their volume mixing ratios constrained to precisions of approximately 0.25 and 0.5 dex, respectively. These constraints are remarkably consistent across three independent retrieval frameworks and are robust to varying assumptions about stellar heterogeneity, cloud and haze coverage, and the atmospheric pressure-temperature profile. The simultaneous presence of \ce{CH4} and \ce{CO2} at the observed abundances can only be explained by either a massive atmosphere with roughly $100\times$ solar metallicity and a bulk \ce{H2O} content of 10–25\% by volume, or a small atmosphere overlaying a liquid-water ocean. Regardless of whether the planet hosts a liquid-water ocean, our results conclusively demonstrate that K2-18~b has a water-rich interior.

Our spectral retrievals using flat priors on gas abundances also yielded stringent upper limits on \ce{H2O}, \ce{NH3}, HCN, and CO. However, when the atmospheric composition is constrained to include higher abundances of \ce{H2O}, \ce{NH3}, or \ce{CO}, alternative solutions that provide visually acceptable fits to the observed spectrum emerge. Some of these solutions require detector-to-detector offsets that differ from those preferred by the unconstrained (free) retrievals. These findings highlight the need for caution in interpreting nondetections and underscore the importance of additional observations to robustly confirm the absence or low abundance of these gases in the atmosphere of K2-18~b.

If the nondetections are confirmed, they offer important insights into the physical processes and internal structure of K2-18~b. (1) The absence of detectable atmospheric water vapor suggests an efficient cold trap that removes water from the observable atmosphere through condensation. Such a process would require the planet to reflect roughly half of the incoming stellar radiation, i.e., a high Bond albedo. This elevated albedo would not only help stabilize the atmosphere against a runaway greenhouse state but also enhance the likelihood that K2-18~b supports a persistent liquid-water layer beneath its atmosphere. (2) Massive-atmosphere models calibrated to reproduce the observed \ce{CH4} and \ce{CO2} abundances significantly overpredict the atmospheric levels of \ce{NH3}, HCN, and CO when compared to constraints from the free retrievals. Photochemical processes alone cannot adequately deplete \ce{NH3}. Instead, explaining the low nitrogen abundance requires invoking either substantial sequestration into the interior or an intrinsically nitrogen-poor bulk composition, both of which push the limits of current cosmochemical and geochemical understanding. The persistent overproduction of CO in these models remains unexplained and further complicates the massive-envelope scenario. In contrast, models of a thin atmosphere overlying a liquid-water ocean naturally account for the low \ce{NH3} abundance and predict lower CO levels, offering better consistency with the retrieval results. While alternative interpretations of the spectrum remain possible, the collective chemical evidence presented here provides strong motivation to continue seeking robust observational signatures of a liquid-water ocean on this exoplanet.

Our spectral retrievals show only marginal signals for dimethyl sulfide (DMS), methyl mercaptan (\ce{CH3SH}), and nitrous oxide (\ce{N2O}), with none exhibiting a model selection preference exceeding $3\sigma$, and all remaining below $\sim2\sigma$ in the absence of a strong super-Rayleigh haze layer. Meanwhile, our self-consistent photochemical models identified novel abiotic pathways for the formation of DMS and \ce{CH3SH} in massive, high-metallicity \ce{H2}-rich atmospheres. These pathways operate through catalytic cycles involving CO and methyl radicals derived from \ce{CH4}, enabling the efficient production of organosulfur species at mixing ratios exceeding $10^{-5}$, sufficient to produce detectable signatures in transmission spectra. These results demonstrate that, in the context of temperate sub-Neptunes with massive envelopes, DMS and \ce{CH3SH} are not definitive biosignatures but can instead arise through abiotic processes under chemically favorable, carbon- and sulfur-rich conditions. Notably, whether the DMS and \ce{CH3SH} signatures are accompanied by spectral features of other gases may help break the degeneracy: organosulfur compounds produced abiotically via photochemistry are typically associated with H$_2$S, SO$_2$, or OCS, whereas biogenic surface fluxes result in co-production of \ce{C2H6} or \ce{C2H2} at detectable levels.

Finally, we updated the original characterization roadmap for temperate sub-Neptunes \citep{hu2021unveiling}, incorporating new insights from advanced atmospheric chemistry models and recent JWST observations. We propose an expanded observational sequence that begins with the detection of key trace gases -- \ce{CH4}, \ce{CO2}, and \ce{NH3} -- to establish baseline atmospheric composition. This should be followed by targeted efforts to constrain additional species such as CO, OCS, and \ce{SO2}, which can help resolve degeneracies in interior structure. For instance, the \ce{CO2}-to-CO ratio emerges as a particularly informative diagnostic: in thin secondary atmospheres, this ratio remains high ($\gtrsim 3$), while in massive \ce{H2}-rich envelopes, it tends to be much lower ($\lesssim 1$). This contrast provides a robust means of distinguishing between atmospheric scenarios, especially in cases where \ce{NH3} is not detected.

Through repeated high-precision JWST transmission spectroscopy, combined with robust atmospheric retrievals and self-consistent climate and chemistry modeling, this work demonstrates that K2-18~b is a water-rich world that may host an effective cold trap and potentially even a liquid-water ocean. K2-18~b stands out as one of the most compelling examples of a temperate sub-Neptune -- a class of planets that offers a uniquely accessible and promising avenue for detecting potentially habitable environments beyond the solar system. Continued characterization of K2-18~b and similar worlds will benefit from deeper observational campaigns that establish and refine upper limits, confirm tentative molecular detections and nondetections, and employ high-resolution spectroscopy to better constrain key trace gases such as CO. In parallel, theoretical modeling must evolve to capture the complex coupling between atmospheric chemistry and interior processes, while three-dimensional atmospheric dynamics models will be essential for understanding cloud formation, cold trap efficiency, and their impact on planetary albedo. Together, these efforts will transform our ability to interpret the atmospheric fingerprints of temperate sub-Neptunes. With K2-18~b as a trailblazer, the discovery of a potentially habitable world is no longer a distant aspiration but an emerging scientific reality.
 
\section*{Acknowledgments}
The authors thank Sara Seager and Jonathan Lunine for valuable discussions on the broad interpretation of the results, Luis Welbanks for helpful discussions on spectral retrievals, Geronimo Villanueva and Tyler Robinson for insightful input on molecular spectroscopy, and Nicholas Wogan for insight into numerical treatments of molecular diffusion. This work is based in part on observations made with the NASA/ESA/CSA James Webb Space Telescope, and the observations are associated with Program 2372 (PI: Renyu Hu) and 2722 (PI: Nikku Madhusudhan). The data were obtained from the Mikulski Archive for Space Telescopes at the Space Telescope Science Institute. Part of the research was carried out at the Jet Propulsion Laboratory, California Institute of Technology, under a contract with the National Aeronautics and Space Administration. NM and SC acknowledge support from the UK Research and Innovation (UKRI) Frontier Research Grant (EP/X025179/1; PI: N. Madhusudhan). 

Author contributions. RYH conceived the study, led the observations and modeling, performed high-level science analyses, and wrote the paper. ABA performed the primary reduction of the NIRSpec data, AT performed the primary spectral retrievals, JY performed the self-consistent modeling, and MD conducted the observation planning and advised on spectral retrievals. PAR provided the secondary reduction of the NIRSpec data and performed the SCARLET spectral retrievals, and LPC performed the reduction of the NIRISS data. NM and SC conducted the AURA spectral retrievals, and BB advised on data reduction and analyses. All authors commented on the general narrative of the paper.

Data availability. All the {\it JWST} data used in this paper can be found in MAST: \dataset[10.17909/9gr6-6388]{http://dx.doi.org/10.17909/9gr6-6388}. The transmission spectra derived from the JWST observations presented in this study are publicly available at https://osf.io/hpu8g/ (DOI: 10.17605/OSF.IO/HPU8G). Additional data supporting the findings of this study are available from the corresponding author upon reasonable request.

\appendix
\renewcommand{\thefigure}{A\arabic{figure}}
\setcounter{figure}{0}
\renewcommand{\thetable}{A\arabic{table}}
\setcounter{table}{0}

\section{Comparison between data reduction methods}
\begin{figure*}[!htbp]
  \centering
  \subfigure{\includegraphics[trim={15 15 45 15},clip, width=0.495\textwidth]{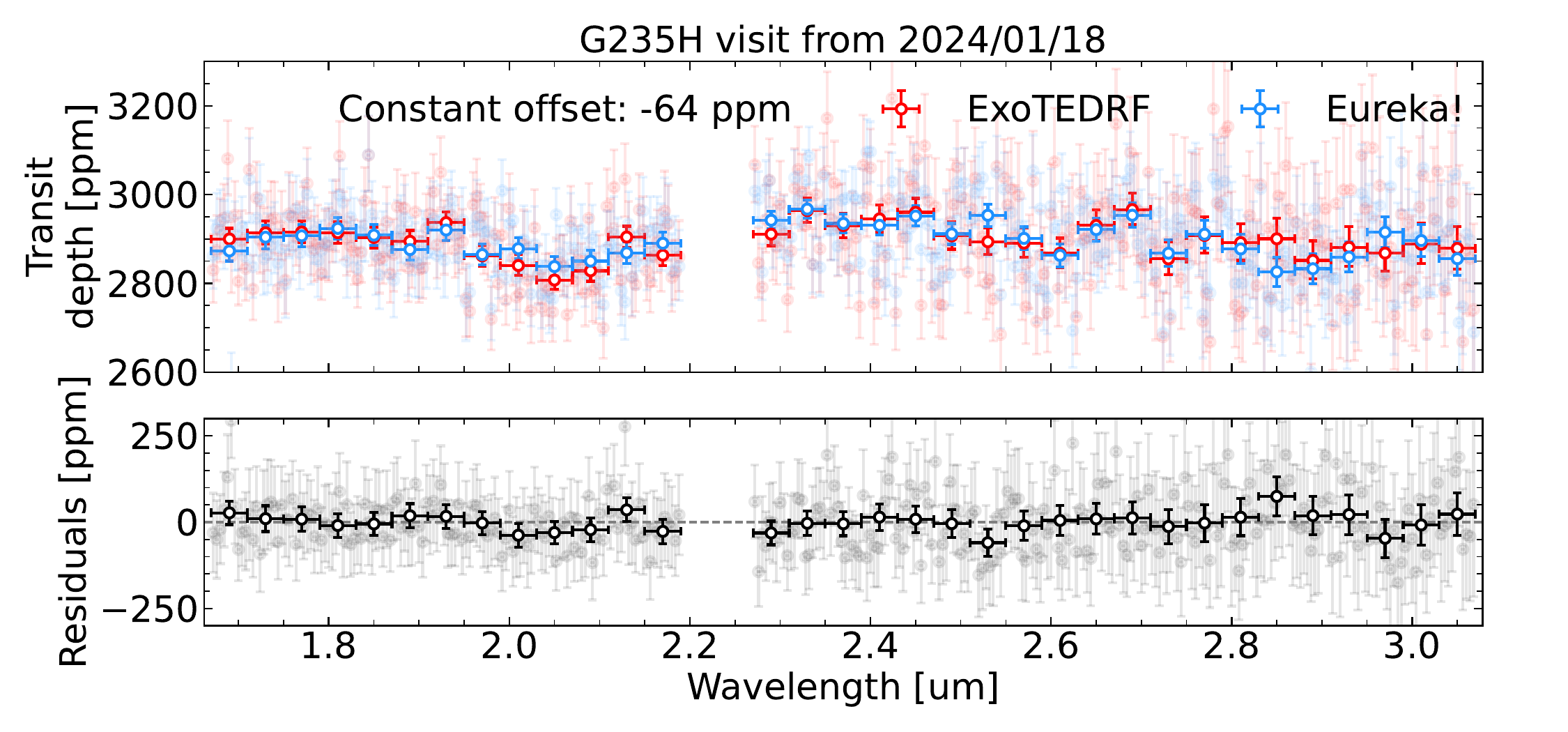}}
  \hfill
  \subfigure{\includegraphics[trim={15 15 45 15},clip, width=0.495\textwidth]{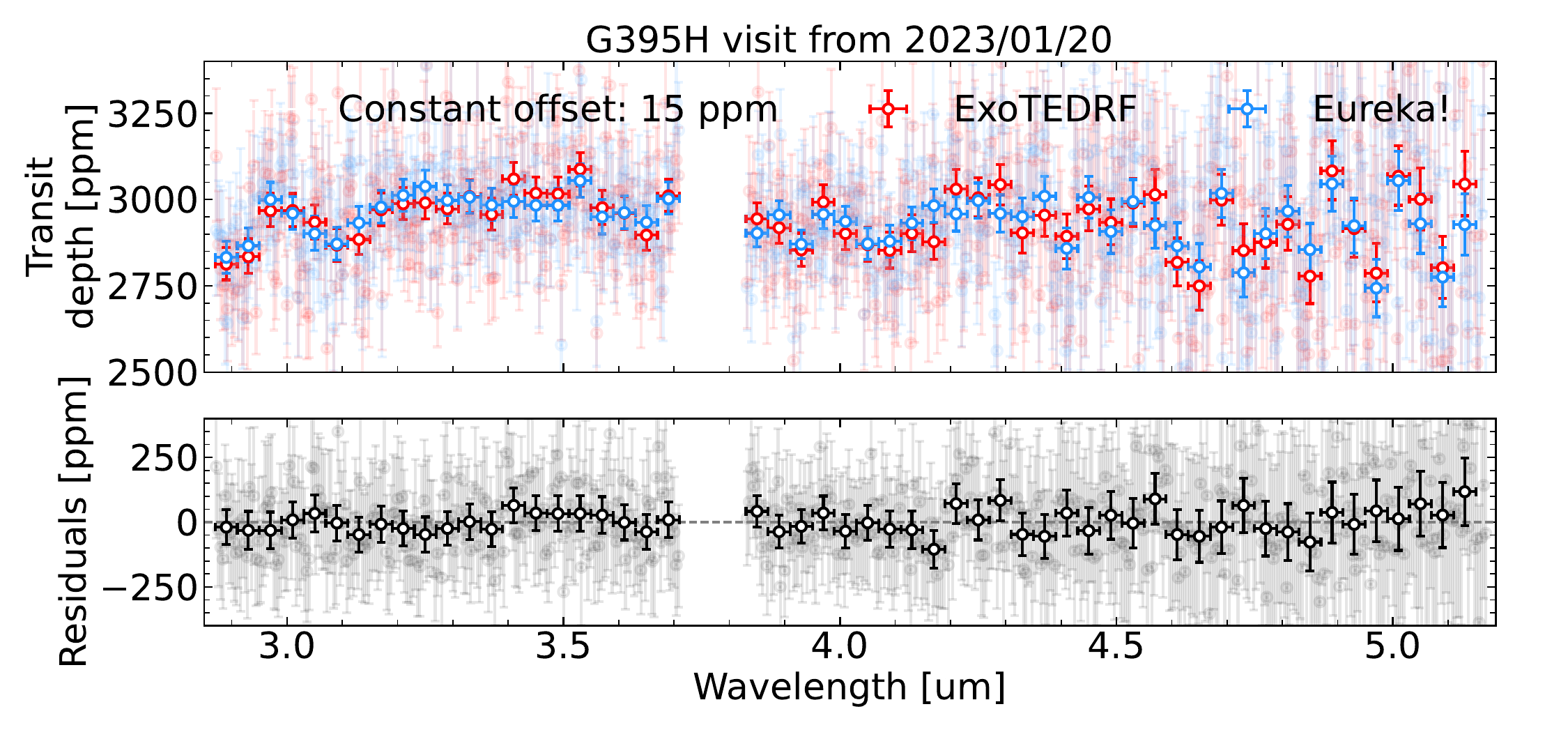}}

  \subfigure{\includegraphics[trim={15 15 45 15},clip, width=0.495\textwidth]{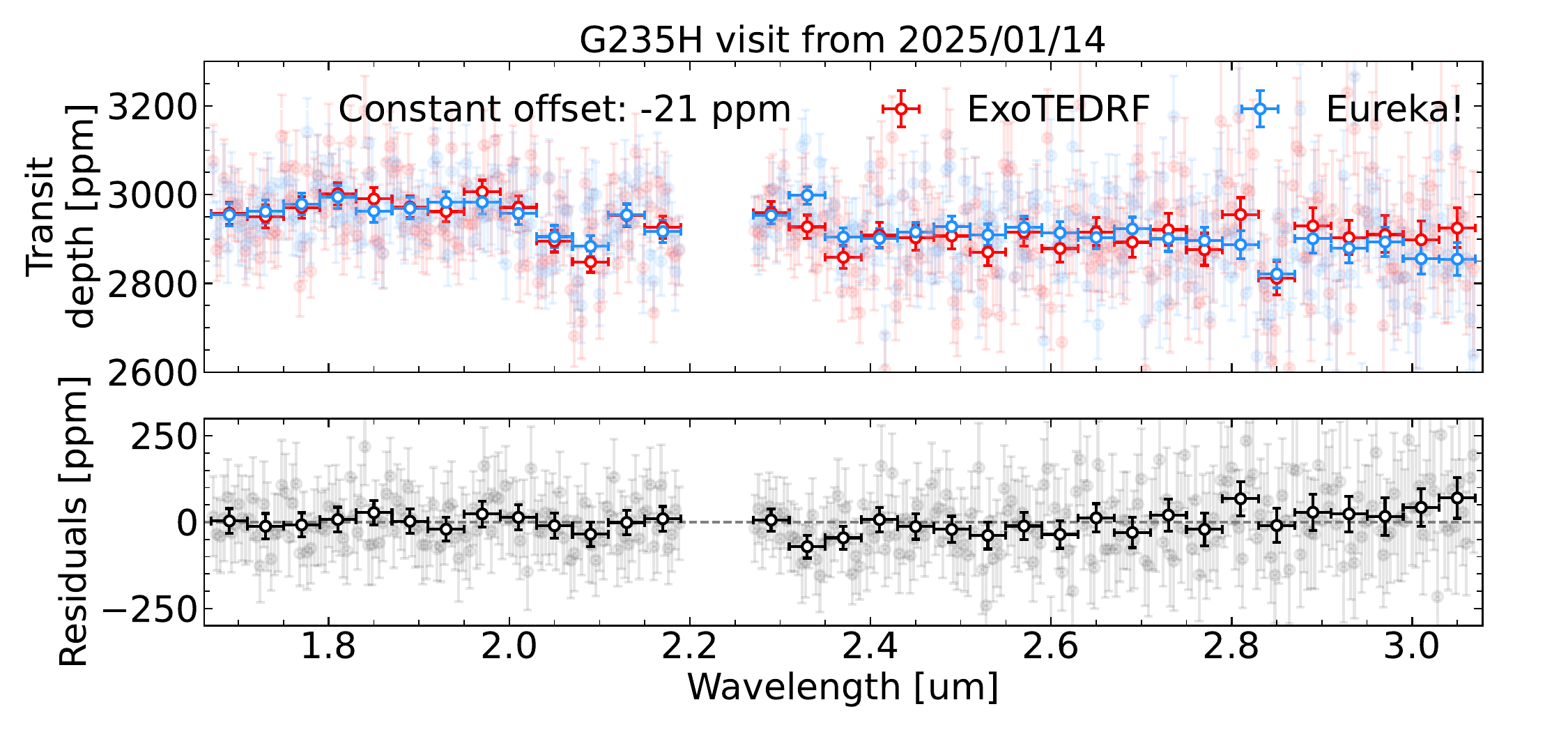}}
  \hfill
  \subfigure{\includegraphics[trim={15 15 45 15},clip, width=0.495\textwidth]{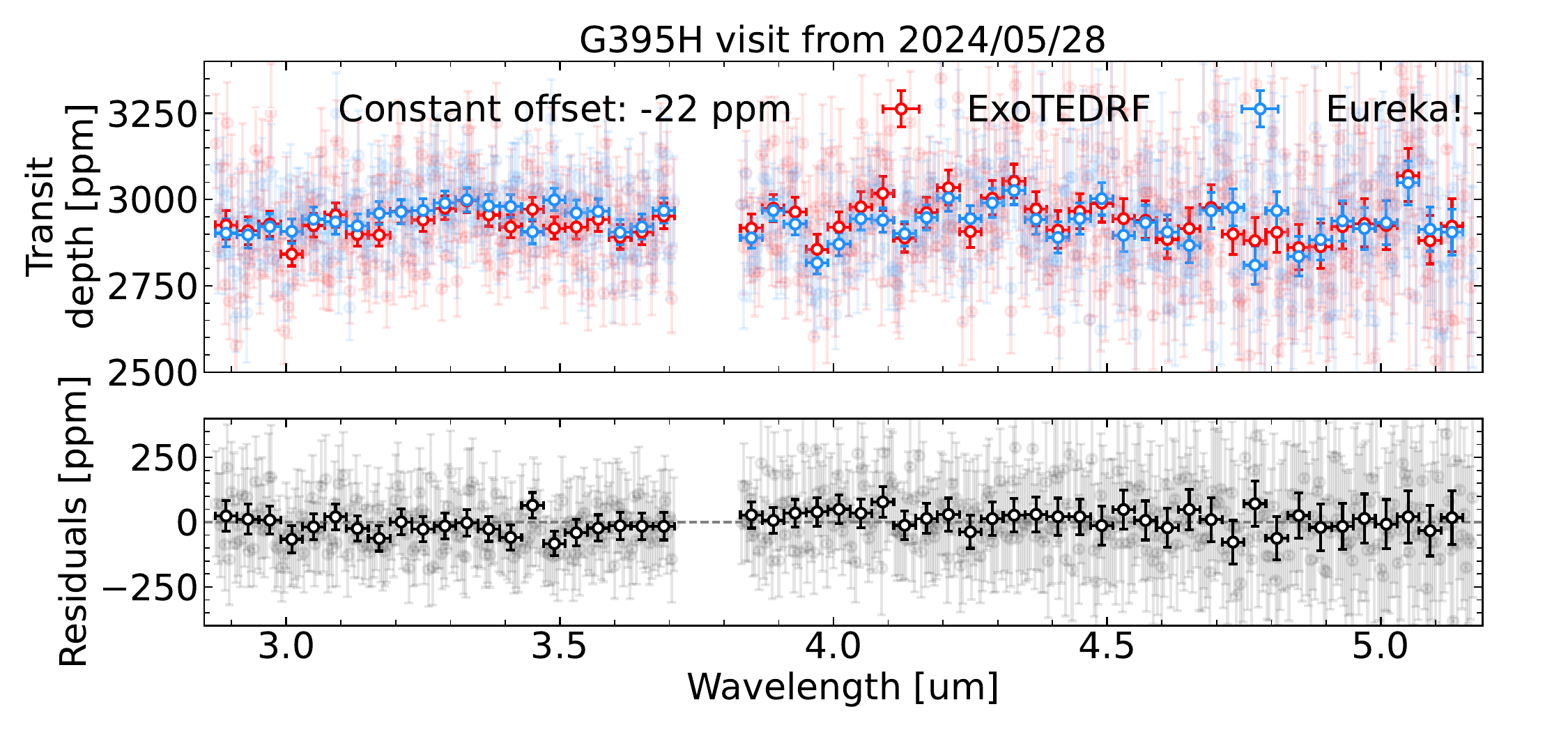}}

  \hfill
  \subfigure{\includegraphics[trim={15 15 45 15},clip, width=0.495\textwidth]{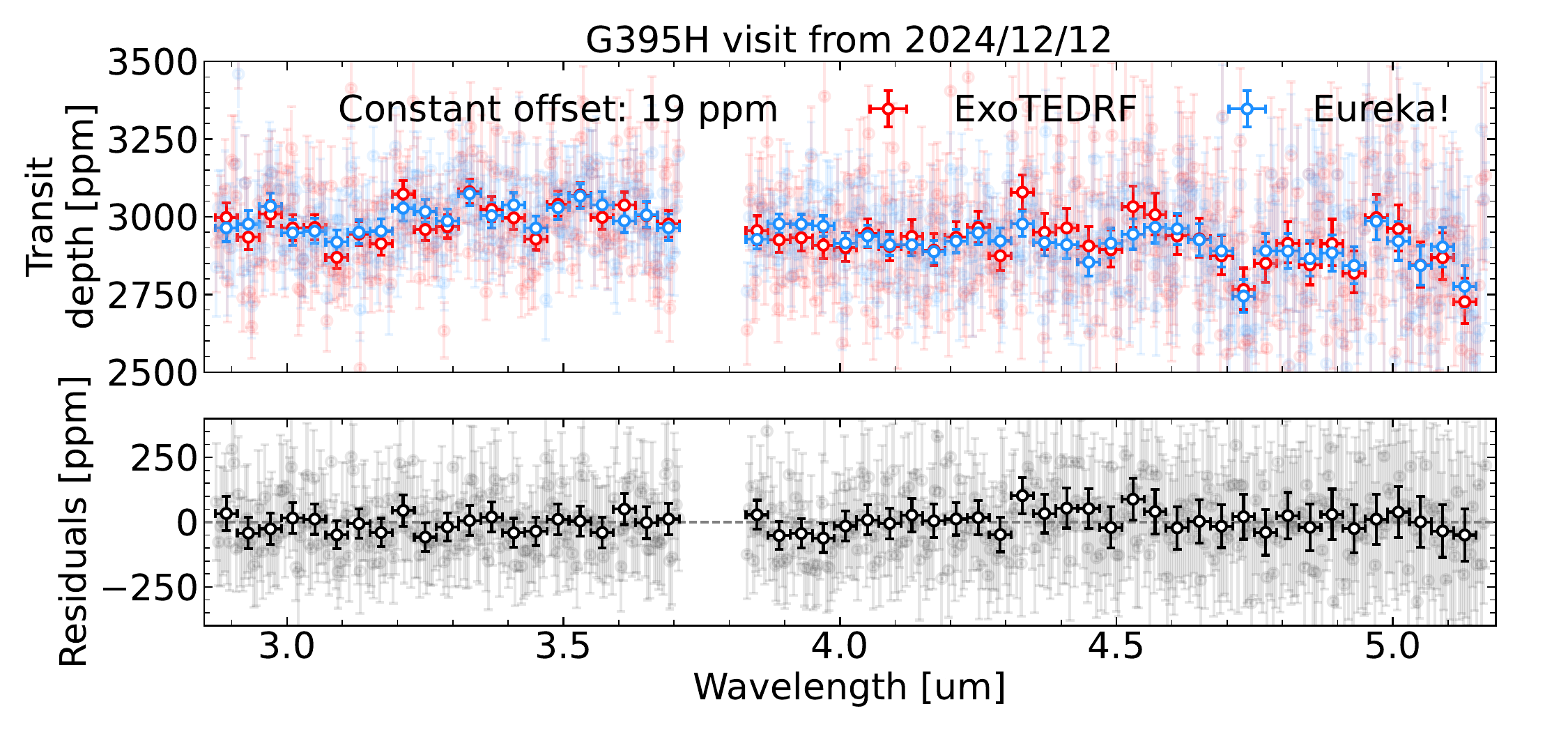}}

\caption{Comparison between the transmission spectra of K2-18~b obtained using the Eureka! and ExoTEDRF data reduction frameworks. The transmission spectra of K2-18~b are shown for the two NIRSpec/G235H visits in the left panels, and for the three NIRSpec/G395H visits in the right panels, with the date of each visit indicated on top of the panels. For all visits, the Eureka! (red) and ExoTEDRF (blue) transmission spectra are shown using the 0.004\,$\mu$m binning scheme (light), and a binned 0.04\,$\mu$m version (bold) is also shown on top for clarity. The corresponding residuals are shown in the bottom panels. A constant offset (due to different orbital parameters) is removed between the spectra displayed, and is indicated in the top left of each panel. The agreement between the transmission spectra obtained using both data analysis methods is strong, and there is no discrepant spectral signals between different reductions of the same visit. The offsets and subtle ($<$1$\sigma$) slopes that exist in the residuals for some of the detectors and visits are likely due to the different orbital parameters, and to the different limb darkening parameterizations.}
\label{fig:compare_reduction}
\end{figure*}

\clearpage

\section{Comparison between data reduction methods for NIRISS observations}
\begin{figure*}[!htbp]
  \centering
  \includegraphics[width=0.8\textwidth]{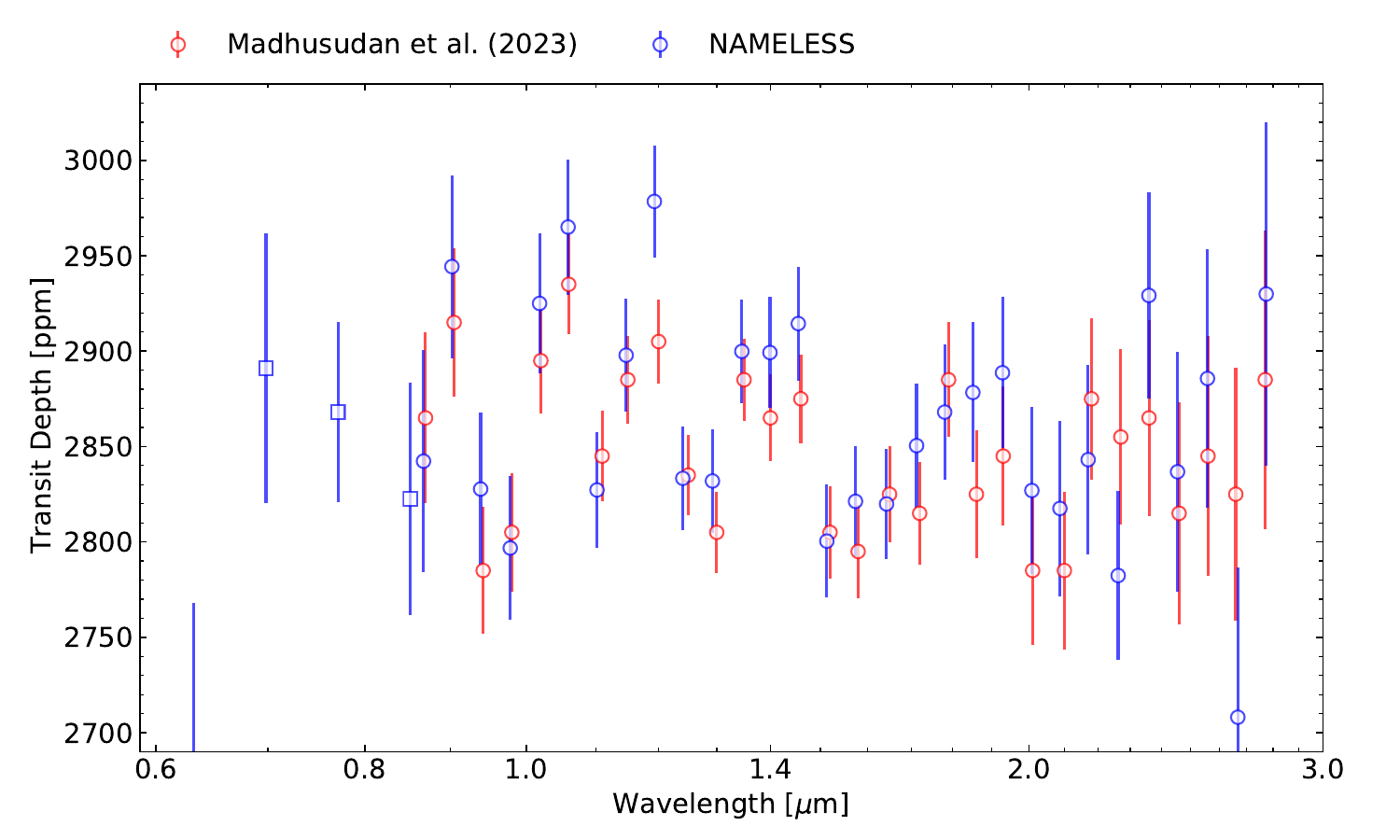}

\caption{Comparison between the transmission spectra of K2-18~b obtained using the JExoRES \citep{Holmberg_2023,madhusudhan2023carbon} and NAMELESS \citep[][this work]{Coulombe2023,Coulombe2025highlyreflectivewhiteclouds} data reduction frameworks. The Order 1 and Order 2 spectra are binned at fixed resolving powers of $R=25$ and $R=15$ for clarity. The spectrum presented in \citet{madhusudhan2023carbon} did not include the Order 2 data and is shifted by -75\,ppm to account for an apparent offset between the two reductions, potentially caused by the difference in the orbital parameters considered for the light curve fits. The two spectra show similar structure, with the NAMELESS spectrum exhibiting slightly larger final uncertainties possibly due to the difference in data reduction methods.}
\label{fig:compare_reduction_NIRISS}
\end{figure*}

\section{Shifted Average of Multiple Visit Transit Spectra}
\label{sec:shiftedaverage}

By visually comparing the spectra between the two NIRSpec/G235H visits and among the three NIRSpec/G395H visits, we noticed potential inconsistencies. Using a simple $\chi^2$ statistical metric, we found that the spectra from Visits B1 and B2 are likely inconsistent. For G395H, the spectrum from Visit C2 is probably inconsistent with those from C1 or C3, while the spectra from Visits C1 and C3 appear to be consistent with each other.

This observation motivated us to consider a more appropriate approach for combining spectra from repeated visits. We first examined whether stellar activity could introduce significant variability into the spectra. To assess this possibility, we performed spectral retrievals for each visit individually. In these retrievals, we fit only the offsets between the two detectors (NRS1 and NRS2) and the parameters associated with stellar heterogeneity -- namely, the heterogeneity fraction, heterogeneity temperature, and photosphere temperature. The results show consistent values for the heterogeneity and photosphere temperatures across all visits, while the heterogeneity fraction varies but remains within the uncertainties (Table~\ref{table:offset_stellar_exotr}). These findings suggest that changes in stellar heterogeneity are unlikely to be responsible for the observed spectral discrepancies between visits.

Meanwhile, we noticed that the offsets between the NRS1 and NRS2 detectors vary significantly across visits (Table~\ref{table:offset_stellar_exotr}). For example, the NRS1–NRS2 offset in Visit C2 appears to differ from those in Visits C1 and C3, which could contribute to the apparent discrepancies among their spectra. To address this, we manually adjusted the NRS2 spectrum of Visit C2 by –50 ppm to align its NRS1–NRS2 offset with those of Visits C1 and C3. Similarly, we applied a +60 ppm offset to the NRS2 spectrum of Visit B2 to match the NRS1–NRS2 offset of Visit B1.

After these corrections, we computed the mean transit depth for each visit, incorporating both NRS1 and NRS2 data, and applied an additional offset to the entire spectrum of each visit to match these mean levels. The required adjustments were approximately 20 ppm between Visits B1 and B2, and less than 10 ppm among Visits C1, C2, and C3. These corrections substantially reduced the $\chi^2$ values between the G235H visits and among the G395H visits, resulting in corrected spectra that are mutually consistent across repeated observations (Table~\ref{table:offset_stellar_exotr}).

       \begin{deluxetable}{cccccccc}
		\tablecaption{Results of fitting individual visits with stellar heterogeneity and offset between detectors.}
		\tablehead{
			\colhead{Instrument} & \colhead{Visit} & \colhead{Offset (ppm)}& \colhead{$\delta_{\text{het}}$}& \colhead{$T_{\text{het}}$ (K)}&\colhead{$T_{\text{phot}}$ (K)} &\colhead{$\chi^2_0$}&\colhead{$\chi^2_{d}$}}
		\startdata
        G235H & B1 & -15.30$^{+9.4}_{-8.8}$&0.22$^{+0.17}_{-0.11}$&3659.14$^{+104.03}_{-80.97}$&3477.82$^{+59.46}_{-65.81}$&1.97&1.05\\
        G235H & B2 & 44.43$^{+4.05}_{-4.61}$&0.12$^{+0.22}_{-0.08}$&3636.45$^{+113.01}_{-73.24}$&3506.12$^{+61.80}_{-47.83}$&--&--\\
        G395H & C1 & 82.98$^{+11.37}_{-19.06}$&0.15$^{+0.18}_{-0.09}$&3705.47$^{+260.07}_{-190.39}$&3450.26$^{+98.84}_{-106.2}$&1.34&1.06\\
        G395H & C2 & 38.12$^{+17.63}_{-18.40}$&0.07$^{+0.19}_{-0.05}$&3629.09$^{+321.82}_{-269.50}$&3464.46$^{+88.36}_{-94.80}$&--&--\\
        G395H & C3 & 90.98$^{+6.49}_{-10.99}$& 0.12$^{+0.17}_{-0.06}$&3756.16$^{+227.09}_{-177.71}$&3468.26$^{+82.78}_{-95.61}$&1.79&1.15\\
		\enddata
       \label{table:offset_stellar_exotr}
       \tablecomments{ The Offset column reports the retrieved offset between NRS1 and NRS2 for each visit. $\chi^2_0$ denotes the initial (pre-shift) reduced chi-squared value comparing the spectrum of that visit to the second visit of the same instrument. $\chi^2_d$ is the chi-squared value after applying shifts to account for differences in NRS1–NRS2 offsets and in the mean transit depth between visits.}
      \end{deluxetable}

Finally, we combined the shifted spectra in the standard fashion. Given the methodology that produced these datasets, we refer to them as ``shifted average'' data. Since we manually applied NRS1–NRS2 offsets to align with one of the visits, we treated each detector separately when using the shifted average data in spectral retrievals, allowing the NRS1–NRS2 offsets in the averaged data to be fit freely together with the atmospheric parameters.

\section{Retrievals using different combinations of visits and model frameworks}
\label{sec:retrievalsensitivity}

\begin{figure*}[!htbp]
\centering
\includegraphics[trim=5 5 5 5, clip,width=\textwidth]{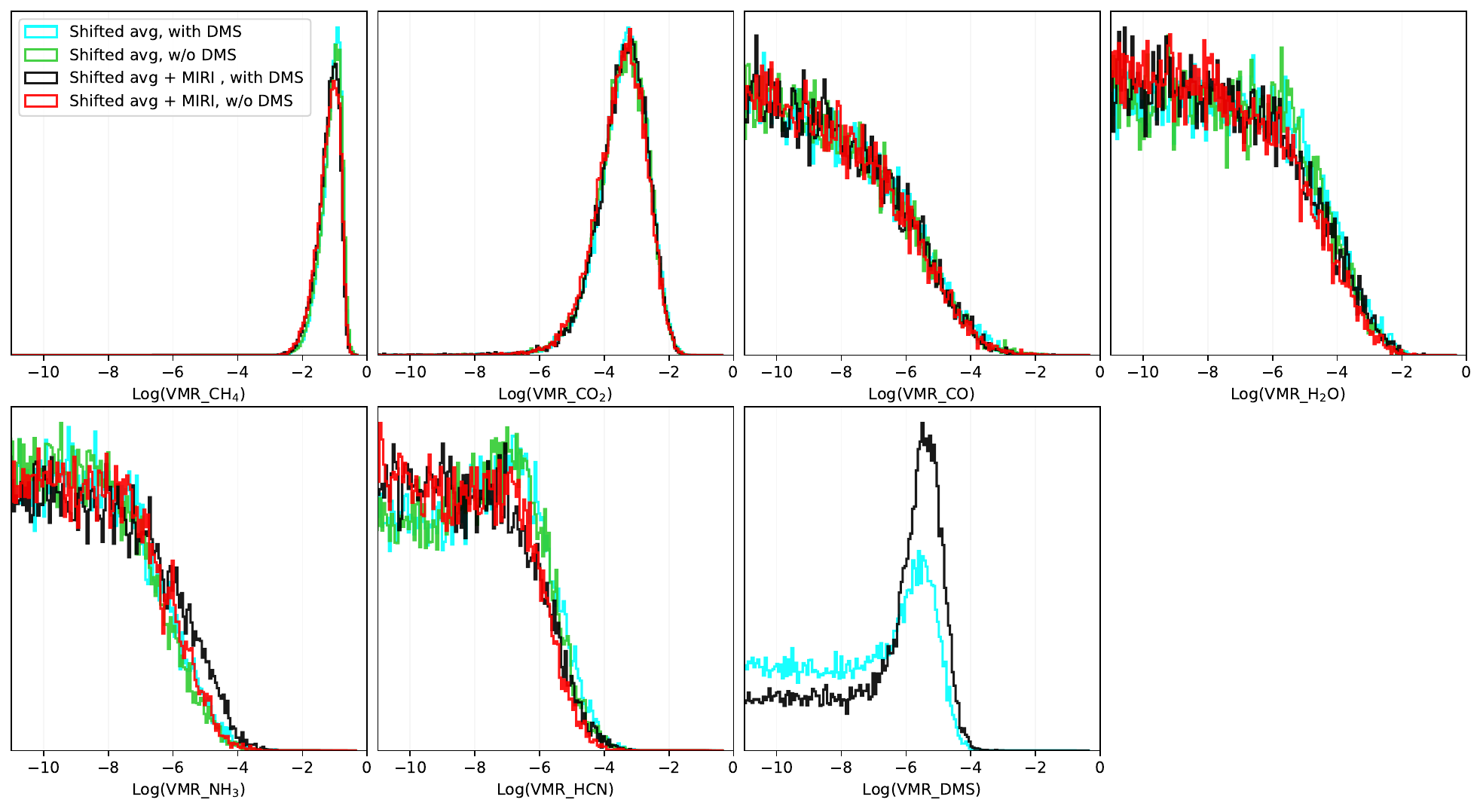}
\caption{
Posterior distributions from \exotr\ spectral retrievals of K2-18~b's transmission spectra, comparing results with and without the inclusion of the MIRI/LRS dataset. Incorporating the MIRI/LRS data yields a weaker indication of HCN and a stronger, though still tentative, signal for DMS, while the posterior distributions of other gases remain largely unchanged. Notably, the DMS posterior retains a long tail toward zero abundance, underscoring the marginal nature of the signal.
}
\label{fig:exotr_posterior_miri}
\end{figure*}

\begin{figure*}[!htbp]
\centering
\includegraphics[trim=5 5 5 5, clip,width=\textwidth]{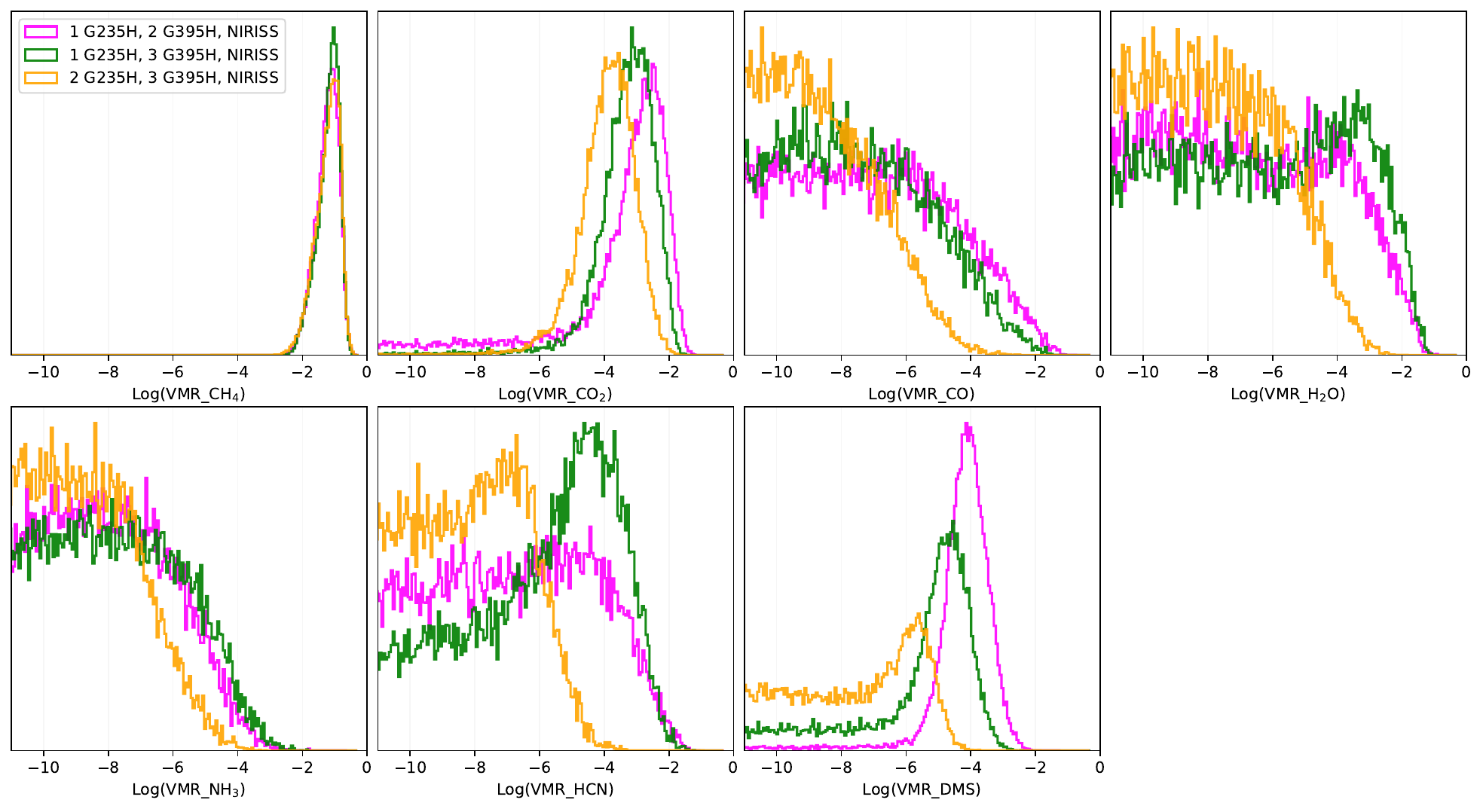}
\caption{Posterior distributions from \exotr\ spectral retrievals performed on different combinations of K2-18~b observations: 4 visits (magenta), 5 visits (green), and 6 visits (orange) as defined in Table~\ref{table:cases_by_visits}. The data are directly combined for these retrievals. As additional visits are included, the constraints on CO$_2$ improve, the upper limits on unconstrained molecules become tighter, and the detection significance decreases for DMS.
}
\label{fig:exotr_posterior_by_visit}
\end{figure*}

\begin{figure*}[!htbp]
\centering
\includegraphics[trim=5 5 5 5, clip,width=\textwidth]{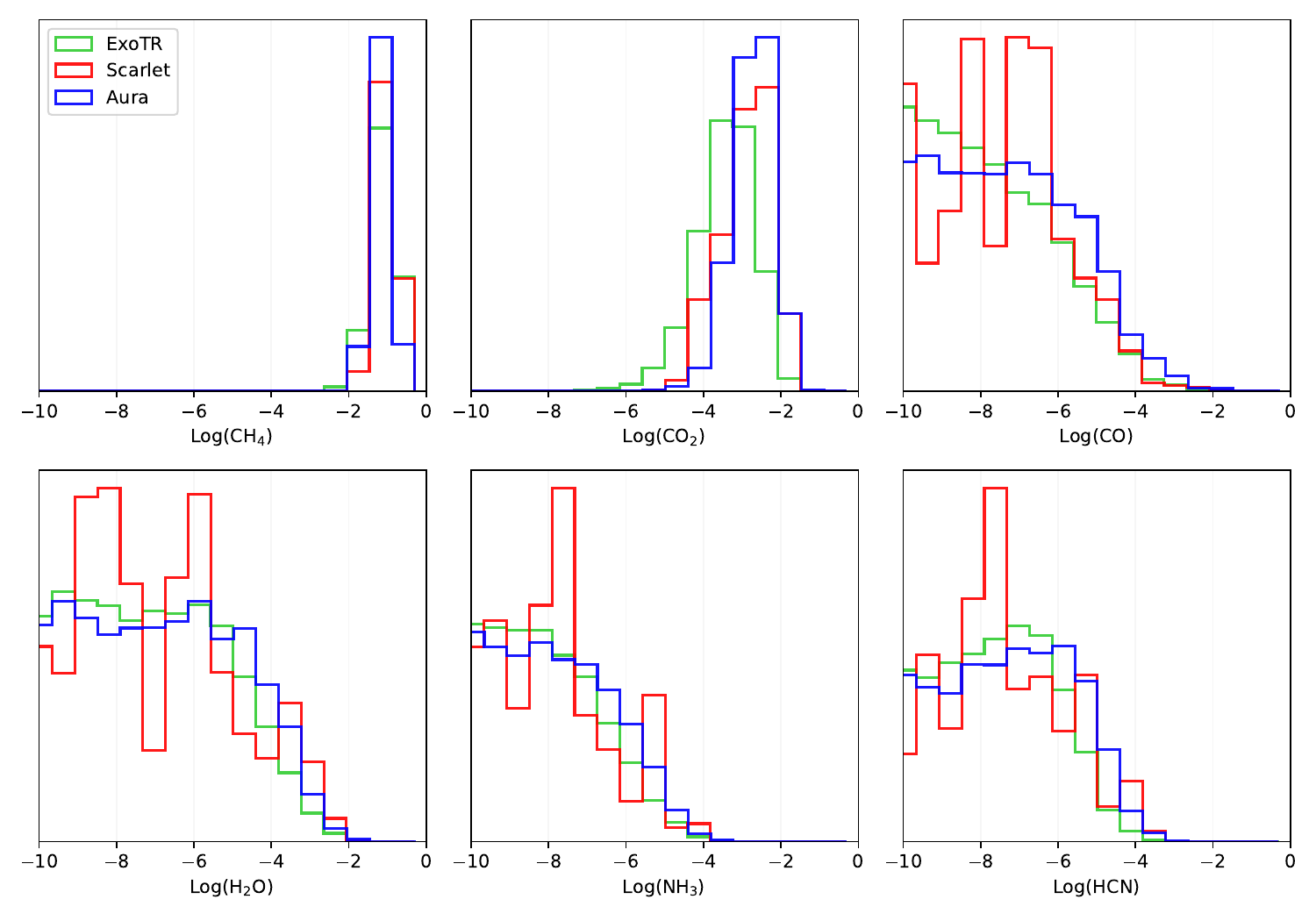}
\caption{Posterior 1D histograms for the baseline retrieval case across three retrieval frameworks used in this study. There is good agreement in the retrieved molecular abundances, confirming the consistency in constraints shown in Figure~\ref{fig:compare_constraints}. 
}
\label{fig:baseline_posteriors_all}
\end{figure*}

\section{$P-T$ and $K_{\rm zz}$ profiles for atmospheric chemistry modeling} \label{sec:photochemical_inputs_profile}

\begin{figure*}[!htbp]
\centering
\includegraphics[width=0.5\textwidth]{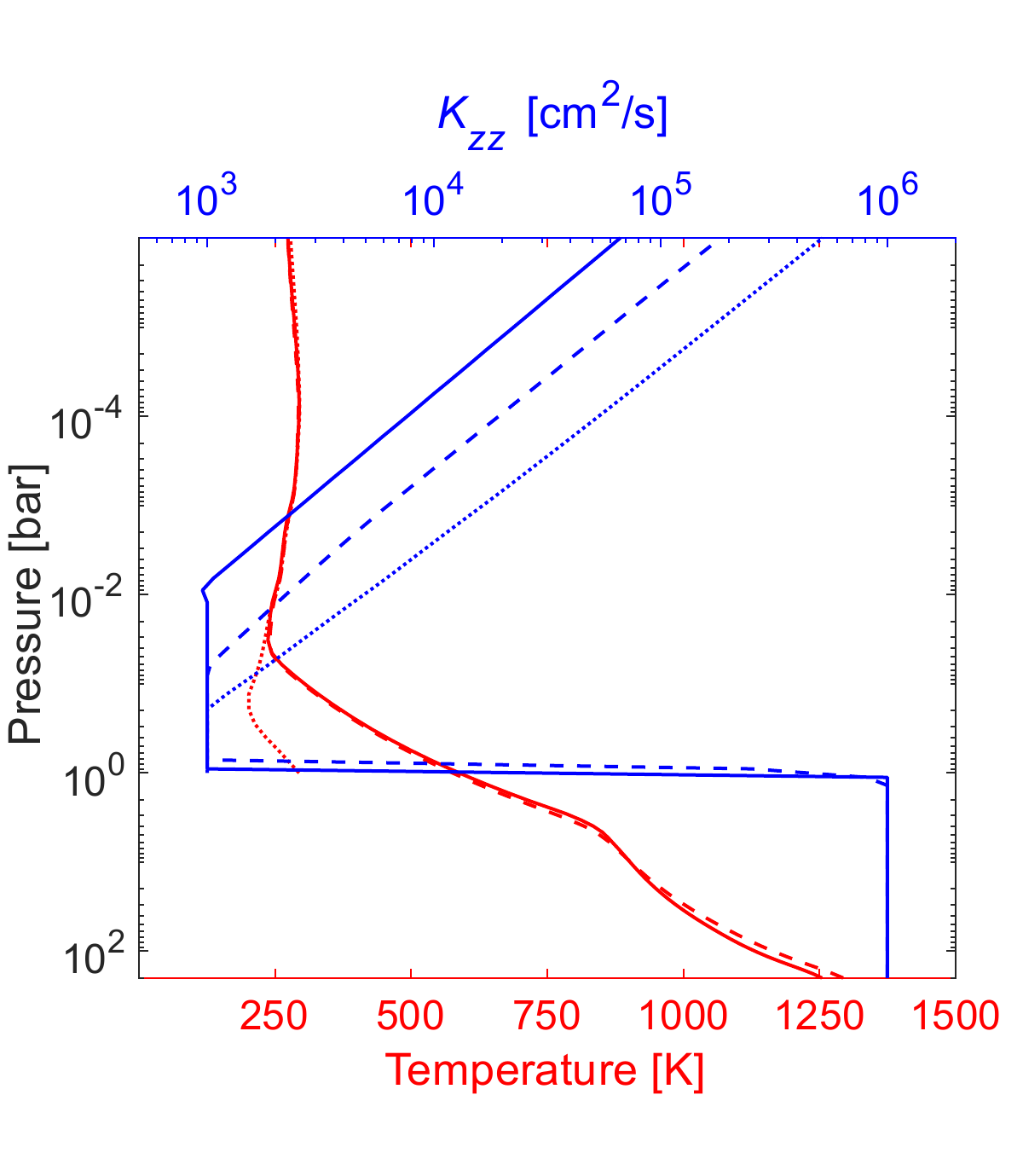}
\caption{Pressure–temperature (\textit{P--T}, red) and eddy diffusion coefficient ($K_{\rm zz}$, blue) profiles used in the atmospheric chemistry modeling of K2-18~b. The solid red line represents the standard $100\times$ solar metallicity atmosphere adopted from \citet{yang2024chemical}, while the dashed red line corresponds to a more water-rich envelope with a \ce{H2O}-to-\ce{H2} ratio of 25:75. The dotted red line shows the \textit{P--T} profile simulated in this work for a thin, 1-bar atmosphere, with \ce{CH4} and \ce{CO2} mixing ratios informed by the transmission spectra. For the $K_{\rm zz}$ profiles, we adopt a deep-atmosphere value of $10^6$ cm$^2$ s$^{-1}$ \citep{zhang2018global1}, a minimum of $10^3$ cm$^2$ s$^{-1}$ near the tropopause, and an increase in the stratosphere following a $n^{-1/2}$ dependence on number density, as described in \citet{hu2021photochemistry}. The solid and dashed blue lines correspond to different assumptions about the onset pressure of increased mixing in the massive atmosphere cases. The blue dotted line shows the $K_{\rm zz}$ profile used for the thin atmosphere scenario.
}
\label{fig:model_inputs}
\end{figure*}

\section{Constrained retrieval tests for H$_2$O, NH$_3$, and CO}

\begin{figure*}[!htbp]
\centering
\includegraphics[width=0.8\textwidth]{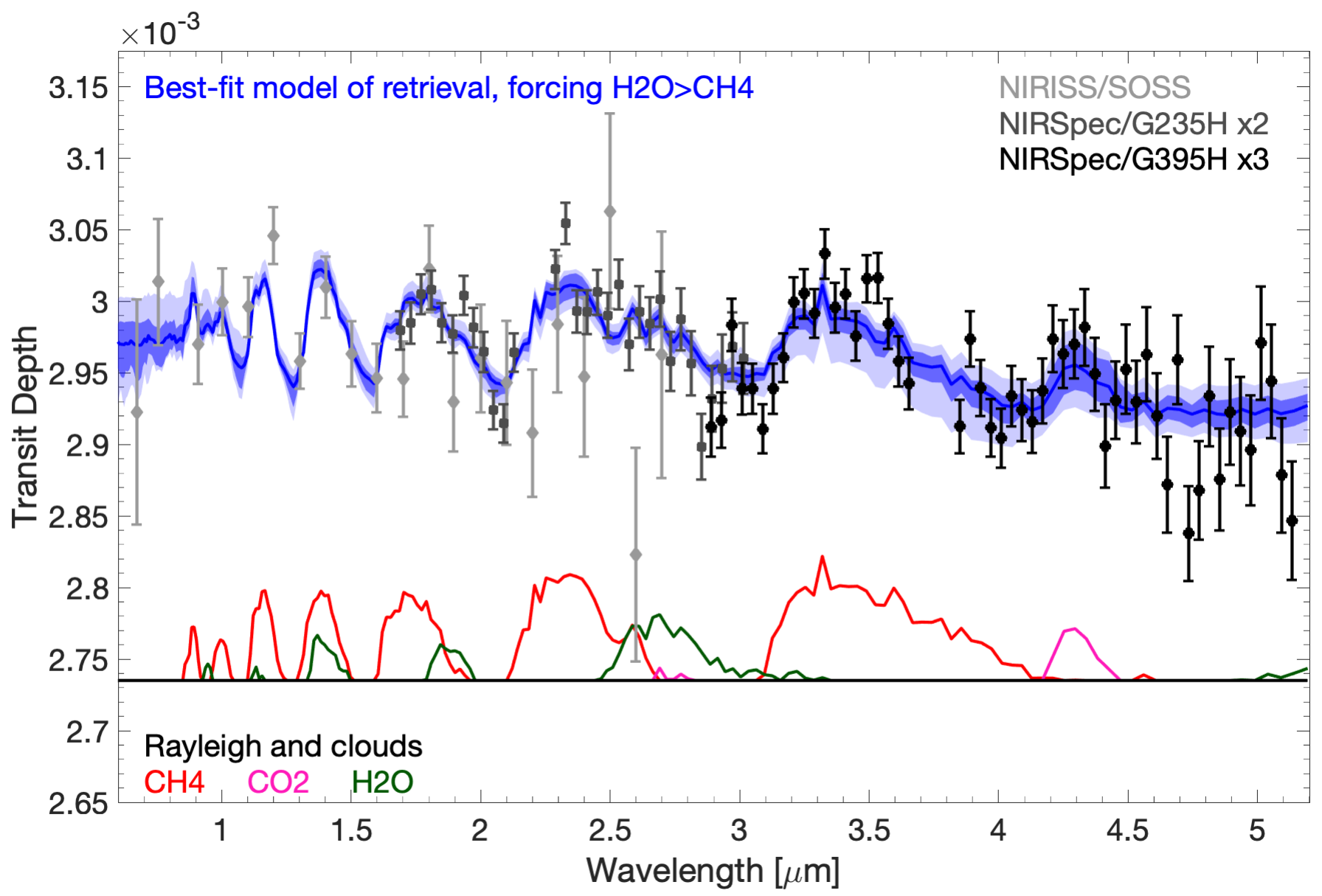}
\caption{Same format as Figure~\ref{fig:model_fit} but showing the best-fit models when H$_2$O is forced to be more than CH$_4$ in the spectral retrieval using \exotr. The best-fit model in this constrained retrieval involves a cloud deck at $\sim0.02$ bar (compared to $>0.03$ bar in the free retrieval) and can achieve a $\chi^2/dof$ of 1503/1264 (compared to 1479/1264 in the free retrieval).}
\label{fig:h2oggch4}
\end{figure*}

\begin{figure*}[!htbp]
\centering
\includegraphics[width=0.8\textwidth]{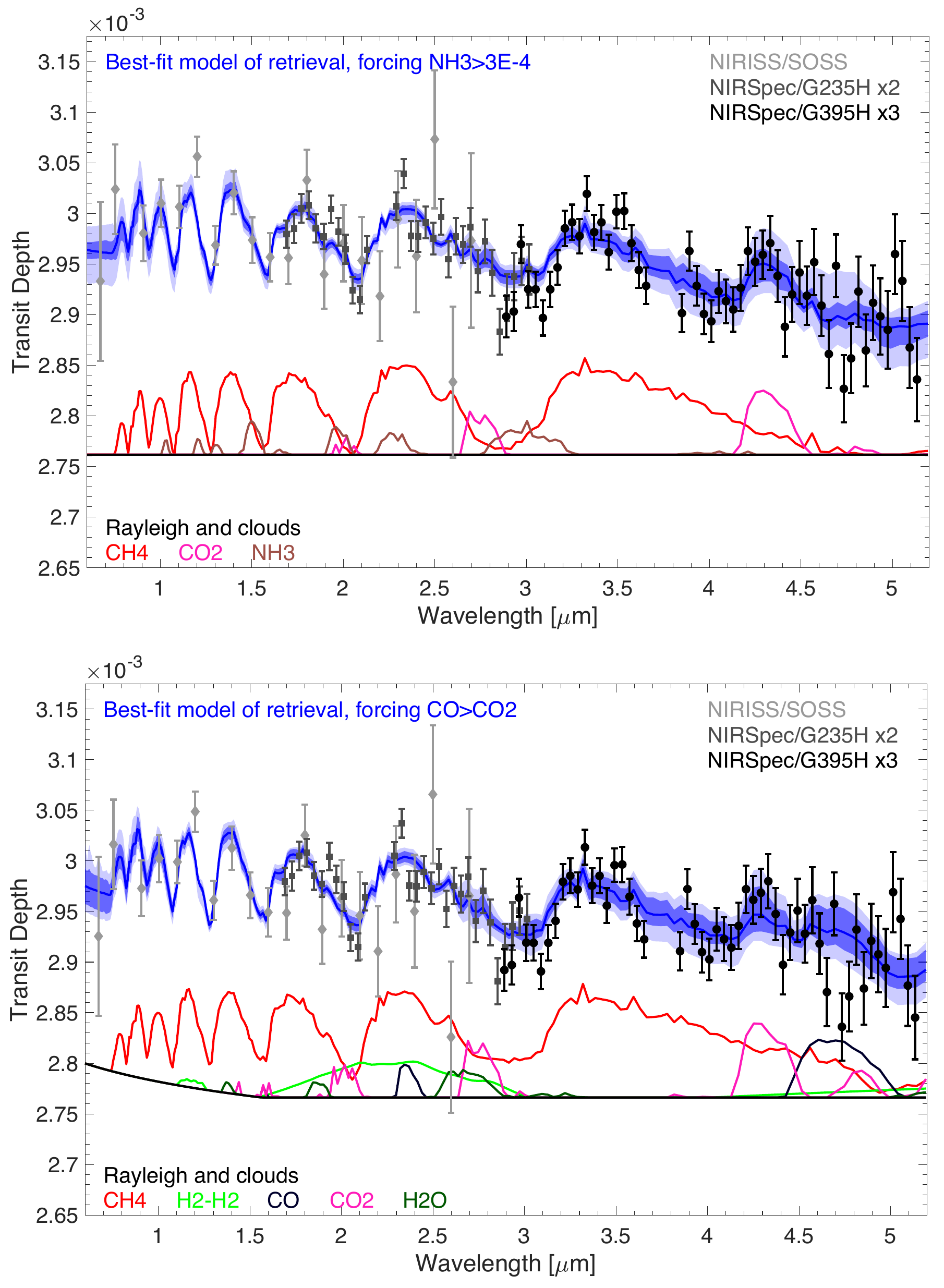}
\caption{Same format as Figure~\ref{fig:model_fit} but showing the best-fit models when NH$_3$ is forced to be more than a $3\times10^{-4}$ mixing ratio (upper panel), or CO is forced to be more than CO$_2$ (lower panel). 
The best-fit model with constrained NH$_3$ puts the NH$_3$ mixing ratio very close to the minimum and incurs a cloud deck at $\sim0.03$ bar. It achieves a $\chi^2/dof$ of 1491/1264 (compared to 1479/1264 in the free retrieval).
The retrieval with constrained CO generates a posterior where the CO mixing ratio traces the CO$_2$ mixing ratio in a nearly 1:1 correlation, and the best-fit model has the CO mixing ratio at $10^{-3.3}$ and CO$_2$ at $10^{-3.6}$. The best-fit model in this case has a deep cloud (like in the free retrieval) and achieves a $\chi^2/dof$ of 1487/1264 (compared to 1479/1264 in the free retrieval).
}
\label{fig:nh3co}
\end{figure*}

\end{document}